\documentclass[aps,preprint,showkeys,superscriptaddress,nofootinbib]{revtex4-1}
\usepackage{bbm}
\usepackage{amsmath}
\usepackage{graphicx}
\usepackage{simplewick}
\newcommand{\Slash}[1]{{\ooalign{\hfil/\hfil\crcr$#1$}}}
\DeclareMathOperator{\re}{Re}

\def\simge{\mathrel{%
       \rlap{\raise 0.511ex \hbox{$>$}}{\lower 0.511ex \hbox{$\sim$}}}}
\def\simle{\mathrel{
       \rlap{\raise 0.511ex \hbox{$<$}}{\lower 0.511ex \hbox{$\sim$}}}}

\begin{document} 

\title{\boldmath Exploring $N_f=2+1$ QCD thermodynamics from the gradient flow}

\preprint{UTHEP-691, UTCCS-P-91, J-PARC-TH-0064, KYUSHU-HET-167}

\author{Yusuke Taniguchi}
\thanks{Corresponding author}
\email{tanigchi@het.ph.tsukuba.ac.jp}
\affiliation{Center for Computational Sciences (CCS), University of Tsukuba,
Tsukuba, Ibaraki 305-8571, Japan}
\author{Shinji Ejiri}
\email{ejiri@muse.sc.niigata-u.ac.jp}
\affiliation{Department of Physics, Niigata University, Niigata 950-2181,
Japan}
\author{Ryo Iwami}
\email{iwami@muse.sc.niigata-u.ac.jp}
\affiliation{Graduate School of Science and Technology, Niigata University,
Niigata 950-2181, Japan}
\author{Kazuyuki Kanaya}
\email{kanaya@ccs.tsukuba.ac.jp}
\affiliation{Center for Integrated Research in Fundamental Science and
Engineering (CiRfSE), University of Tsukuba, Tsukuba, Ibaraki 305-8571, Japan}
\author{Masakiyo Kitazawa}
\email{kitazawa@phys.sci.osaka-u.ac.jp}
\affiliation{Department of Physics, Osaka University, Osaka 560-0043, Japan}
\affiliation{J-PARC Branch, KEK Theory Center, Institute of Particle and Nuclear Studies,
KEK, 203-1, Shirakata, Tokai, Ibaraki, 319-1106, Japan}
\author{Hiroshi Suzuki}
\email{hsuzuki@phys.kyushu-u.ac.jp}
\affiliation{Department of Physics, Kyushu University, 744 Motooka, Fukuoka
819-0395, Japan}
\author{Takashi Umeda}
\email{tumeda@hiroshima-u.ac.jp}
\affiliation{Graduate School of Education, Hiroshima University,
Higashihiroshima, Hiroshima 739-8524, Japan}
\author{Naoki Wakabayashi}
\email{wakabayashi@muse.sc.niigata-u.ac.jp}
\affiliation{Graduate School of Science and Technology, Niigata University,
Niigata 950-2181, Japan}
\collaboration{WHOT-QCD Collaboration}
\noaffiliation

\date{\today}

\begin{abstract}
The energy-momentum tensor plays an important role in QCD thermodynamics.
Its expectation value contains information of the pressure and the
energy density as its diagonal part. Further properties like viscosity and
specific heat can be extracted from its correlation function. A
nonperturbative evaluation of it on the lattice is called. 
Recently a new method based on the gradient flow 
was introduced to calculate the energy-momentum tensor on the lattice, and 
has been successfully applied to quenched QCD. 
In this paper, we apply the gradient flow method to calculate the energy-momentum tensor in $(2+1)$-flavor QCD 
adopting a nonperturbatively 
$O(a)$-improved Wilson quark action and the renormalization group-improved Iwasaki gauge action. 
As the first application of the method with dynamical quarks, we study at a single but fine lattice
spacing $a\simeq0.07\,\mathrm{fm}$ 
with heavy $u$ and $d$ quarks ($m_\pi/m_\rho\simeq0.63$) and approximately physical $s$ quark ($m_{\eta_{ss}}/m_\phi\simeq0.74$).
With the fixed-scale approach, temperature is varied by the temporal lattice size $N_t$ at a fixed lattice spacing. 
Performing simulations on lattices with $N_t=16$ to 4, the temperature
 range of $T\simeq174$--$697$ MeV is covered. 
We find that the results of the pressure and the energy density by the gradient flow
method are consistent with the previous results using the $T$-integration
method at $T \simle 280\,\mathrm{MeV}$ ($N_t\simge10$), while the results show disagreement 
at $T\simge350\,\mathrm{MeV}$ ($N_t \simle 8$), 
presumably due to the small-$N_t$ lattice artifact of $O\left((aT)^2\right)=O\left(1/N_t^2\right)$.

We also apply the gradient flow method to evaluate the chiral condensate 
taking advantage of the gradient flow method that renormalized quantities can be directly computed 
avoiding the difficulty of explicit chiral violation with lattice quarks. 
We compute the renormalized chiral condensate in the $\overline{\mathrm{MS}}$ scheme
at renormalization scale $\mu=2\,\mathrm{GeV}$ with a high precision to study the
temperature dependence of the chiral condensate and its disconnected susceptibility. 
Even with the Wilson-type quark action which violates the chiral symmetry explicitly, we obtain 
the chiral condensate and its disconnected susceptibility showing a clear signal of pseudocritical temperature at $T\sim190\,\mathrm{MeV}$ related to the chiral restoration crossover.
\end{abstract}

\maketitle

\section{Introduction}
\label{sec:intro}
Precise determination of thermodynamic properties of the quark matter is a key
step towards understanding the early evolution of the Universe as well as the
nature of neutron/quark stars. Numerical simulation of QCD on the lattice
provides us with the only way to study the nature of the strongly coupled
quark matter directly from the first principles of QCD. Recently, the
Yang-Mills gradient flow~\cite{Luscher:2009eq,Luscher:2010iy,Luscher:2011bx,%
Luscher:2013cpa,Narayanan:2006rf} has introduced big advances in numerical
determination of various observables in lattice QCD~\cite{Luscher:2013vga,%
Ramos:2015dla,Lat16suzuki}. Fields at positive flow time, $t>0$, can be viewed as smeared
fields averaged over a mean physical radius of~$\sqrt{8t}$ in four dimensions. Salient features
of the gradient flow are the UV finiteness and the absence of short-distance
singularities in the operators constructed by flowed fields at~$t>0$. This enables
us to directly construct renormalized quantities in terms of the flowed fields, \textit{i.e.},
a new renormalization scheme can be introduced by the gradient flow. 
Because the flowed fields are defined nonperturbatively, we can evaluate their 
nonperturbative expectation values directly on the lattice.
This opened us a large variety of possibilities to significantly simplify the
determination of physical observables on the lattice. 
In this paper, we determine the equation of state (EOS) as well as the chiral condensate 
in $(2+1)$-flavor QCD at finite temperatures, by applying the
methods of Refs.~\cite{Makino:2014taa,Hieda:2016lly} using the gradient flow.
The EOS in $(2+1)$-flavor QCD has been calculated at the physical point and extrapolated to the continuum limit using staggered-type lattice quarks ~\cite{Borsanyi:2013bia,Bazavov:2014pvz}.
To avoid theoretical uncertainties with staggered-type lattice quarks, calculations using Wilson-type lattice quarks have been also attempted~\cite{Umeda:2008bd,Burger:2014xga}.
See Ref.~\cite{Lat16HTDing} for the recent status and related developments on the lattice. 

In this study, we extract the EOS from the diagonal elements of the energy-momentum tensor, $T_{\mu\nu}(x)$. 
The energy-momentum tensor is the generator of continuous
coordinate translations and thus, is not uniquely given on discrete lattices as
a conserved current. An approach to overcome this problem is to use finite
observables, which are independent of the regularization in the continuum limit.
The gradient flow enables us to define such finite observables. Unfortunately
these finite renormalized tensor operators are not necessarily equal to the
conserved energy-momentum tensor but an appropriate combination of them can be
the energy-momentum tensor in a small flow time limit~\cite{Makino:2014taa,Suzuki:2013gza}.
In~Refs.~\cite{Makino:2014taa,Suzuki:2013gza}, coefficients needed to extract the
energy-momentum tensor which satisfies the Ward-Takahashi identity associated
with the translational invariance from appropriate flowed tensor operators were
calculated, by using a small flow time expansion of flowed
operators~\cite{Luscher:2011bx}. In this method, we observe several lattice
operators at small~$t$ and take their vanishing $t$ limit. The coefficients
relating these limiting values with the energy-momentum tensor is calculated
in a renormalized theory. They can be computed by perturbation theory using the
asymptotic freedom at small~$t$, and those for the quenched case are computed
in~Ref.~\cite{Suzuki:2013gza}. Finally, the EOS is given by $\epsilon=-\langle T_{00}\rangle$
and~$p=\sum_i \langle T_{ii}\rangle/3$, where $\epsilon$ and~$p$ are the energy
density and the pressure, respectively. Some other thermodynamic quantities, such
as the bulk and shear viscosities etc., can also be extracted from the
energy-momentum tensor. 
Here, we stress that, though these coefficients are computed in perturbation theory, 
they are used just to guide the $t\to0$ extrapolation. 
We thus consider that our evaluation of the energy-momentum tensor is essentially nonperturbative.\footnote{This idea has been tested in solvable
models~\cite{Makino:2014cxa,Suzuki:2015fka}.}

The method was tested in quenched QCD by the FlowQCD
Collaboration in~Ref.~\cite{Asakawa:2013laa}.  The resulting
EOS from the gradient flow shows a good agreement with the previous results of
the conventional integration and $T$-integration
methods~\cite{Boyd:1996bx,Okamoto:1999hi,Namekawa:2001ih,Levkova:2006gn,%
Umeda:2008bd,Borsanyi:2012ve}.

In this paper, we extend the study of the energy-momentum tensor and EOS to QCD with dynamical quarks, adopting the method
of~Ref.~\cite{Makino:2014taa}. The gradient flow in full QCD was
investigated by L\"uscher in Ref.~\cite{Luscher:2013cpa}. Because the
{\it raison d'etre} of the gradient flow in our study is the semilocal smearing of the
fields, it is not mandatory to introduce quarks in the dynamics of the flow
in $t$. A numerically easier way is to keep the quenched flow equations for the
gauge fields and combine them with a gauge-covariant quenched flow equation for
the quark fields~\cite{Luscher:2013cpa}. Fermionic operators, however, require
additional wave function renormalization of quark fields, which can be carried
out by normalizing the flowed quark fields by the vacuum expectation value of a
flowed quark kinetic operator at zero temperature~\cite{Makino:2014taa}. The
coefficients required to compute the energy-momentum tensor and EOS in full QCD
were computed in~Ref.~\cite{Makino:2014taa}.

We note that the calculation of EOS by the gradient flow method does not require the information of beta functions.
In a conventional calculation of EOS using the derivative method, the integration method, or $T$-integration method, evaluation of the nonperturbative beta functions is a big numerical task, in particular in full QCD for which we first have to determine a line of constant physics in a multidimensional coupling parameter space on zero-temperature lattices and then measure the beta functions defined as the slopes of each coupling parameter under a variation of the lattice spacing $a$ along the line of constant physics.
With the fixed-scale approach using the $T$-integration method, the same set of zero-temperature configurations can be used to compute EOS at all temperatures, provided that the beta functions are available.
If the beta functions are not available, we have to carry out a series of systematic zero-temperature simulations in a multidimensional parameter space to determine the beta functions, and thus the benefit of the fixed-scale approach is reduced. 
The gradient flow method in part removes the weak point of the fixed-scale approach.

In this study, we also calculate the chiral condensate. Using the
gradient flow method of~Ref.~\cite{Hieda:2016lly}, the proper chiral condensate
which satisfies the partially conserved axial vector current (PCAC) relation is
extracted through a similar idea as the energy-momentum tensor.\footnote{A different method to compute the chiral condensate by using the gradient flow has been discussed in~Ref.~\cite{Luscher:2013cpa}.} The temperature
dependence of the chiral condensate as well as its disconnected susceptibility is studied 
and a signal of chiral crossover is observed at $T\sim190\,\mathrm{MeV}$.

The gradient flow method of~Ref.~\cite{Hieda:2016lly} was also applied to study the topological susceptibility in finite temperature QCD~\cite{WHOT2017}. 
Preliminary results of our study was reported in Refs.~\cite{Lat16WHOTa,Lat16WHOTb}.
This paper is organized as follows: In Sec.~\ref{sec:definitions},
we define our gradient flow equations and give explicit
formulas for the energy-momentum tensor and the chiral condensate. Our
simulation parameters are summarized in Sec.~\ref{sec:parameters}, and the
results of the numerical simulation for the energy-momentum tensor and chiral 
condensate are presented in Secs.~\ref{sec:EMTresults} and~\ref{sec:chiralcond}, respectively.
Section~\ref{sec:conclusion} is devoted to our conclusions and discussions. 
In Appendixes~\ref{sec:setup} and \ref{sec:flowalgo}, we introduce our
simulation algorithms for the gradient flow and measurements with quarks. Definitions of our
running coupling and running masses, which are necessary in the evaluation of
conversion coefficients, are given in~Appendix~\ref{sec:running}. 
Several additional tests on our data for the energy-momentum tensor are
presented in~Appendix~\ref{sec:tests}.

\section{Definition of observables}
\label{sec:definitions}
\subsection{Flow equations}
\label{sec:2.1}
Our flow equations are identical to those given in Refs.~\cite{Luscher:2010iy}
and \cite{Luscher:2013cpa}. That is, for the gauge field, we set\footnote{In what follows, 
the sum over repeated Lorentz indices, $\mu$, $\nu$, $\rho$, \dots, over $0$, $1$, $2$, and~$3$, and the sum of the adjoint indices, $a$, $b$, \dots, are always understood. On the other hand, without indicated otherwise, the summation over repeated flavor indices, $f$, $f'=u$, $d$, $s$ is not assumed.}
\begin{equation}
   \partial_tB_\mu(t,x)=D_\nu G_{\nu\mu}(t,x),\qquad
   B_\mu(t=0,x)=A_\mu(x),
\label{eq:(2.1)}
\end{equation}
where the field strength and the covariant derivative of the flowed gauge field
are 
\begin{equation}
   G_{\mu\nu}(t,x)
   =\partial_\mu B_\nu(t,x)-\partial_\nu B_\mu(t,x)
   +[B_\mu(t,x),B_\nu(t,x)],
\label{eq:(2.2)}
\end{equation}
and
\begin{equation}
   D_\nu G_{\nu\mu}(t,x)
   =\partial_\nu G_{\nu\mu}(t,x)+[B_\nu(t,x),G_{\nu\mu}(t,x)],
\label{eq:(2.3)}
\end{equation}
respectively. For the quark fields, we set
\begin{align}
   &\partial_t\chi_f(t,x)=\Delta\chi_f(t,x),\qquad
   \chi_f(t=0,x)=\psi_f(x),
\label{eq:(2.4)}
\\
   &\partial_t\Bar{\chi}_f(t,x)
   =\Bar{\chi}_f(t,x)\overleftarrow{\Delta},
   \qquad\Bar{\chi}_f(t=0,x)=\Bar{\psi}_f(x),
\label{eq:(2.5)}
\end{align}
where $f=u$, $d$, $s$, denotes the flavor index, and
\begin{align}
   &\Delta\chi_f(t,x)\equiv D_\mu D_\mu\chi_f(t,x),\qquad
   D_\mu\chi_f(t,x)\equiv\left[\partial_\mu+B_\mu(t,x)\right]\chi_f(t,x),
\label{eq:(2.6)}
\\
   &\Bar{\chi}_f(t,x)\overleftarrow{\Delta}
   \equiv\Bar{\chi}_f(t,x)\overleftarrow{D}_\mu\overleftarrow{D}_\mu,
   \qquad\Bar{\chi}_f(t,x)\overleftarrow{D}_\mu
   \equiv\Bar{\chi}_f(t,x)\left[\overleftarrow{\partial}_\mu-B_\mu(t,x)\right].
\label{eq:(2.7)}
\end{align}
Note that our flow equations are independent of the flavor.

\subsection{Energy-momentum tensor}
\label{sec:2.2}
We follow the proposal
of~Refs.~\cite{Suzuki:2013gza,Asakawa:2013laa,Makino:2014taa} which employs the
gradient flow and the fermion flow and their small flow time
expansion~\cite{Luscher:2011bx} to define the energy-momentum tensor.
According to the reasoning of~Refs.~\cite{Suzuki:2013gza,Makino:2014taa}, in
terms of composite operators made out from the flowed fields, the correctly
normalized energy-momentum tensor is given by\footnote{In this definition, we
subtract the vacuum expectation value of the operator which might be
divergent.}
\begin{align}
   T_{\mu\nu}(x)
   &=\lim_{t\to0}\biggl\{c_1(t)\left[
   \Tilde{\mathcal{O}}_{1\mu\nu}(t,x)
   -\frac{1}{4}\Tilde{\mathcal{O}}_{2\mu\nu}(t,x)
   \right]
\notag\\
   &\qquad{}
   +c_2(t)\left[
   \Tilde{\mathcal{O}}_{2\mu\nu}(t,x)
   -\left\langle\Tilde{\mathcal{O}}_{2\mu\nu}(t,x)\right\rangle_{\! 0}
   \right]
\notag\\
   &\qquad{}
   +c_3(t)\sum_{f=u,d,s}
   \left[
   \Tilde{\mathcal{O}}_{3\mu\nu}^f(t,x)
   -2\Tilde{\mathcal{O}}_{4\mu\nu}^f(t,x)
   -\left\langle
   \Tilde{\mathcal{O}}_{3\mu\nu}^f(t,x)
   -2\Tilde{\mathcal{O}}_{4\mu\nu}^f(t,x)
   \right\rangle_{\! 0}
   \right]
\notag\\
   &\qquad{}
   +c_4(t)\sum_{f=u,d,s}
   \left[
   \Tilde{\mathcal{O}}_{4\mu\nu}^f(t,x)
   -\left\langle\Tilde{\mathcal{O}}_{4\mu\nu}^f(t,x)\right\rangle_{\! 0}
   \right]
\notag\\
   &\qquad{}
   +\sum_{f=u,d,s}c_5^f(t)\left[
   \Tilde{\mathcal{O}}_{5\mu\nu}^f(t,x)
   -\left\langle\Tilde{\mathcal{O}}_{5\mu\nu}^f(t,x)\right\rangle_{\! 0}
   \right]\biggr\},
\label{eq:(2.8)}
\end{align}
where $\langle\cdots\rangle_{0}$ stands for the vacuum expectation value
(VEV), i.e., the expectation value at zero temperature.
The operators in the right-hand side of~Eq.~\eqref{eq:(2.8)} are
defined by
\begin{align}
   \Tilde{\mathcal{O}}_{1\mu\nu}(t,x)&\equiv
   G_{\mu\rho}^a(t,x)\,G_{\nu\rho}^a(t,x),
\label{eq:(2.9)}
\\
   \Tilde{\mathcal{O}}_{2\mu\nu}(t,x)&\equiv
   \delta_{\mu\nu}\,G_{\rho\sigma}^a(t,x)\,G_{\rho\sigma}^a(t,x),
\label{eq:(2.10)}
\\
   \Tilde{\mathcal{O}}_{3\mu\nu}^f(t,x)&\equiv
   \varphi_f(t)\,\Bar{\chi}_f(t,x)
   \left(\gamma_\mu\overleftrightarrow{D}_\nu
   +\gamma_\nu\overleftrightarrow{D}_\mu\right)
   \chi_f(t,x),
\label{eq:(2.11)}
\\
   \Tilde{\mathcal{O}}_{4\mu\nu}^f(t,x)&\equiv
   \varphi_f(t)\,\delta_{\mu\nu}\,
   \Bar{\chi}_f(t,x)
   \overleftrightarrow{\Slash{D}}
   \chi_f(t,x),
\label{eq:(2.12)}
\\
   \Tilde{\mathcal{O}}_{5\mu\nu}^f(t,x)&\equiv
   \varphi_f(t)\,\delta_{\mu\nu}\,
   \Bar{\chi}_f(t,x)\,
   \chi_f(t,x),
\label{eq:(2.13)}
\end{align}
where
\begin{equation}
   \overleftrightarrow{D}_\mu\equiv D_\mu-\overleftarrow{D}_\mu,
\label{eq:(2.14)}
\end{equation}
and for the $(2+1)$-flavor QCD, the normalization factor~$\varphi_f(t)$ is
given by~\cite{Makino:2014taa},
\begin{equation}
   \varphi_f(t)\equiv
  - \frac{6}
   {(4\pi)^2\,t^2
   \left\langle\Bar{\chi}_f(t,x)\overleftrightarrow{\Slash{D}}\chi_f(t,x)
   \right\rangle_{\! 0}}.
\label{eq:(2.15)}
\end{equation}
Note that, from above definitions, it follows that
\begin{equation}
   2\left\langle\Tilde{\mathcal{O}}_{3\mu\nu}^f(t,x)\right\rangle_{\! 0}
   =\left\langle\Tilde{\mathcal{O}}_{4\mu\nu}^f(t,x)\right\rangle_{\! 0}
   = - \frac{6}{(4\pi)^2\,t^2}\delta_{\mu\nu}.
\label{eq:(2.16)}
\end{equation}
The coefficients in Eq.~\eqref{eq:(2.8)} for $(2+1)$-flavor QCD are given
as~\cite{Makino:2014taa},
\begin{align}
   c_1(t)&
   =\frac{1}{\Bar{g}\!\left(1/\sqrt{8t}\right)^2}
   -\frac{1}{(4\pi)^2}\left[
   9(\gamma-2\ln2)+\frac{19}{4}\right],
\label{eq:(2.17)}
\\
   c_2(t)&
   =\frac{1}{(4\pi)^2}\frac{33}{16},
\label{eq:(2.18)}
\\
   c_3(t)&
   =\frac{1}{4}\left\{1+\frac{\Bar{g}\!\left(1/\sqrt{8t}\right)^2}{(4\pi)^2}
   \left[2+\frac{4}{3}\ln(432)\right]\right\},
\label{eq:(2.19)}
\\
   c_4(t)&=\frac{1}{(4\pi)^2}\Bar{g}\!\left(1/\sqrt{8t}\right)^2,
\label{eq:(2.20)}
\\
   c_5^f(t)&=-\Bar{m}_f\!\left(1/\sqrt{8t}\right)
   \left\{1+\frac{\Bar{g}\!\left(1/\sqrt{8t}\right)^2}{(4\pi)^2}
   \left[4(\gamma-2\ln2)+\frac{14}{3}+\frac{4}{3}\ln(432)\right]\right\},
\label{eq:(2.21)}
\end{align}
where $\gamma$ denotes the Euler constant and $\Bar{g}(\mu)$
and~$\Bar{m}_f(\mu)$ are the running gauge coupling and the running quark mass
of the flavor~$f$ in the $\overline{\mathrm{MS}}$ scheme at the scale $\mu$, respectively.

In principle, one may use any lattice transcription for the composite
operators in Eqs.~\eqref{eq:(2.9)}-\eqref{eq:(2.13)} as well as for the
flow equations~\eqref{eq:(2.1)}, \eqref{eq:(2.4)}, and~\eqref{eq:(2.5)}; in
this sense, the above formula for the energy-momentum tensor is ``universal''.
This universality follows from the fact that any composite operator of the
flowed fields becomes a renormalized
operator~\cite{Luscher:2011bx,Luscher:2013cpa} under the multiplicative
renormalization of the flowed quark fields (see also~Ref.~\cite{Hieda:2016xpq}). The normalization factor~\eqref{eq:(2.15)} takes
care of this multiplicative renormalization of the flowed quark
fields~\cite{Makino:2014taa}. Such a renormalized composite operator must be
independent of the regularization, i.e., the way of lattice discretization, for
example, after taking the continuum limit.

\subsection{Scalar density and the chiral condensate}
\label{sec:2.3}

In Ref.~\cite{Hieda:2016lly}, the small flow time behavior of a composite
operator of flowed quark fields is related to the quark scalar density. For the
renormalized scalar density of the form (suppressing the flavor
indices)
\begin{equation}
   \left\{\Bar{\psi}\{\{t^A,M\},t^B\}\psi\right\}\!(x),
\label{eq:(2.22)}
\end{equation}
where $t^A$ and $t^B$ denote the (antihermitian) generators of the flavor
group~$SU(3)$ and $M$ is the renormalized quark mass matrix of the form
\begin{equation}
   M=\begin{pmatrix}
   m_{ud}&0&0\\
   0&m_{ud}&0\\
   0&0&m_s\\
   \end{pmatrix},
\label{eq:(2.23)}
\end{equation}
one has
\begin{align}
   &\left\{\Bar{\psi}\{\{t^A,M\},t^B\}\psi\right\}\!(x)
\notag\\
   &=\lim_{t\to0}\left\{1+
   \frac{\Bar{g}\!\left(1/\sqrt{8t}\right)^2}{(4\pi)^2}
   \left[4(\gamma-2\ln2)+8+\frac{4}{3}\ln(432)\right]\right\}
\notag\\
   &\qquad\times
   \left[
   \sum_{f,f'=u,d,s}\!
   \sqrt{\varphi_f(t)}\sqrt{\varphi_{f'}(t)}\,
   \Bar{\chi}_f(t,x)\,\{\{t^A,\Bar{M}\!\left(1/\sqrt{8t}\right)\},t^B\}_{ff'}\,
   \chi_{f'}(t,x)
   -\mathrm{VEV}\right].
\label{eq:(2.24)}
\end{align}
In the last line, the vacuum expectation value (VEV) of the first term on the same line is
subtracted. In the right-hand side, the running
coupling $\Bar{g}(\mu)$ and the running masses in the matrix $\Bar{M}$,
\begin{equation}
   \Bar{M}(\mu)=\begin{pmatrix}
   \Bar{m}_{ud}(\mu)&0&0\\
   0&\Bar{m}_{ud}(\mu)&0\\
   0&0&\Bar{m}_s(\mu)\\
   \end{pmatrix},
\label{eq:(2.25)}
\end{equation}
are renormalized in the $\overline{\mathrm{MS}}$ scheme at the scale $\mu$.

The relation~\eqref{eq:(2.24)} is obtained in the following way~\cite{Hieda:2016lly}:
We define the scalar density~\eqref{eq:(2.22)} as the chiral rotation of 
the pseudoscalar density $\left\{\Bar{\psi}\,\gamma_5\{t^A,M\}\psi\right\}\!(x)$, 
where the chiral rotation is defined by 
\begin{equation}
   \psi(x)\to e^{\alpha\gamma_5t^B}\psi(x),\qquad
   \Bar{\psi}(x)\to\Bar{\psi}(x)\,e^{\alpha\gamma_5t^B},
\label{eq:(2.26)}
\end{equation}
and, correspondingly, for the flowed quark fields,
\begin{equation}
   \chi(x)\to e^{\alpha\gamma_5t^B}\chi(x),\qquad
   \Bar{\chi}(x)\to\Bar{\chi}(x)\,e^{\alpha\gamma_5t^B}.
\label{eq:(2.27)}
\end{equation}
The normalization of the pseudoscalar density is uniquely fixed by the PCAC
relation. The small flow time representation of the pseudoscalar density was
obtained in Ref.~\cite{Endo:2015iea}. 
Because the composite operators of the flowed quark fields transform under the chiral transformation as if they are simple products of elementary quark fields (i.e., no nontrivial renormalization is required under the transformation), one obtains the relation~\eqref{eq:(2.24)} by the chiral transformation~\eqref{eq:(2.27)} of the pseudoscalar density.

This argument based on the chiral transformation does not necessarily
require the subtraction of the VEV in Eq.~\eqref{eq:(2.24)}. 
We have to note, however, that the flavor
singlet part of the scalar density possesses the quantum number identical to
the vacuum and, when quarks are massive, its expectation value can have terms 
proportional to $M^2/t$ depending on the prescription, 
though the more conventional divergence $M^2/a^2$ is prohibited by the finiteness of flowed operators.\footnote{We would like to thank Tetsuya
Onogi and Hidenori Fukaya for a discussion on this point.} 
In fact, the small flow time behavior of the operator in the right-hand side
of Eq.~\eqref{eq:(2.24)} is estimated as
\begin{align}
   &\sum_{f,f'=u,d,s}\!\sqrt{\varphi_f(t)}\sqrt{\varphi_{f'}(t)}
   \, \Bar{\chi}_f(t,x) \, \{\{t^A,M\},t^B\}_{ff'}\,\chi_{f'}(t,x)
\notag\\
   &\stackrel{t\to0}{\sim}
   \left[
   -\frac{12}{(4\pi)^2}
   \sum_{f=u,d,s}\!
   \left(
   \{\{t^A,M\},t^B\}\,M
   \left\{
   \frac{1}{2t}
   +M^2\left[\gamma+\ln(2M^2t)\right]+O(t)\right\}
   \right)_{\! ff}
   +O(g^2)\right]
   \mathbbm{1}
\notag\\
   &\qquad{}
   +\left[1+O(g^2)\right]\Bar{\psi}(x)\,\{\{t^A,M\},t^B\}\,\psi(x)+O(t).
\label{eq:(2.28)}
\end{align}
Therefore, when quarks are massive, the first term with the identity
operator $\mathbbm{1}$ diverges as~$t\to0$. To remove such term, we 
subtract the VEV in Eq.~\eqref{eq:(2.24)}. 
We may alternatively calculate the scalar density in the chiral limit, by first taking
the chiral limit~$M\to0$ and then taking the small flow time limit~$t\to0$. We leave
this possibility for future study.

Now, by setting
\begin{equation}
   t^A=t^B=\frac{i}{2}\begin{pmatrix}
   0&1&0\\
   1&0&0\\
   0&0&0\\
   \end{pmatrix}
\label{eq:(2.29)}
\end{equation}
in Eq.~\eqref{eq:(2.24)} and dividing the both sides by $m_u$ or $m_d$, we get
\begin{align}
   &\left\{\Bar{\psi}_u\psi_u\right\}\!(x)
   +\left\{\Bar{\psi}_d\psi_d\right\}\!(x)
\notag\\
   &=\lim_{t\to0}\left\{1+
   \frac{\Bar{g}\!\left(1/\sqrt{8t}\right)^2}{(4\pi)^2}
   \left[4(\gamma-2\ln2)
   +8
   +\frac{4}{3}\,\ln(432)\right]\right\}
\notag\\
   &\qquad\qquad{}\times
   \frac{\Bar{m}_{ud}\!\left(1/\sqrt{8t}\right)}{m_{ud}}
   \left[\varphi_u(t)\,\Bar{\chi}_u(t,x)\,\chi_u(t,x)
   +\varphi_d(t)\,\Bar{\chi}_d(t,x)\,\chi_d(t,x)-\mathrm{VEV}\right],
\label{eq:(2.30)}
\end{align}
while by setting
\begin{equation}
   t^A=t^B=\frac{i}{2}\,\frac{1}{\sqrt{3}}\begin{pmatrix}
   1&0&0\\
   0&1&0\\
   0&0&-2\\
   \end{pmatrix}
\label{eq:(2.31)}
\end{equation}
and using Eq.~\eqref{eq:(2.30)}, we get
\begin{align}
   \left\{\Bar{\psi}_s\psi_s\right\}\!(x)
   &=\lim_{t\to0}\left\{1+
   \frac{\Bar{g}\!\left(1/\sqrt{8t}\right)^2}{(4\pi)^2}
   \left[4(\gamma-2\ln2)
   +8
   +\frac{4}{3}\,\ln(432)\right]\right\}
\notag\\
   &\qquad\qquad{}\times \,
   \frac{\Bar{m}_s\!\left(1/\sqrt{8t}\right)}{m_s} \,
   \left[\varphi_s(t)\,
   \Bar{\chi}_s(t,x)\,\chi_s(t,x)-\mathrm{VEV}\right].
\label{eq:(2.32)}
\end{align}
For clarity, let us denote the chiral condensate at $t\ne0$ without the VEV subtraction as
$\left\{\Bar{\psi}_f\psi_f\right\}^{(0)}\!(t,x)$,
\begin{align}
   \left\{\Bar{\psi}_f\psi_f\right\}^{(0)}\!(t,x)
   &=\left\{
   1+\frac{\Bar{g}\!\left(1/\sqrt{8t}\right)^2}{(4\pi)^2}
   \left[4(\gamma-2\ln2)+8+\frac{4}{3}\ln(432)\right]\right\}
\notag\\
   &\qquad{}\times
   \frac{\Bar{m}_f\!\left(1/\sqrt{8t}\right)}{m_f}
   \left[\varphi_f(t)\,\Bar{\chi}_f(t,x)\,\chi_f(t,x)\right] .
\label{eqn:bpsipsi}
\end{align}

\subsection{Finite flow time effects and lattice artifacts}
\label{sec:finitet}

To avoid boundary effects due to oversmearing, the smeared range of the
gradient flow  $\sqrt{8t}$ should not exceed $\min(N_t/2,N_s/2)\times a$. 
Thus, the measurements should be performed within flow times
\begin{equation}
 \frac{t}{a^2} \le  t_{1/2} \equiv \frac{1}{8}\left[
   \min\left(\frac{N_t}{2},\frac{N_s}{2}\right)\right]^2 .
\label{eq:t-half}
\end{equation}

We then take the $t\to0$ limit as required in Eq.~\eqref{eq:(2.8)}.
A typical form of small flow time effects in the energy-momentum tensor would be
\begin{equation}
   T_{\mu\nu}(t,x)=T_{\mu\nu}(x)+t\,S_{\mu\nu}(x)+O(t^2),
\label{eq:o-t-eff}
\end{equation}
where $T_{\mu\nu}(t,x)$ corresponds to that in~Eq.~\eqref{eq:(2.8)} before
taking the $t\to0$ limit. $S_{\mu\nu}(x)$ is a sum of dimension-six operators
with the same quantum number and $O(t^2)$ is contribution from higher
dimensional operators. $T_{\mu\nu}(x)$ is our target conserved energy-momentum
tensor.

On finite lattices, however, we also have lattice artifacts due to finite lattice spacing $a$.
Since we adopt the nonperturbatively $O(a)$-improved Wilson fermion, 
the lattice artifact would start with $O(a^2)$.
Then, small lattice spacing corrections to $T_{\mu\nu}(t,x)$ at $t>0$ would be 
\begin{align}
   T_{\mu\nu}(t,x,a)
   &=T_{\mu\nu}(t,x)
   +A_{\mu\nu}(x)\,\frac{a^2}{t}+\sum_fB_{f\mu\nu}(x)\,(am_f)^2+C_{\mu\nu}(x)\,(aT)^2
 \notag\\
   &\qquad{}
    +D_{\mu\nu}(x)\left(a\Lambda_{\mathrm{QCD}}\right)^2
    +a^2S'_{\mu\nu}(x)+O(a^4),
\label{eq:a2overt}
\end{align}
where $T_{\mu\nu}(t,x,a)$ is the flowed tensor operator on the lattice. $A_{\mu\nu}$,
$B_{f\mu\nu}$, $C_{\mu\nu}$, and~$D_{\mu\nu}$ are contributions from dimension-four operators and $S'_{\mu\nu}$ is that from dimension-six operators. 
We note that the $a^2/t$ term can appear to the lowest order in $a^2$ through mixing with 
dimension-four operators.
In the higher orders in $a^2$, more singular terms like $1/t^2$ can enter.

When we take the continuum limit before taking the $t\to0$ limit, 
the $O(a^2)$ terms in~Eq.~\eqref{eq:a2overt}, including all the singular terms at $t=0$, are removed, and we can carry out the $t\to0$ extrapolation safely. 
In numerical simulations, however, it is sometimes favorable to take the continuum extrapolation at a later stage of analyses.
This exchange of the order of limiting procedures is allowed if we can remove the singular terms at $t=0$.
We come back to this issue in the actual $t\to0$ extrapolations in Sec.~\ref{sec:EMTresults}.

Similar to the case of the energy-momentum tensor, the chiral condensate on finite lattices 
is expected to be 
\begin{align}
   \left\{\Bar{\psi}_f\psi_f\right\}^{(0)}\!(t,x,a)
   &=\left\{\Bar{\psi}_f\psi_f\right\}^{(0)}\!(t,x)
   +A(x)\,\frac{a^2}{t}
   +\sum_fB_f(x)\,(am_f)^2
\notag\\
   &\qquad{}
   +C(x)\,(aT)^2
   +D(x)\left(a\Lambda_{\mathrm{QCD}}\right)^2
   +a^2S(x)+O(a^4)
\label{eqn:bpsipsi-a}
\end{align}
to the lowest order of $a^2$, where $\{\Bar{\psi}_f\psi_f\}^{(0)}(t,x,a)$ is the flowed operator at finite lattice
spacing before the VEV subtraction. $A$, $B_f$, $C$, and $D$ are contributions from dimension-three
operators and $S$ is that from dimension-five operators. After taking the
continuum limit, the scalar density should be given by
\begin{equation}
   \left\{\Bar{\psi}_f\psi_f\right\}^{(0)}\!(t,x)=
   \left\{\Bar{\psi}_f\psi_f\right\}_{\overline{\mathrm{MS}}}(x)
   +\frac{m_f}{t}N(x)
   +t\,S'(x)
   +O(t^2),
\end{equation}
where $\left\{\Bar{\psi}_f\psi_f\right\}_{\overline{\mathrm{MS}}}(x)$ is the
renormalized chiral condensate in ${\overline{\mathrm{MS}}}$ scheme. $N$~is a
contribution of dimensionless operators and $S'$ is that from dimension-five
operators.
Thus,  $\{\Bar{\psi}_f\psi_f\}^{(0)}(t,x,a)$ at finite lattice spacing and finite quark mass has both $m_f/t$ and $a^2/t$ singularities around the origin. 

When we take the chiral and continuum limits before taking the $t\to0$ limit, 
these singular terms of the chiral condensate at $t=0$ are removed, and we can do the $t\to0$ extrapolation safely.
Conversely, when we can remove the singular terms at $t=0$, we can exchange the order of the three limiting procedures. 
The $m_f/t$ singularity can be removed by the VEV subtraction discussed in Sec.~\ref{sec:2.3}.
Because the lattice spacing is the same in the VEV,
we expect that the $a^2/t$ singularity is also in part removed by the VEV subtraction. 
We study this issue with the actual data in Sec.~\ref{sec:chiralcond}.

\section{Simulation parameters and numerical procedures}
\label{sec:parameters}

Measurements of the energy-momentum tensor are performed on $N_f=2+1$ gauge
configurations generated for Ref.~\cite{Umeda:2012er}. In these calculations,
we need to subtract the zero-temperature values of the operators. The zero
temperature gauge configurations are also prepared which were generated
for~Ref.~\cite{Ishikawa:2007nn}. These configurations are open to the public on
ILDG/JLDG~\cite{Maynard:2010wi}.

The nonperturbatively $O(a)$-improved Wilson quark
action~\cite{Sheikholeslami:1985ij} and the renormalization-group improved Iwasaki gauge
action~\cite{Iwasaki:2011np,Iwasaki:1985we} are adopted. The bare coupling
constant is set to $\beta=2.05$, which corresponds
to~$a=0.0701(29)\,\mathrm{fm}$ ($1/a\simeq2.79\,\mathrm{GeV}$) with an input
of~$r_0=0.5\,\mathrm{fm}$~\cite{Maezawa:2009di}. The nonperturbative clover
coefficient is $c_{\mathrm{SW}}=1.628$ at~$\beta=2.05$, which is determined by
the Schr\"odinger functional method~\cite{Aoki:2005et}. The hopping parameters
are set to $\kappa_u=\kappa_d\equiv\kappa_{ud}=0.1356$ and $\kappa_s=0.1351$,
which correspond to heavy $u$ and $d$ quarks, $m_\pi/m_\rho\simeq0.63$, and
almost physical $s$ quark, $m_{\eta_{ss}}/m_\phi\simeq0.74$, 
where $\eta_{ss}$ is the strange pseudoscalar meson whose mass is phenomenologically estimated as 
$m_{\eta_{ss}} \approx \sqrt{2m_K^2 - m_\pi^2}$. 
The bare PCAC quark masses are
\begin{equation}
   a\,m_{ud}=0.02105(17),\qquad
   a\,m_s=0.03524(26),
\label{eqn:PCACmass}
\end{equation}
where $m_{ud}=m_u=m_d$ is the degenerate mass of $u$ and $d$ quarks \cite{Ishikawa:2007nn}.

In this study, we adopt the fixed-scale approach~\cite{Levkova:2006gn,Umeda:2008bd} 
in which the temperature $T=1/(aN_t)$ is varied by changing the
temporal lattice size~$N_t$ with a fixed lattice spacing~$a$.
This enables us to use one common zero-temperature simulation  
to subtract zero-temperature contributions at all temperatures. 
The equation of state using the $T$-integration method~\cite{Umeda:2008bd} was obtained previously using the same set of configurations \cite{Umeda:2012er}.

The values of temperature at each $N_t$ are given in Table~\ref{table:parameters}.
In the table,  $T/T_{\mathrm{pc}}$ assuming the pseudocritical temperature
to be $T_{\mathrm{pc}}=190\,\mathrm{MeV}$~\cite{Umeda:2012er} is also listed. 
The spatial box size is $32^3$ for $T>0$ and $28^3$ for $T=0$. 
The values of $t_{1/2}$ defined by Eq.~\eqref{eq:t-half} are also given in the Table. 

\begin{table}[htb]
\centering
\caption{Parameters for the numerical simulation: Temperature in MeV,
$T/T_{\mathrm{pc}}$ assuming $T_{\mathrm{pc}}=190$ MeV, the temporal lattice size $N_t$, $t_{1/2}$ defined by Eq.~(\ref{eq:t-half}), and the number of configurations used in gauge and fermion measurements. 
The bare gauge coupling parameter and the hopping parameters are set to $\beta=2.05$, $\kappa_{ud}=0.1356$, and $\kappa_s=0.1351$. 
Spatial box size is $32^3$ for $T>0$ and $28^3$ for $T=0$.}
\label{table:parameters}
\begin{tabular}{cccccc}
 $T$[MeV] & $T/T_{\mathrm{pc}}$ & $N_t$ & $t_{1/2}$ & Gauge configurations & Fermion configurations \\
\hline
 $0$   & $0$    & $56$ &$24.5$ & $650$ & $65$ \\
 $174$ & $0.92$ & $16$ &$8$ & $1440$ & $144$ \\
 $199$ & $1.05$ & $14$ &$6.125$ & $1270$ & $127$ \\
 $232$ & $1.22$ & $12$ &$4.5$ & $1290$ & $129$ \\
 $279$ & $1.47$ & $10$ &$3.125$ & $780$ & $78$ \\
 $348$ & $1.83$ &  $8$ &$2$ & $510$ & $51$ \\
 $464$ & $2.44$ &  $6$ &$1.125$ & $500$ & $50$ \\
 $697$ & $3.67$ &  $4$ &$0.5$ & $700$ & $70$ \\
\hline
\end{tabular}
\end{table}

The gauge observables \eqref{eq:(2.9)} and~\eqref{eq:(2.10)} are measured every
five trajectories at $T>0$ and every ten trajectories
at $T=0$. The fermionic observables~\eqref{eq:(2.11)},
\eqref{eq:(2.12)}, and~\eqref{eq:(2.13)} are measured every $50$~trajectories
at $T>0$ and every $100$~trajectories at $T=0$. 
Number of configurations used for gauge and fermion
measurements are summarized in~Table~\ref{table:parameters}. 

Our numerical procedures to compute the fermionic observables \eqref{eq:(2.11)}, \eqref{eq:(2.12)},
and~\eqref{eq:(2.13)} at $t>0$ are given in Appendix~\ref{sec:setup}. 
To evaluate fermionic observables, we use the noisy estimator method. 
The number of noise vectors is 20 for each color.
To reduce correlation among data points at different values of $t$, we generate independent noise vectors at each $t$.
The statistical errors are estimated by the standard jackknife analysis.
After a study of the bin size dependence, we choose the bin size of $100$~trajectories 
for the energy-momentum tensor and $300$~trajectories for the chiral condensate and susceptibility.

To compute observables at $t>0$, we need flowed gauge and quark fields. 
Our numerical algorithm for gradient flow of gauge and quark fields is summarized in 
Appendix~\ref{sec:flowalgo}.
We adopt the third order Runge-Kutta method~\cite{Luscher:2010iy,Luscher:2013cpa} with the step
size of~$\epsilon=0.02$ to solve the differential equation for both the gauge
and quark fields. 

For the flowed operators
$\Tilde{\mathcal{O}}_{i\mu\nu}(t,x)$ in~Eqs.~\eqref{eq:(2.9)}-\eqref{eq:(2.13)},
we adopt the lattice symmetric covariant differential. 
For the quadratic terms of the field strength tensor $G_{\mu\nu}(x)$ in~Eqs.~\eqref{eq:(2.9)} and \eqref{eq:(2.10)}, 
there are several alternative choices of lattice operators. 
In this study, we combine clover operator with four plaquette Wilson loops 
and that with eight $1\times2$ rectangle Wilson loops such that 
the tree-level improved field strength squared is obtained~\cite{AliKhan:2001ym}.

\section{Results for the energy-momentum tensor}
\label{sec:EMTresults}

The pressure and the energy density are given by an averaged spatial component of the energy-momentum tensor 
and the temporal component of the energy-momentum tensor, 
\begin{equation}
p/T^4  = \sum_i \langle T_{ii}\rangle /(3T^4), \hspace{5mm}
\epsilon/T^4 = - \langle T_{00}\rangle/T^4 .
\end{equation}
In Figs.~\ref{fig5} and \ref{fig7}, we show the results of the entropy density
\begin{equation}
   \frac{\epsilon+p}{T^4}
   =-\frac{4}{3 T^4}\left\langle T_{00}-\frac{1}{4}T_{\mu\mu}\right\rangle 
\label{eq:(4.3)}
\end{equation}
and the trace anomaly
\begin{equation}
   \frac{\epsilon-3p}{T^4}
   =-\frac{1}{T^4}\langle T_{\mu\mu}\rangle
\label{eq:(4.4)}
\end{equation}
as functions of $t/a^2$.
Seven subplots in each figure are for the results at 
$T\simeq174$, $199$, $232$, $279$, $348$, $464$, and $697\,\mathrm{MeV}$
($N_t=16$, 14, 12, 10, 8, 6 and 4, respectively) from the top left to the bottom. 
The errors shown are statistical only.

\subsection{Extrapolation to $t\to0$}
\label{sec:t0extrapolation}

We extract physical results for the energy-momentum tensor by extrapolating the data to $t\to0$.
As discussed in Sec.~\ref{sec:finitet}, on finite lattices, we have to take care of unphysical singularities like $a^2/t$ around the origin.
On the other hand, our data shown in the figures indicates that, except for the case of the highest temperature $T\simeq697$ MeV ($N_t=4$), we do have ranges of $t/a^2$ in which the data show well linear behavior.
This suggests that the singular terms like $a^2/t$ are numerically negligible when $t/a^2$ is not so small. 

\begin{figure}[ht]
\centering
\includegraphics[width=6.5cm]{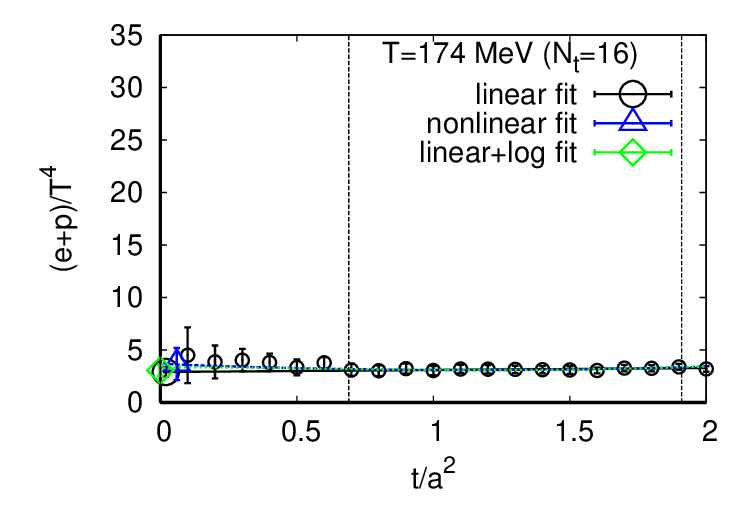}
\includegraphics[width=6.5cm]{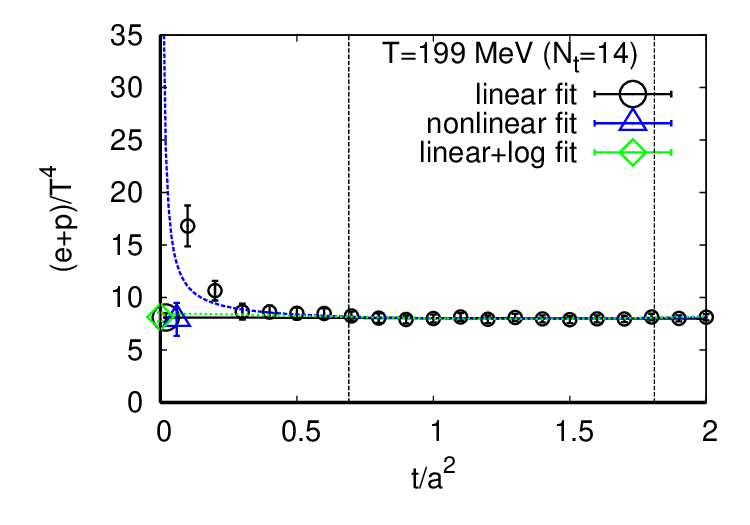}
\includegraphics[width=6.5cm]{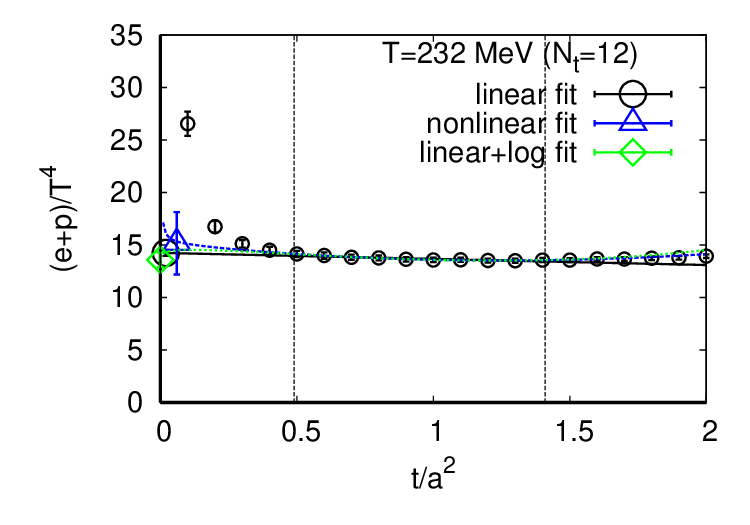}
\includegraphics[width=6.5cm]{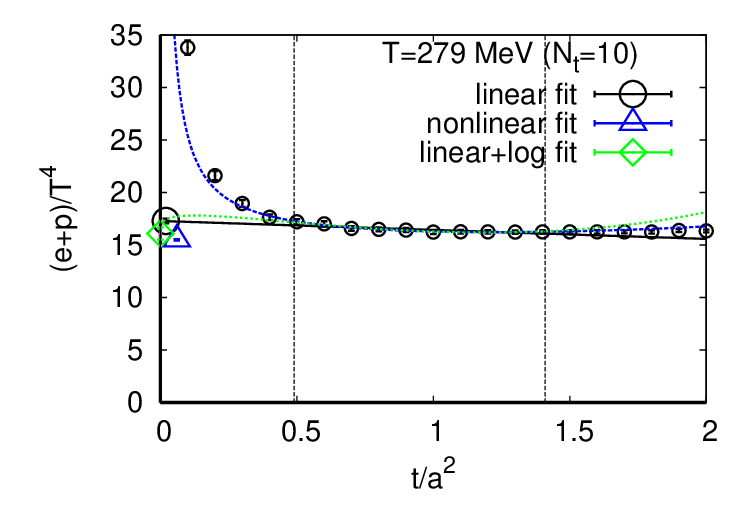}
\includegraphics[width=6.5cm]{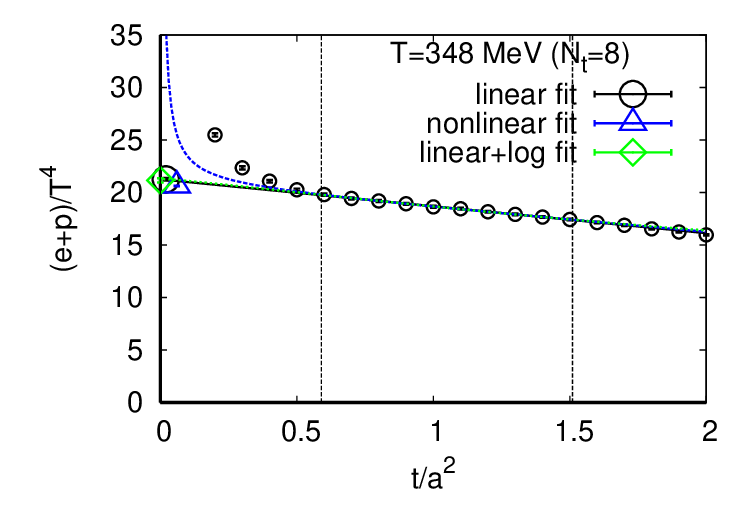}
\includegraphics[width=6.5cm]{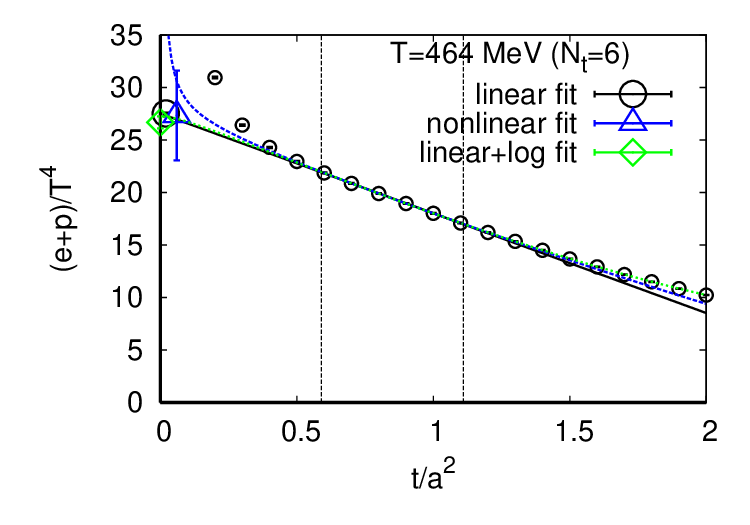}
\includegraphics[width=6.5cm]{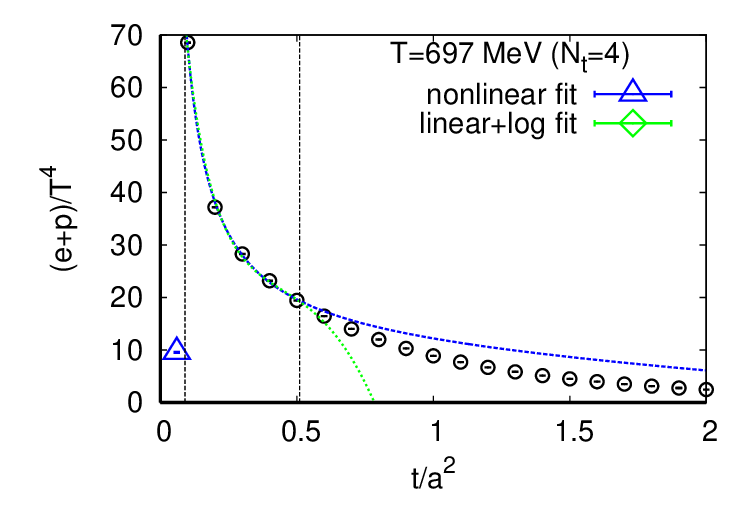}
\caption{Entropy density $(\epsilon+p)/T^4$ as a function of the flow time. From the
top left to the bottom: $T\simeq174$, $199$, $232$, $279$, $348$, $464$, $697\,\mathrm{MeV}$
($N_t=16$, 14, 12, 10, 8, 6, and 4, respectively). 
The pair of dashed vertical lines indicates the window used for the fit at each $T$.
Black solid lines are the fit results with the linear fit
 Ansatz~\eqref{eqn:5.3}, and the big open circles at~$t=0$ are the
 entropy density extracted from the fits.
 Blue and green dashed curves together with blue upward triangles and green diamonds at $t\sim0$ are the fit results with the nonlinear Ansatz~\eqref{eqn:5.4} and linear+log Ansatz~\eqref{eqn:5.4l}, respectively. 
  Errors are statistical only.}
\label{fig5}
\end{figure}

\begin{figure}[ht]
\centering
\includegraphics[width=6.5cm]{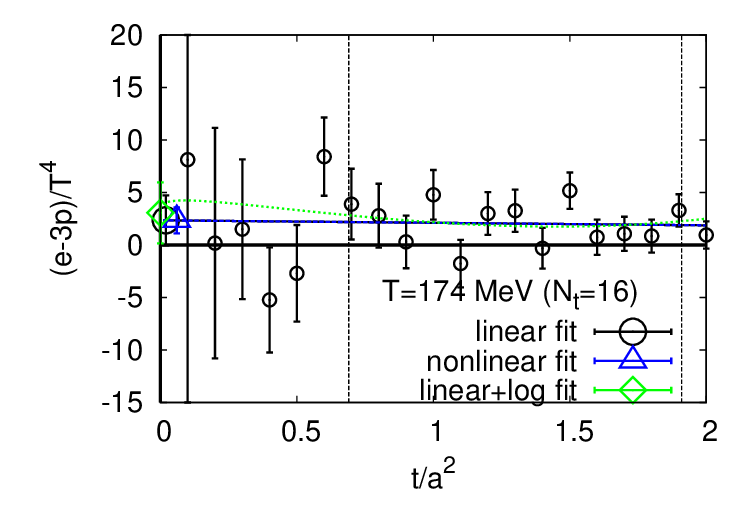}
\includegraphics[width=6.5cm]{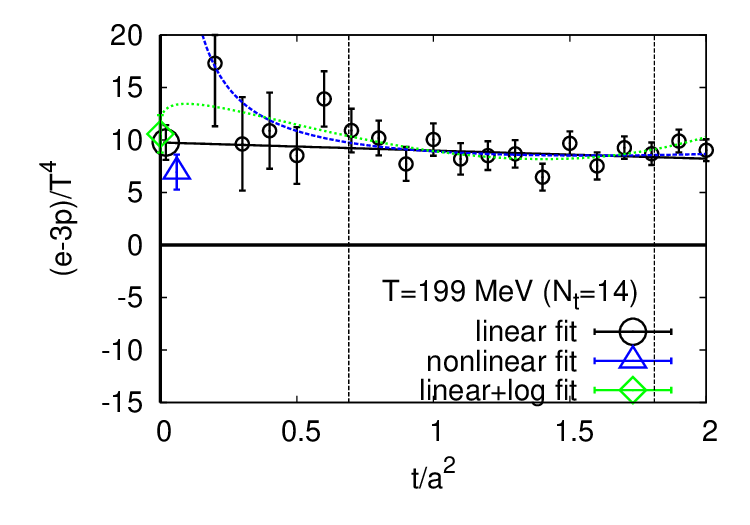}
\includegraphics[width=6.5cm]{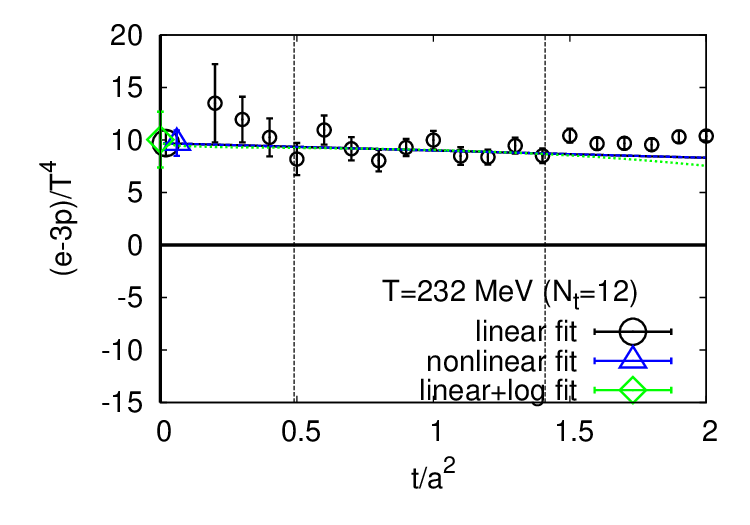}
\includegraphics[width=6.5cm]{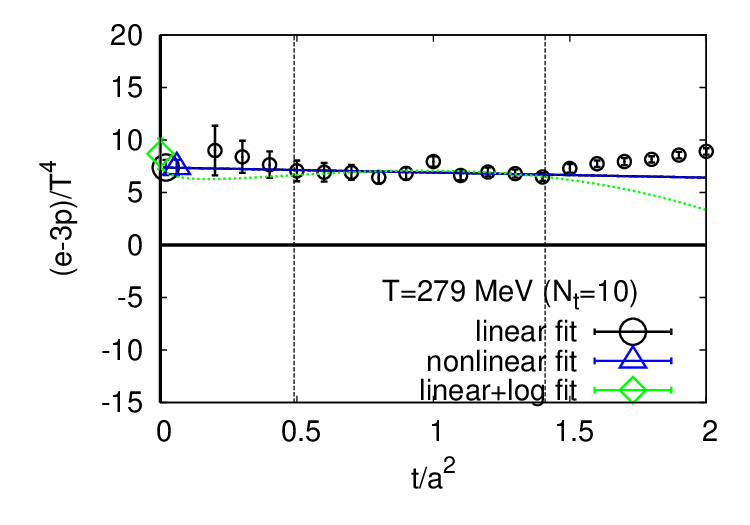}
\includegraphics[width=6.5cm]{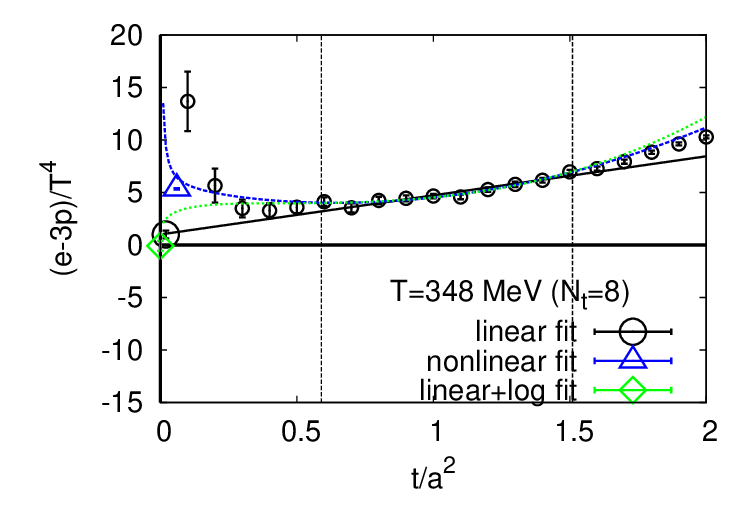}
\includegraphics[width=6.5cm]{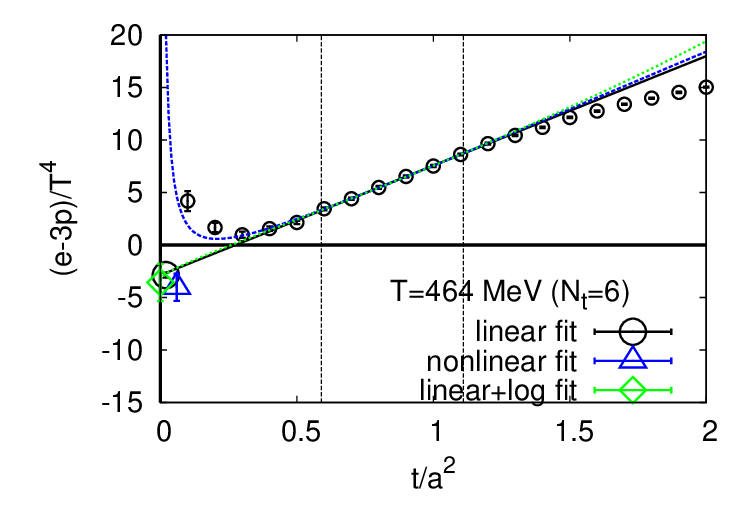}
\includegraphics[width=6.5cm]{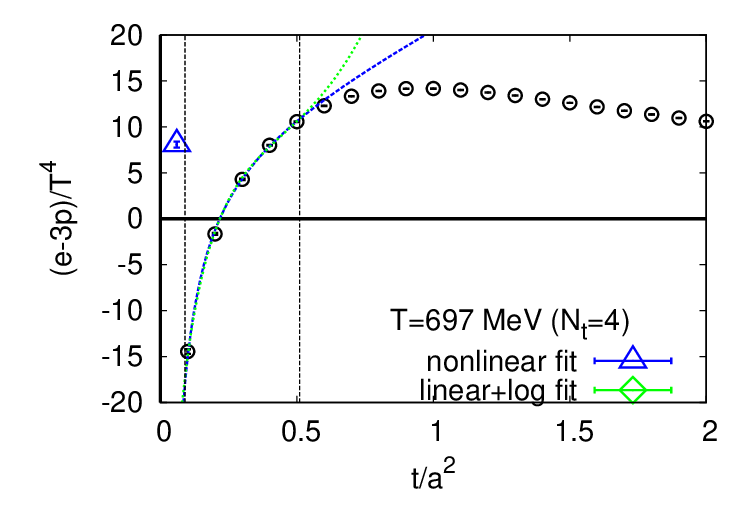}
\vspace{1cm}
\caption{The same as Fig.~\ref{fig5} but for the trace anomaly~$(\epsilon-3p)/T^4$.}
\label{fig7}
\end{figure}

We first identify linear windows from the data shown in Figs.~\ref{fig5} and \ref{fig7}
as ranges in $t/a^2$ in which the data are well linear 
under the condition that $t/a^2<t_{1/2}$. 
The windows are selected such that the linear fit discussed in the following leads to $\chi^2/N_\textrm{dof} \le O(1)$.
We also require that the window is common to all components of the energy-momentum tensor on each lattice.
The results for the linear window are shown by a pair of dashed vertical lines in Figs.~\ref{fig5} and \ref{fig7},
except for the case of $T\simeq697$ MeV ($N_t=4$) for which no clear linear window is visible below $t_{1/2}=0.5$.
We note that the case of $T\simeq464$ MeV ($N_t=6$) may be marginal to clearly identify a wide linear window because $t_{1/2}=1.125$ for this lattice is also small. 

At $T\simle464$ MeV, we perform a linear extrapolation 
\begin{equation}
 \langle T_{\mu\nu}(t,a) \rangle = \langle T_{\mu\nu}\rangle +
  t\,S_{\mu\nu}+O(a^2,t^2)
\label{eqn:5.3}
\end{equation}
adopting the linear windows of Figs.~\ref{fig5} and \ref{fig7}, 
to obtain the physical results $\langle T_{\mu\nu}\rangle$ for the energy-momentum tensor.
Our linear fits and the results of extrapolation are shown by black solid lines and big open circles at $t=0$ in Figs.~\ref{fig5} and \ref{fig7}.%
\footnote{As in the previous study in quenched QCD~\cite{Asakawa:2013laa}, we disregard correlation among different flow times in this study.
Our introduction of independent noise vectors at each $t$ should reduce the correlation in fermionic contributions. 
The jagged behavior visible, e.g., in Fig.~\ref{fig7} may be suggesting that the correlation is small in several observables. 
However, we find that our statistics is not high enough to discuss the correlation conclusively.
We leave the study of the correlation for the next step.}
We note that the data at $T\simle232\,\mathrm{MeV}$ are well flat within the window.
In fact, a constant fit leads to results consistent with the linear fit within statistical errors.

To confirm the validity of the linear window and to estimate a systematic error due to the fit Ansatz, we also make 
additional fits adopting two different fit Ans\"atze using the data within the same window.
One is
a nonlinear fit inspired from Eq.~\eqref{eq:a2overt},
\begin{equation}
 \langle T_{\mu\nu}(t,a) \rangle
  =\langle T_{\mu\nu} \rangle +A_{\mu\nu}\frac{a^2}{t}+t\,S_{\mu\nu}
  + t^2 R_{\mu\nu}  .
\label{eqn:5.4}
\end{equation}
Another is a linear+log fit including an additional $1/\log^2(\sqrt{8t}/a)$ term, 
\begin{equation}
 \langle T_{\mu\nu}(t,a) \rangle
  =\langle T_{\mu\nu} \rangle + t\,S_{\mu\nu}
  +  \frac{Q_{\mu\nu}}{\log^2(\sqrt{8t}/a)} .
\label{eqn:5.4l}
\end{equation}
The latter is inspired from possible higher order corrections to the matching coefficients $c_i(t)$ in~Eqs.~\eqref{eq:(2.17)}--\eqref{eq:(2.21)}, which are computed in one-loop perturbation theory~\cite{Makino:2014taa} in our study. 
As discussed in~Ref.~\cite{DelDebbio:2013zaa} [Eq.~(7.14)], for a small
but finite flow time~$t$, those perturbative one-loop coefficients may contain 
error of the order $\bar{g}(1/\sqrt{8t})^4/(4\pi)^4\sim 1/\log^2(t)$
associated with neglected higher-order loop corrections. 
Though higher-order perturbative corrections should be subdominant at small $t$ because of the asymptotic freedom, the
formula~\eqref{eq:o-t-eff} thus may in principle be modified by $O\left(1/\log^2(t)\right)$ terms. 
A fit including all the correction terms in (\ref{eqn:5.4}) and (\ref{eqn:5.4l}) turned out to be unstable due to too many fitting parameters.

The results of the nonlinear 
and liner+log fits for the entropy density and the trace anomaly at $T\simle464$ MeV ($N_t\ge6$) are shown by blue and green dashed curves in Figs.~\ref{fig5} and \ref{fig7}, respectively. 
In these figures, physical results $\langle T_{\mu\nu} \rangle$ extracted from these fits are shown by blue upward triangles and green diamonds at $t\sim0$.
We find that all the three fits are almost indistinguishable in the windows and describe the data within the windows well.
We also note that the nonlinear fit frequently fails to reproduce the singular behavior at small $t/a^2$ out of the linear window. 
On the other hand, the linear+log fit stays close to the linear fit down to small $t$ in most cases, but can slightly deviate when the data are noisy, as seen in Fig.~\ref{fig7}.

At $T\simle464$ MeV ($N_t\ge6$), we adopt the results of the linear fit for our central values and 
take the difference between the linear fit and the nonlinear or linear+log fits as an estimate of the systematic error due to the choice of the fit Ansatz.
We find that the differences are at most a few times of the statistical
error at $T\simle232$ MeV ($N_t\ge12$), while a larger difference can appear at higher temperatures.%
\footnote{We should, however, notice that the lattice artifacts
$B_{f\mu\nu}(am_f)^2+C_{\mu\nu}(aT)^2+D_{\mu\nu}(a\Lambda_{\mathrm{QCD}})^2+a^2S'_{\mu\nu}(x)$ 
of Eq.~\eqref{eq:a2overt} still remain and can be settled only after taking the continuum limit.}

Finally, we estimate the systematic error from the one-loop perturbative coefficients themselves.
For the perturbative coefficients, Eqs.~\eqref{eq:(2.17)}--\eqref{eq:(2.21)}, 
we need to know the running gauge coupling~$\Bar{g}\!\left(1/\sqrt{8t}\right)$ and the running quark 
masses~$\Bar{m}_f\!\left(1/\sqrt{8t}\right)$.
Definitions of these running coupling and running masses are given in Appendix~\ref{sec:running}. 
Inputs for $\Bar{g}(\mu)$ and $\Bar{m}_f(\mu)$ are the QCD scale $\Lambda_{\mathrm{QCD}}$ and the bare quark masses. 
For the QCD scale, we refer the value quoted in the Particle Data Group~\cite{Agashe:2014kda}
\begin{equation}
   \Lambda_{\overline{\mathrm{MS}}}^{(3)}=332(19)\;\mathrm{MeV}.
\end{equation}
Since $1/\sqrt{8t}$ plays a role of the renormalization scale, the QCD scale appears
with the form $a\sqrt{8(t/a^2)}\Lambda_{\mathrm{QCD}}$ in the perturbative
coefficients, where $t/a^2$ is a dimensionless flow time used on the lattice.
In this combination of the QCD scale, we should take into account the statistical error in the lattice spacing. 
As the bare quark masses, we use the PCAC masses of Eq.~\eqref{eqn:PCACmass} obtained on the same zero-temperature configuration as ours~\cite{Ishikawa:2007nn}.
In the running quark masses, the bare quark masses appear in the combination of renormalization group invariant masses, for which we should take into account the error in the renormalization factor too. 
The values as well as the errors for $\Bar{g}(\mu)$ and $\Bar{m}_f(\mu)$ are estimated in Appendix~\ref{sec:running}.

Our results of the equation of state in the $t\to0$ limit are summarized
in Table~\ref{table:EM} and \ref{table:EM2}.
In this table, we give the values of statistical error as well as 
the systematic errors due to the perturbative coefficients and the fit Ansatz, separately.

At $T\simeq697$ MeV ($N_t=4$), because a clear linear window is not available, we attempt a fit of the form~\eqref{eqn:5.4} adopting a fit range $t/a^2=[0.1,0.5=t_{1/2}]$ shown by dashed vertical lines in the bottom plots of Figs.~\ref{fig5} and \ref{fig7}, but with dropping the $t^2$ term to keep a nonvanishing DOF.
The results of $\langle T_{\mu\nu}\rangle$ are shown by blue upward triangles at $t\sim0$ in these plots.
As seen from the resulting fits shown by dashed curves, although the nonlinear fit Ansatz describes the data at small $t$ well, the lattice artifact term is completely dominating over the linear term which contain physical information. 
We thus consider that the results at $T\simeq697$ MeV  ($N_t=4$) are not reliable and disregard them in the followings.

\begin{table}[tbh]
\centering
\caption{Equation of state (pressure and energy density)
evaluated with the gradient flow method in the $t\to0$ limit.
The first parenthesis is for the statistical error 
estimated by a jackknife method.
The second and the third are for systematic errors due to
$\Lambda_{\overline{\mathrm{MS}}}^{(3)}$ and the bare quark masses in
 the perturbative coefficients.
The last parenthesis is for the systematic error due to the fit Ansatz 
estimated using Eqs.~(\ref{eqn:5.4}) and  (\ref{eqn:5.4l}).
}
\label{table:EM}
\begin{tabular}{c|ll}
 $T$[MeV] & $p/T^{4}$ & $\epsilon/T^{4}$ 
 \cr
\hline
 $174$ 
& $0.13(60)(^{+4}_{-1})(1)(^{+0}_{-15})$ 
& $2.75(68)(^{+8}_{-14})(1)(^{+30}_{-89})$ 
 \cr
 $199$ 
& $-0.42(41)(^{+5}_{-0})(4)(^{+66}_{-19})$ 
& $8.54(57)(^{+15}_{-24})(4)(^{+21}_{-70})$ 
 \cr
 $232$ 
& $1.12(30)(^{+5}_{-4})(5)(^{+0}_{-23})$ 
& $13.07(38)(^{+11}_{-14})(5)(^{+54}_{-36})$ 
 \cr
 $279$ 
& $2.46(19)(^{+6}_{-5})(3)(^{+0}_{-52})$ 
& $14.74(25)(^{+14}_{-17})(3)(^{+0}_{-1.68})$ 
 \cr
 $348$ 
& $5.00(10)(^{+4}_{-3})(2)(^{+31}_{-2.63})$ 
& $16.15(13)(_{-23}^{+19})(2)(^{+1.36}_{-31})$ 
 \cr
$464$ 
& $7.596(65)(^{+11}_{-4})(9)(^{+1}_{-33})$ 
& $19.92(8)(14)(1)(^{+42}_{-77})$ 
 \cr
\hline
\end{tabular}
\end{table}
\begin{table}[tbh]
\centering
\caption{The same as Table \ref{table:EM} but for the entropy density
 and trace anomaly evaluated with the gradient flow method in the
 $t\to0$ limit.
}
\label{table:EM2}
\begin{tabular}{c|ll}
 $T$[MeV] & $(\epsilon+p)/T^{4}$ & $(\epsilon-3p)/T^{4}$ \cr
\hline
 $174$ 
& $2.90(43)(^{+7}_{-11})(0)(^{+76}_{-0})$ 
& $2.4(2.4)(^{+1}_{-2})(0)(^{+7}_{-0})$ \cr
 $199$ 
& $8.09(41)(^{+15}_{-20})(0)(^{+5}_{-17})$ 
& $9.8(1.7)(^{+1}_{-4})(1)(^{+8}_{-2.8})$ \cr
 $232$ 
& $14.25(28)(^{+16}_{-17})(0)(^{+91}_{-67})$ 
& $9.7(1.2)(0)(2)(^{+3}_{-0})$ \cr
 $279$ 
& $17.29(23)(^{+19}_{-21})(0)(^{+0}_{-1.80})$ 
& $7.38(73)(^{+0}_{-3})(14)(^{+1.31}_{-0})$ \cr
 $348$ 
& $21.25(12)(^{+21}_{-24})(0)(^{+0}_{-63})$ 
& $1.00(37)(^{+8}_{-14})(7)(^{+4.33}_{-1.08})$ \cr
$464$ 
& $27.53(8)(^{+15}_{-14})(0)(^{+0}_{-85})$ 
& $-2.87(23)(^{+10}_{-13})(4)(^{+0}_{-1.16})$ \cr
\hline
\end{tabular}
\end{table}

\subsection{Additional tests}

To confirm the validity of the results, we made a couple of additional tests on our numerical data.
The results of the tests are summarized in Appendix~\ref{sec:tests}.

Off diagonal components of the energy-momentum tensor correspond to the momentum and stress density, 
which should vanish on our lattices without external sources.
As discussed in~Appendix~\ref{sec:tests:od}, we confirm that they are consistent with zero within $2\sigma$ in the window adopted in the fits in Sec.~\ref{sec:t0extrapolation}.

We also study the gauge and quark contributions in Eq.~\eqref{eq:(2.8)} separately, and find that both contributions are equally important in the equation of state, 
while the singular term~$a^2/t$ comes dominantly from the quark contributions.
See Appendix~\ref{sec:tests:gq} for details.

Finally, we examine if the results depend on the choice of lattice operators for the field strength squared in~Eqs.~\eqref{eq:(2.9)} and~\eqref{eq:(2.10)} in~Appendix~\ref{sec:tests:op}.
We confirm that the dependence is small.

\subsection{Equation of state}
\label{sec:eos}

\begin{figure}[ht]
\centering
\includegraphics[width=10cm]{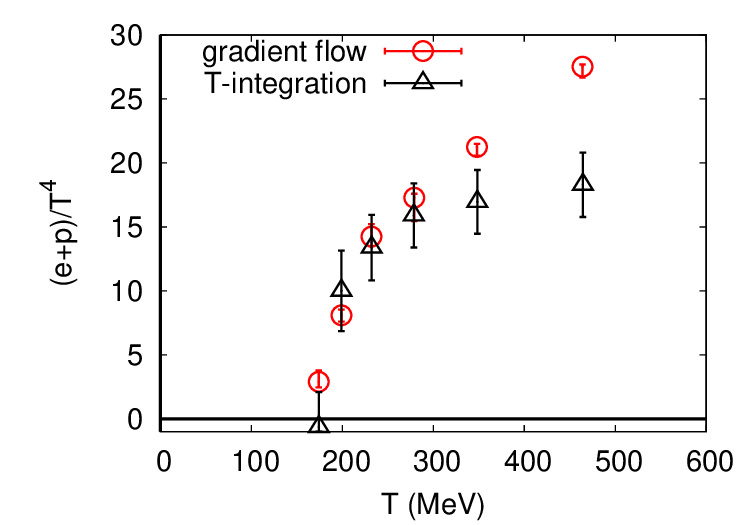}
\vspace{1cm}
\caption{Entropy density~$(\epsilon+p)/T^4$ as a function of temperature. Red circles are our
result with the gradient flow method.  
Errors include both statistical and systematic errors.
Black triangles are previous results obtained 
by the $T$-integration method~\cite{Umeda:2012er}.
}
\label{fig6}
\end{figure}

\begin{figure}[ht]
\centering
\includegraphics[width=10cm]{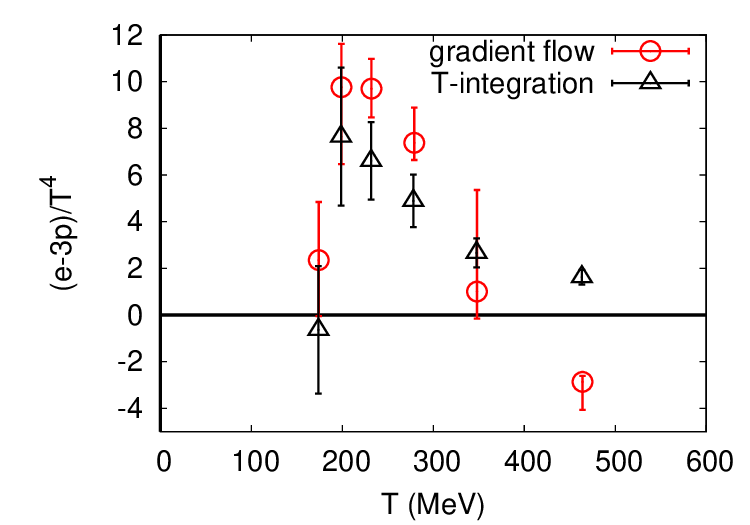}
\vspace{1cm}
\caption{The same as Fig.~\ref{fig6} but for the trace anomaly~$(\epsilon-3p)/T^4$.}
\label{fig8}
\end{figure}

\begin{figure}[ht]
\centering
\includegraphics[width=10cm]{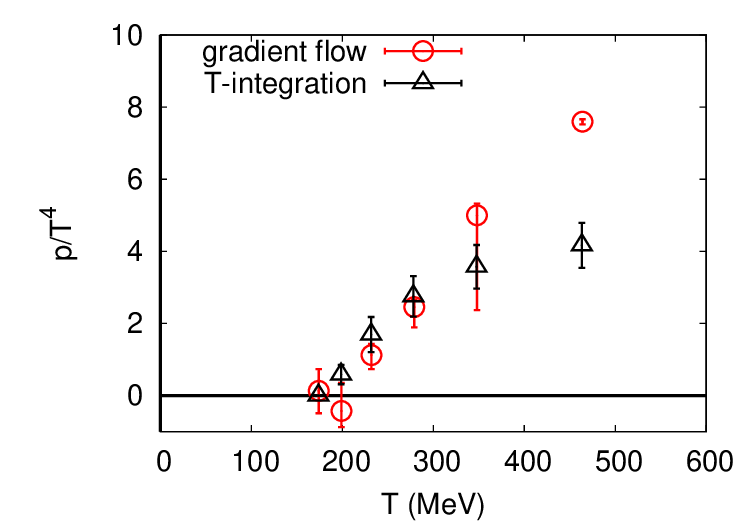}
\vspace{1cm}
\caption{The same as Fig.~\ref{fig6} but for the pressure $p/T^4$.
In the $T$-integration method, the pressure is set to be zero at $T\simeq174\,\mathrm{MeV}$.}
\label{fig2}
\end{figure}

\begin{figure}[ht]
\centering
\includegraphics[width=10cm]{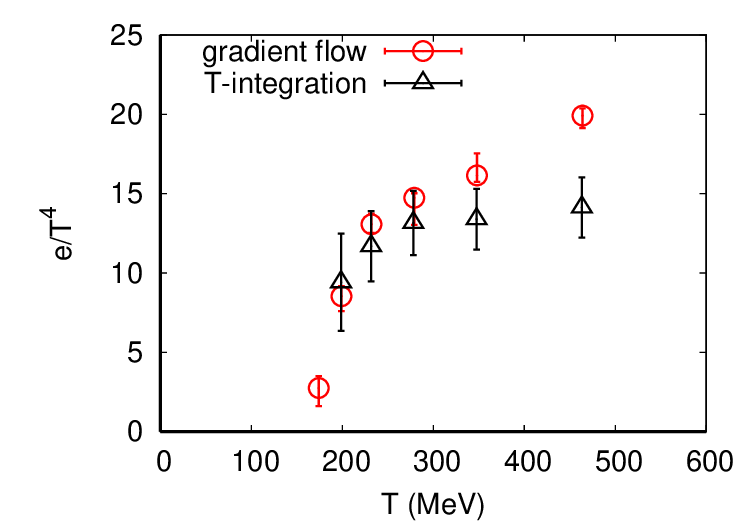}
\vspace{1cm}
\caption{The same as Fig.~\ref{fig6} but for the energy density~$\epsilon/T^4$.}
\label{fig4}
\end{figure}

Our results for the equation of state with the gradient flow method are plotted in 
Figs.~\ref{fig6}, \ref{fig8}, \ref{fig2}, and \ref{fig4} as functions of temperature.
For the pressure and the energy density, we have repeated the same set of analyses from the results of the energy-momentum tensor at $t>0$.
Errors of our data (red open circles) include the statistical error and
the systematic errors from the perturbative coefficients and fit Ansatz.

Also shown in these figures by open triangles are the results obtained previously 
by the $T$-integration method using the same set of configurations~\cite{Umeda:2012er}.
We find that our result of the gradient flow method is well consistent with the result of the conventional method at~$T\simle279\,\mathrm{MeV}$. 
On the other hand, the two results show a deviation at $T\simge348\,\mathrm{MeV}$. 
This may be due to a lattice artifact of $O\left((aT)^2\right)=O\left(1/N_t^2\right)$ from the discretization of thermal modes. 
Our data suggest that such an artifact is not negligible for $N_t\simle8$.

It should be kept in mind that a definite comparison is possible only after taking the continuum limit. 
Nevertheless, besides the results at $N_t\simle8$ which suffer from the small-$N_t$ artifact, 
we obtain good agreement with a conventional method at $N_t \simge 10$ on our finite lattices. 
This may be suggesting that our $a \simeq 0.07$ fm with improved gauge and quark actions is already quite close to the continuum limit.

Here, we emphasize that the values of the beta functions --- $a(d\beta/da)$, $a(d\kappa_{ud}/da)$, and $a(d\kappa_s/da)$ for the present case --- are not required with the gradient flow method.
This will help much to evaluate the equation of state with dynamical quarks in future.

\section{Results for the chiral condensate and disconnected  susceptibility}
\label{sec:chiralcond}

\subsection{Chiral condensate}
\label{sec:chiralcond-t}

\begin{figure}[ht]
\centering
\includegraphics[width=6.cm]{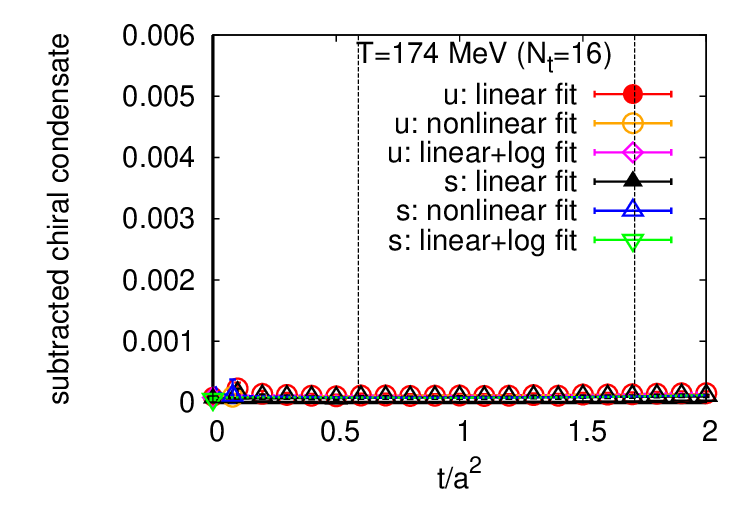}
\includegraphics[width=6.cm]{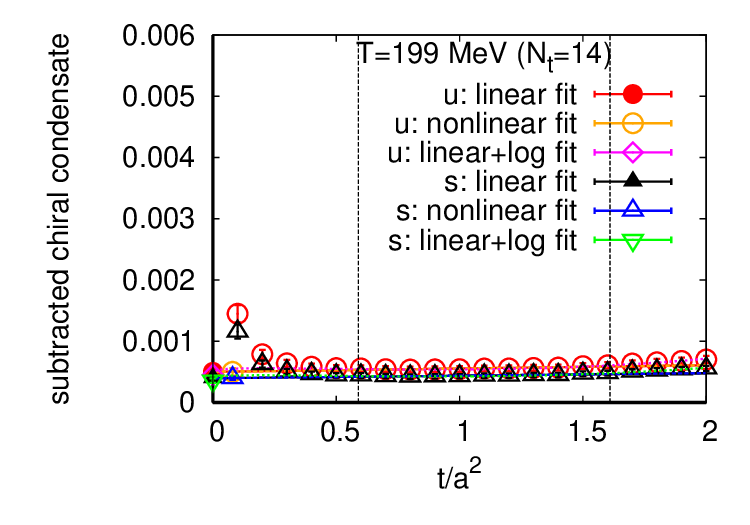}
\includegraphics[width=6.cm]{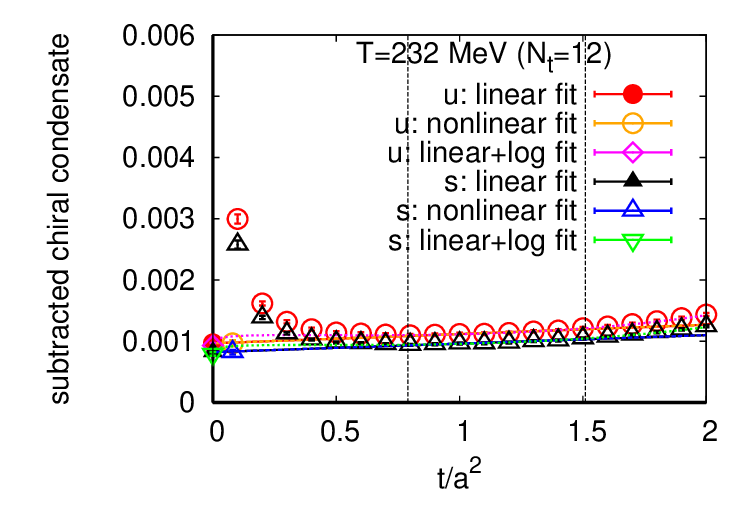}
\includegraphics[width=6.cm]{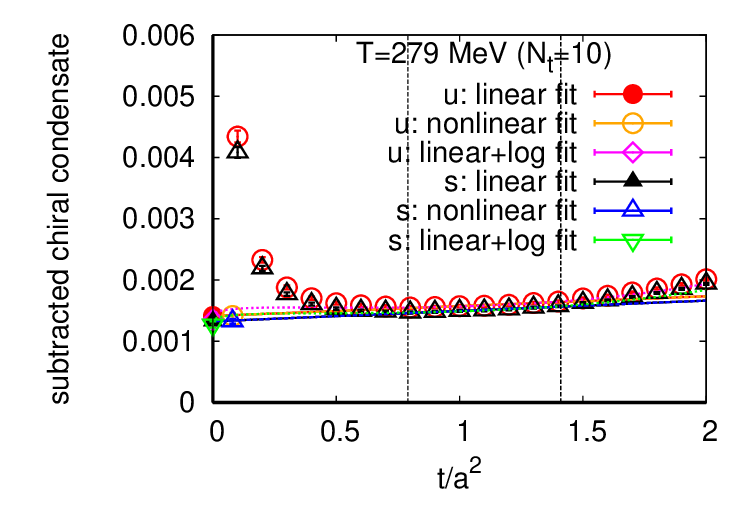}
\includegraphics[width=6.cm]{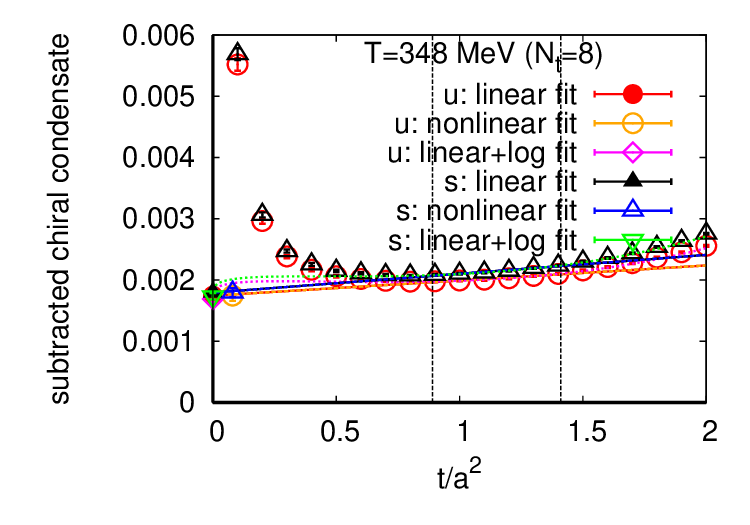}
\includegraphics[width=6.cm]{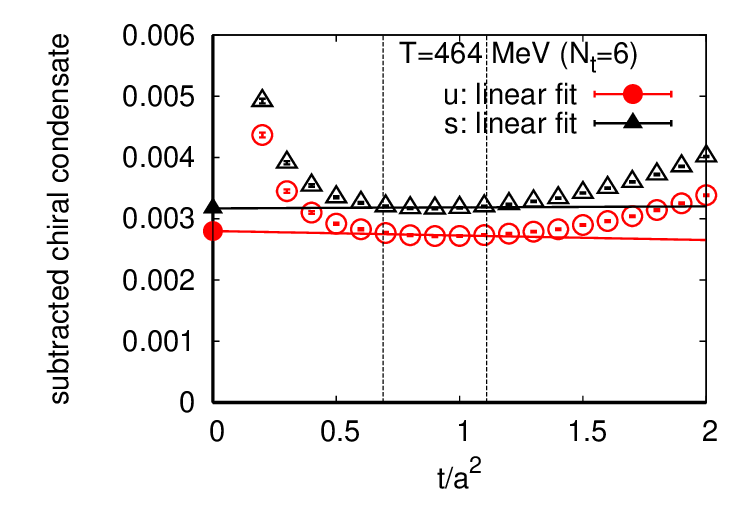}
\includegraphics[width=6.cm]{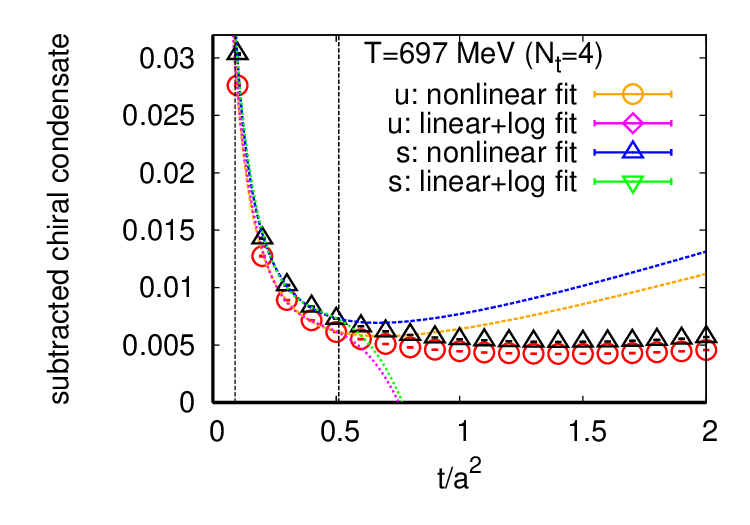}
\vspace*{-0.5cm}
\caption{
 Chiral condensate~$\left\langle\{\Bar{\psi}_f\psi_f\}\right\rangle$ with VEV
 subtraction as a function of the flow time.
 The vertical axis is in lattice unit.  
 Red open circles and black open triangles are for $f=u$ (or $d$) and $s$, respectively. 
  From the top left to the bottom: $T\simeq174$, 199, 232, 279, 348, 464, and 697 MeV 
 ($N_t=16$, 14, 12, 10, 8, 6, and 4, respectively).
 The filled symbols at~$t=0$ are the renormalized chiral condensate given
 by taking the $t\to0$~limit with the linear fit.
 Orange and blue dashed curves with open symbols at
 $t\sim0$ are the results of the nonlinear fit for $u$ and $s$ quark, respectively.
 Magenta and green dashed curves with open symbols at
 $t\sim0$ are the results of the liner+log fit for $u$ and $s$ quark, respectively.
 Pair of dashed vertical lines shows the window used for the fits.
 Errors are statistical only.
}
\label{fig:bpsipsi_subvev}
\end{figure}

In~Fig.~\ref{fig:bpsipsi_subvev}, we show the VEV subtracted chiral condensate at $T>0$ as a function of the flow time. 
We note that the singularity at small $t$ is quite mild in the subtracted chiral condensate at least at low temperatures. 
This suggests that the VEV subtraction not only removes the $m_f/t$ singularity but also reduces the $a^2/t$ singularity.

We adopt the same strategy as that for the energy-momentum tensor to extract renormalized chiral condensate with the VEV subtraction. 
In~Fig.~\ref{fig:bpsipsi_subvev}, results of the linear fits using windows shown by a pair of vertical dashed lines, are given by red and black solid lines for $T\simle464$ MeV. 
Here, the windows are chosen so that the linear fit gives $\chi^2/N_\textrm{dof} \le O(1)$.
The filled red circles and black triangles at $t=0$ are the results of their $t\to0$ extrapolations.
We also perform nonlinear fits similar to Eq.~\eqref{eqn:5.4}
and liner+log fits similar to Eq.~\eqref{eqn:5.4l}, 
adopting the same window.
The results of nonlinear fits are shown by orange and blue dashed curves, and corresponding renormalized chiral condensates are shown by orange and blue open symbols at $t\sim0$.
The results of linear+log fits are shown by magenta and green dashed curves associated with open symbols at $t\sim0$.
At $T\simeq464$ MeV, the nonlinear and linear+log fits are not applicable because we do not have enough number of data points in the window ($N_\textrm{dof} \le 1$).

\begin{figure}[ht]
\centering
\includegraphics[width=10cm]{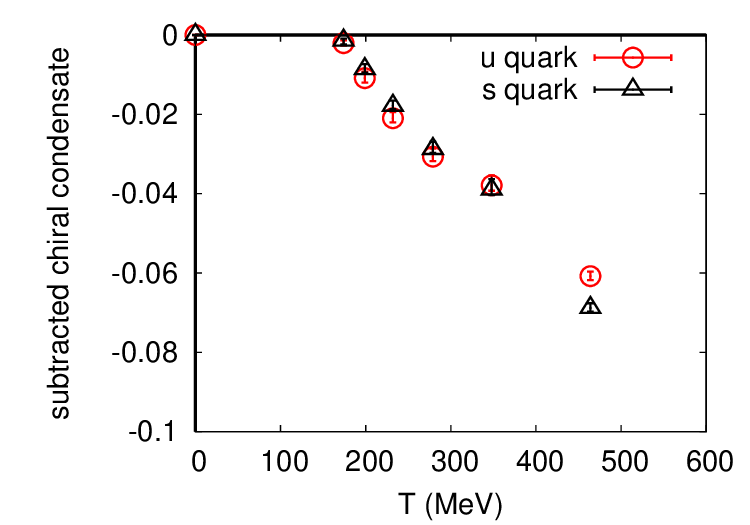}
\vspace{1cm}
\caption{Renormalized chiral condensate with the VEV subtraction,
$-\left\langle\{\Bar{\psi}_f\psi_f\}\right\rangle_{\overline{\mathrm{MS}}}(\mu\!=\!2\,\mathrm{GeV})$, 
in $\overline{\mathrm{MS}}$ scheme as a function of temperature. 
Following a convention, the sign is flipped in the figure.
The vertical axis is in unit of ${\rm GeV}^3$.
Red circles are $u$ (or $d$) quark condensate and black triangles are that for $s$ quark. 
Errors include the statistical error and the systematic error from the perturbative coefficients and fit Ansatz, 
except for the data at $T\simeq464$MeV for which the systematic error due to fit Ansatz was not estimated.
}
\label{fig:bpsipsi_subvev_T}
\end{figure}

From Fig.~\ref{fig:bpsipsi_subvev}, we find that the results of the nonlinear 
and the linear+log fits at $T\simle 348$ MeV 
are consistent with those of the linear fits within 2$\sigma$
of the statistical error. 
We adopt the results of the linear fit for our central values and 
take the deviation due to the nonlinear or linear+log fits as an estimate of the systematic error due to the fit Ansatz.
Final results for the renormalized chiral condensate with the VEV subtraction, 
$
\left\langle\{\Bar{\psi}_f\psi_f\}(x)\right\rangle_{\overline{\mathrm{MS}}}(\mu\!=\!2{\mathrm{GeV}})
$
with $f=u$ (or $d$) and $s$, evaluated in the $t\to0$ limit, are summarized in Tables~\ref{table:bpsipsi-u} and \ref{table:bpsipsi-s}.
In Fig.~\ref{fig:bpsipsi_subvev_T}, we show the renormalized chiral condensates with the VEV subtraction in physical units as a function of the temperature.
Following a convention, the sign is flipped in the figure.
We find that the condensates start to decrease just below $T\sim199$ MeV. 
This is consistent with a previous estimation of the pseudocritical temperature $T_{\mathrm{pc}}\sim190$ MeV~\cite{Umeda:2012er}. 
We also find that the valence quark mass dependence is small in Fig.~\ref{fig:bpsipsi_subvev_T}.  
This suggests that the difference between these two condensates are mostly subtracted out by that at zero
temperature, i.e.\  the mass dependent part of the chiral condensate is almost temperature independent.

\begin{table}[tbh]
\centering
\caption{Renormalized chiral condensate with the VEV subtraction and 
disconnected chiral susceptibility for $u$ (or, equivalently, $d$) quark, evaluated in the $t\to0$ limit.
The values are in lattice unit.
The susceptibility is given in a unit of $10^{-8}$.
The first parenthesis is for the statistical error. 
The second is for systematic error due to
 $a\Lambda_{\overline{\mathrm{MS}}}^{(3)}$ in the perturbative coefficients.
The last  parenthesis is that due to the fit Ansatz estimated using nonlinear and linear+log fits.
At $T\simeq464$MeV, the systematic error due to fit Ansatz was not estimated. See text.
}
\label{table:bpsipsi-u}
\begin{tabular}{c|ll}
 $T$ [MeV] 
  & $a^{3}\left\langle\{\Bar{\psi}_u\psi_u\}(x)\right\rangle_{\overline{\mathrm{MS}}}$
  & $a^{6}\chi_{\Bar{u}u}^{\mathrm{disc.}}\times10^8$ \cr
\hline
 $0$ 
 & $0$ & $0.46(15)(^{+4}_{-10})(^{+2}_{-0})$ \cr
 $174$ 
 & $0.000094(28)(^{+28}_{-5})(^{+0}_{-12})$
 & $2.19(80)(^{4}_{15})(^{+0}_{-23})$ \cr  
 $199$ 
 & $0.000500(53)(^{+9}_{-19})(^{+0}_{-47})$
 & $5.0(1.7)(^{+1}_{-4})(^{+0}_{-5})$ \cr
$232$ 
 & $0.000967(40)(^{+26}_{-51})(^{+0}_{-34})$
 & $1.35(30)(^{+7}_{-19})(^{+1}_{-0})$ \cr
 $279$ 
 & $0.001413(42)(^{+29}_{-58})(^{+0}_{-62})$
 & $1.04(32)(^{+0}_{-3})(^{+2}_{-0})$ \cr
 $348$ 
 & $0.001744(44)(^{+46}_{-90})(^{+0}_{-55})$
 & $1.07(24)(^{+4}_{-5})(^{+0}_{-10})$ \cr
 $464$ 
 & $0.002800(44)(^{+9}_{-31})$(--)
 & $1.27(13)(7)$(--) \cr
\hline
\end{tabular}
\end{table}

\begin{table}[tbh]
\centering
\caption{
The same as Table~\ref{table:bpsipsi-u} but for $s$  quark.
}
\label{table:bpsipsi-s}
\begin{tabular}{c|ll}
 $T$ [MeV] 
  & $a^{3}\left\langle\{\Bar{\psi}_s\psi_s\}(x)\right\rangle_{\overline{\mathrm{MS}}}$
  & $a^{6}\chi_{\Bar{s}s}^{\mathrm{disc.}}\times10^8$ \cr
\hline
 $0$ 
 & $0$ & $0.320(88)(^{+20}_{-51})(^{+17}_{-0})$ \cr
 $174$ 
 & $0.000066(21)(^{+2}_{-4})(^{+38}_{-9})$
 & $1.41(43)(^{+0}_{-6})(^{+0}_{-16})$ \cr
 $199$ 
 & $0.000396(41)(^{+7}_{-15})(^{+0}_{-38})$
 & $3.3(1.0)(^{+0}_{-2})(^{+0}_{-4})$ \cr
 $232$ 
 & $0.000823(31)(^{+23}_{-44})(^{+0}_{-28})$
 & $1.04(19)(^{+5}_{-14})(^{+1}_{-0})$ \cr
 $279$ 
 & $0.001325(35)(^{+29}_{-57})(^{+0}_{-52})$
 & $0.90(25)(^{+0}_{-4})(^{+4}_{-0})$ \cr
 $348$ 
 & $0.001794(38)(^{+52}_{-99})(^{+0}_{-60})$
 & $1.07(26)(^{+4}_{-5})(^{+0}_{-11})$ \cr
 $464$ 
 & $0.003170(38)(^{+14}_{-41})$(--)
 & $1.30(15)(7)$(--) \cr
\hline
\end{tabular}
\end{table}

At $T\simeq697$ MeV ($N_t=4$),  because no clear linear window can be identified below $t_{1/2}=0.5$, 
we attempt a nonlinear fit without the $t^2$ term adopting
the same fit range $t=[0.1,0.5]$ as in Sec.~\ref{sec:t0extrapolation}.
The results are shown in the last panel of Fig.~\ref{fig:bpsipsi_subvev}.
However, since the lattice artifact terms are dominating in the fit range, we disregard the data at $T\simeq697$ MeV in the followings.

\subsection{Disconnected chiral susceptibility}
\label{sec:chiralsuscept}

As a by-product of the chiral condensate calculation, we study disconnected chiral
susceptibility defined by 
\begin{equation}
   \chi_{\Bar{f}f}^{\mathrm{disc.}}
   = \left\langle \left[ \frac{1}{N_\Gamma} \sum_x \{\Bar{\psi}_f\psi_f\}(x) \right]^2 \right\rangle_{\!\mathrm{disconnected}}
   - \left[ \left\langle\frac{1}{N_\Gamma} \sum_x \{\Bar{\psi}_f\psi_f\}(x)\right\rangle \right]^2,
\end{equation}
where the connected quark loop contribution is dropped from the scalar density two
point function. Though this quantity is not the physical susceptibility, it is easy to be 
measured and may be used as a guide to detect the chiral restoration transition.
Because the VEV subtraction has no effect on this quantity, we can compute it also at $T=0$.

\begin{figure}[ht]
\centering
\includegraphics[width=6.5cm]{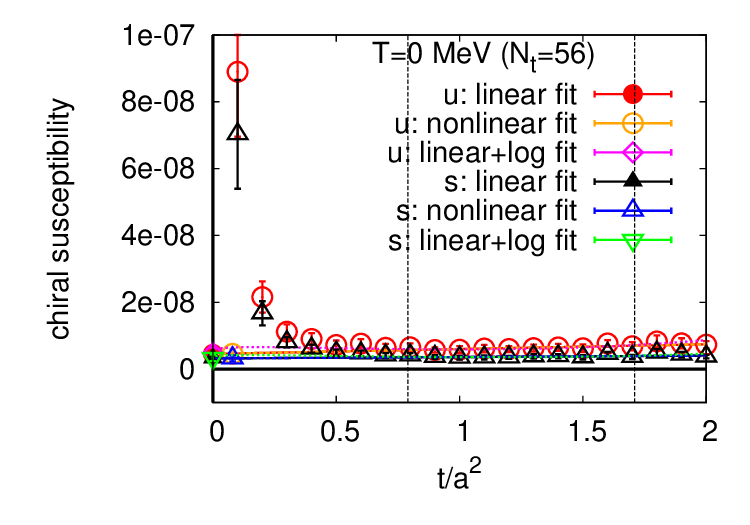}
\includegraphics[width=6.5cm]{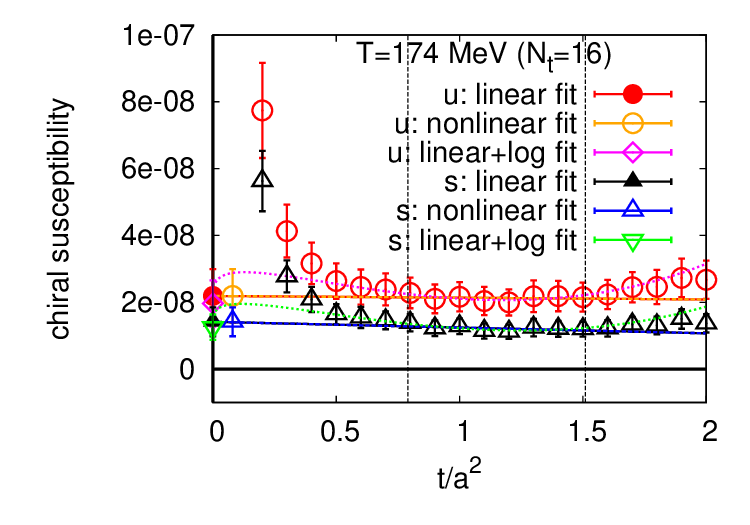}
\includegraphics[width=6.5cm]{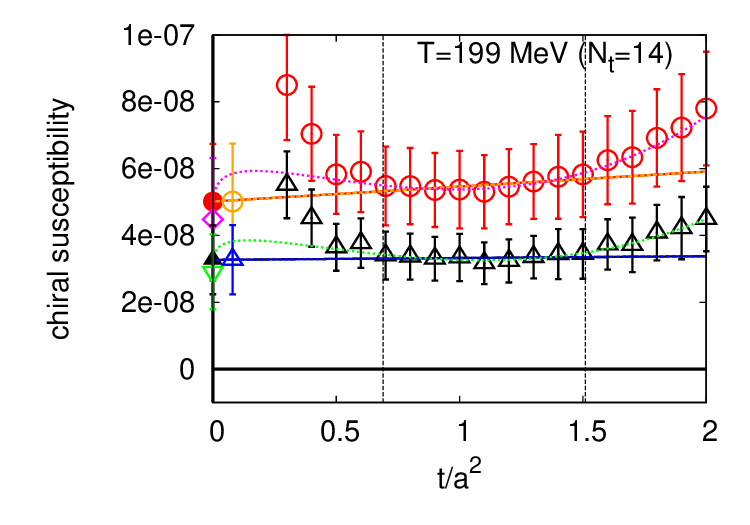}
\includegraphics[width=6.5cm]{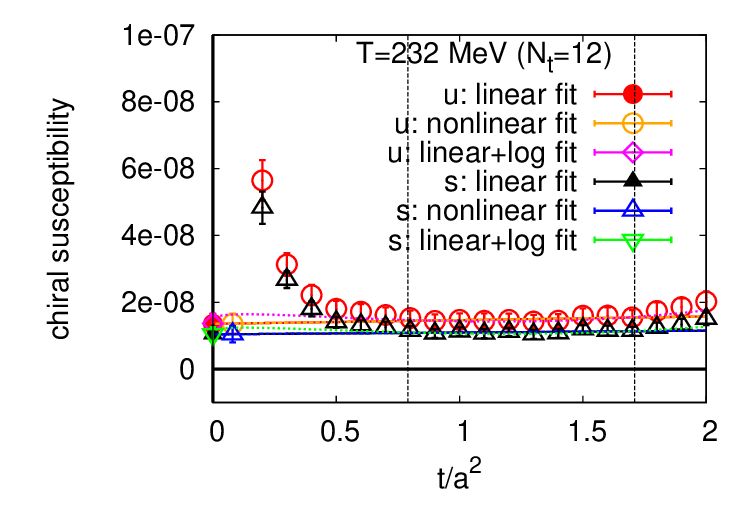}
\includegraphics[width=6.5cm]{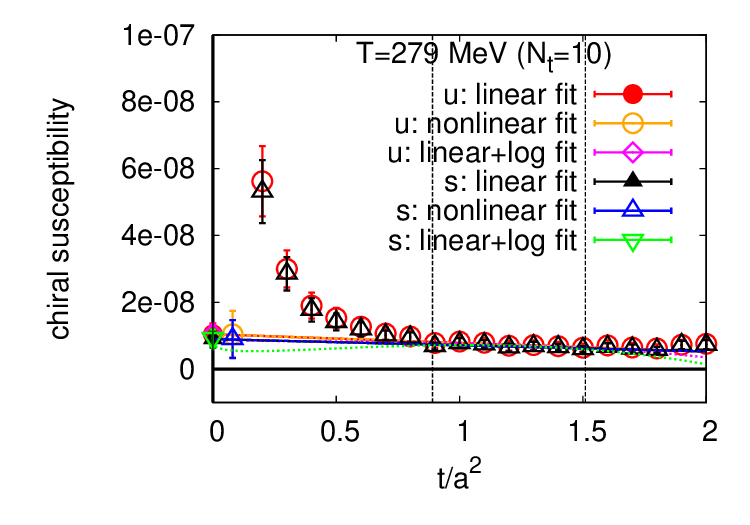}
\includegraphics[width=6.5cm]{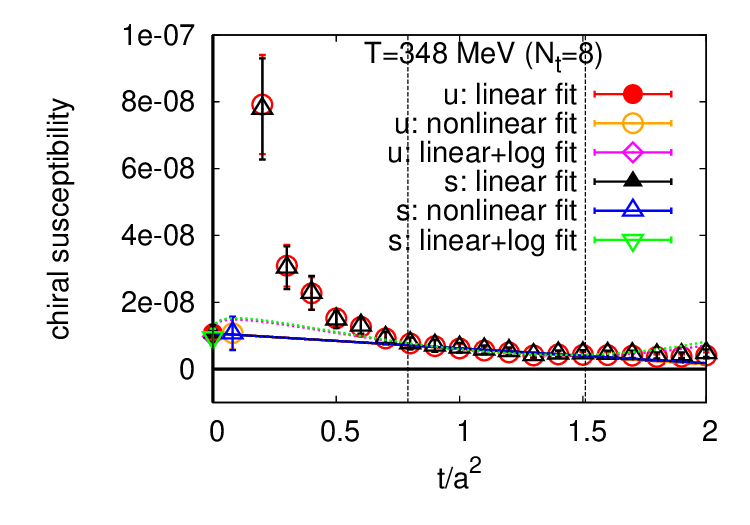}
\includegraphics[width=6.5cm]{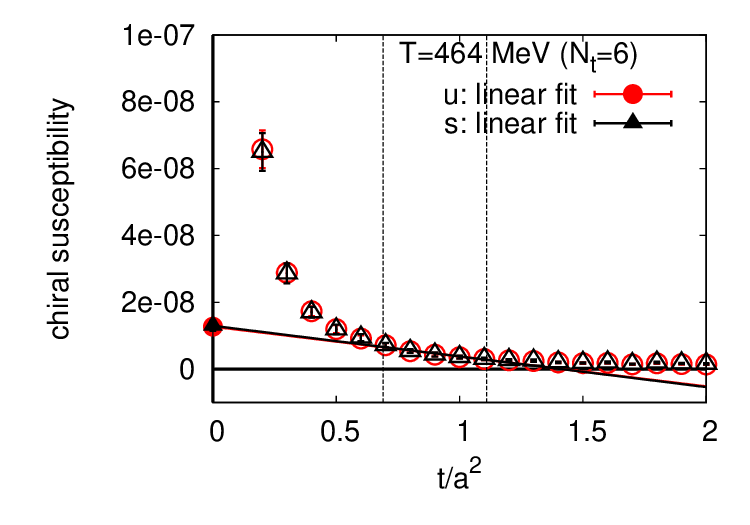}
\includegraphics[width=6.5cm]{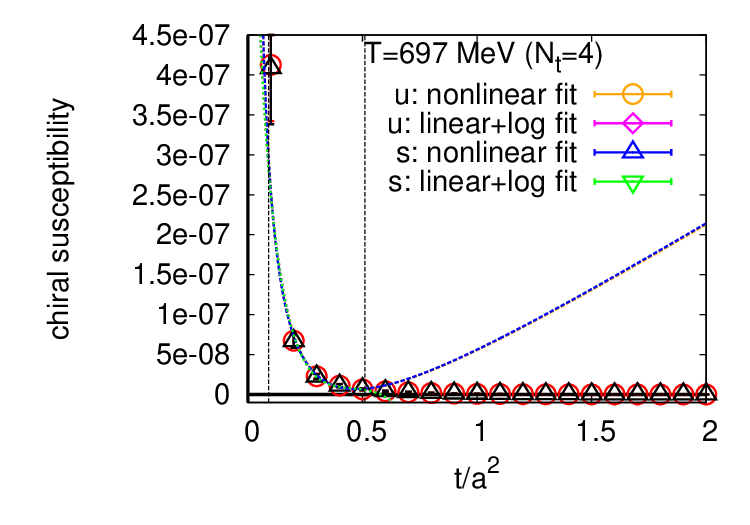}
\vspace{1cm}
\caption{The same as Fig.~\ref{fig:bpsipsi_subvev} but for the 
disconnected chiral susceptibility $\chi_{\Bar{f}f}^{\mathrm{disc.}}(\mu=2\,\mathrm{GeV})$. 
 From the top left to the bottom: $T\simeq0$, 174, 199, 232, 279, 348, 464, and 697 MeV 
 ($N_t=54$, 16, 14, 12, 10, 8, 6, and 4, respectively).
}
\label{fig:chiral_susceptibility}
\end{figure}

In Fig.~\ref{fig:chiral_susceptibility}, we plot the disconnected chiral
susceptibility as a function of the flow time. 
We find good linear windows below $t_{1/2}$ at $T\simle348$ MeV ($N_t\ge8$) 
and a marginal window at $T\simeq464$ MeV ($N_t=6$),
while at $T\simeq697\,\mathrm{MeV}$ ($N_t=4$) no linear window can be identified below $t_{1/2}=0.5$. 
We find that the linear and nonlinear fits give completely consistent results for $T\simle348$ MeV, 
while the linear+log fit sometimes deviate but maximally by about $1\sigma$ of the statistical error. 
At $T\simeq464$ MeV, the number of data points in the window is not enough to carry out the nonlinear as well as the linear+log fits.
At $T\simeq697\,\mathrm{MeV}$, though we test the nonlinear fit adopting the same fit range 
$t= [0.1,0.5]$ as in Secs.~\ref{sec:t0extrapolation} and \ref{sec:chiralcond-t}, 
because the lattice artifact term is dominating in the fit as shown in the last plot of Fig.~\ref{fig:chiral_susceptibility}, we do not take the result as reliable and just disregard it.

\begin{figure}[ht]
\centering
\includegraphics[width=10cm]{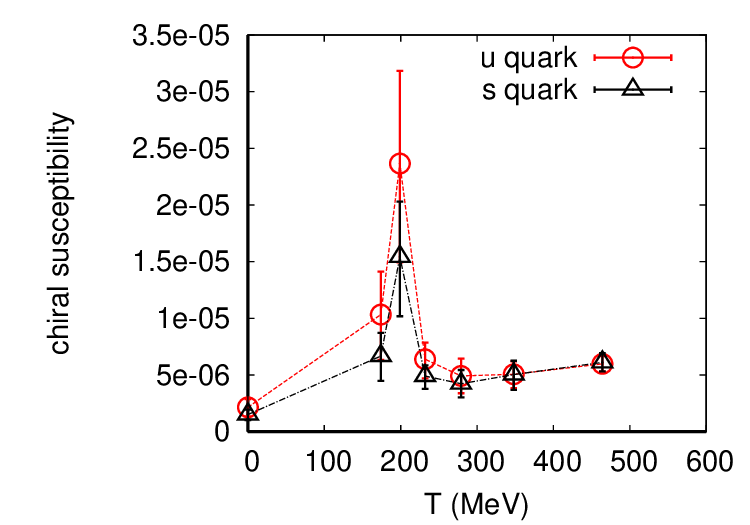}
\vspace{1cm}
\caption{Disconnected chiral
susceptibility~$\chi_{\Bar{f}f}^{\mathrm{disc.}}(\mu=2\,\mathrm{GeV})$
renormalized in $\overline{\mathrm{MS}}$ scheme as a function of temperature.
The vertical axis is in unit of $\mathrm{GeV}^6$.
Red circles are those of $u$ (or $d$) quark and black triangles are those for $s$
quark.
Errors include the statistical error and the systematic errors from the perturbative coefficients and fit Ansatz.
}
\label{fig:chiral_susceptibility_T}
\end{figure}

Results of the renormalized disconnected chiral susceptibility are summarized in the last columns of 
Tables~\ref{table:bpsipsi-u} and \ref{table:bpsipsi-s}, and 
shown in Fig.~\ref{fig:chiral_susceptibility_T} as a function of temperature. 
Errors include the statistical error and the systematic errors from the perturbative coefficients 
and the fit Ansatz.
In Fig.~\ref{fig:chiral_susceptibility_T}, we find a clear peak 
at $T\simeq199$ MeV, which may be indicating the pseudocritical point around this temperature. 
This is consistent with a previous estimate
of $T_{\mathrm{pc}}\sim190$ MeV for the chiral restoration crossover~\cite{Umeda:2012er}. 
We also note that, although the errors are large, the height of the peak increases as we decrease the
valence quark mass from that of  $s$ to $u$ (or $d$). 
Since the sea quark masses are not varied, we do not attempt to extrapolate the results to the chiral limit, 
but the tendency is consistent with our expectation.

\subsection{Chiral condensate without the VEV subtraction}

\begin{figure}[ht]
\centering
\includegraphics[width=6.5cm]{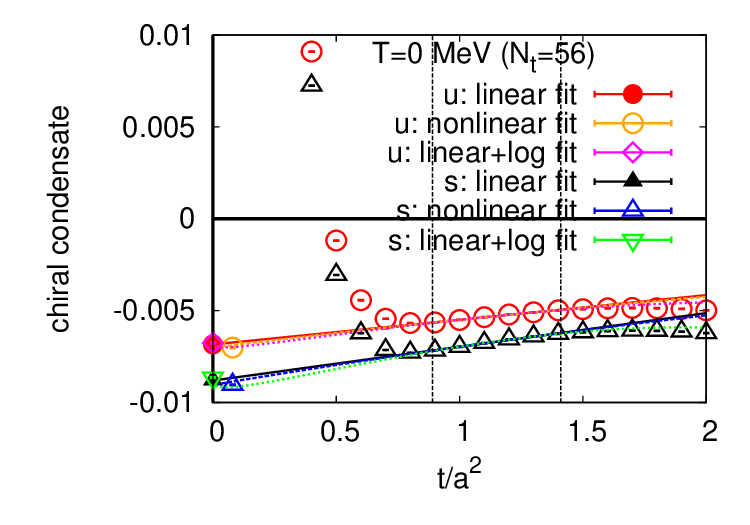}
\includegraphics[width=6.5cm]{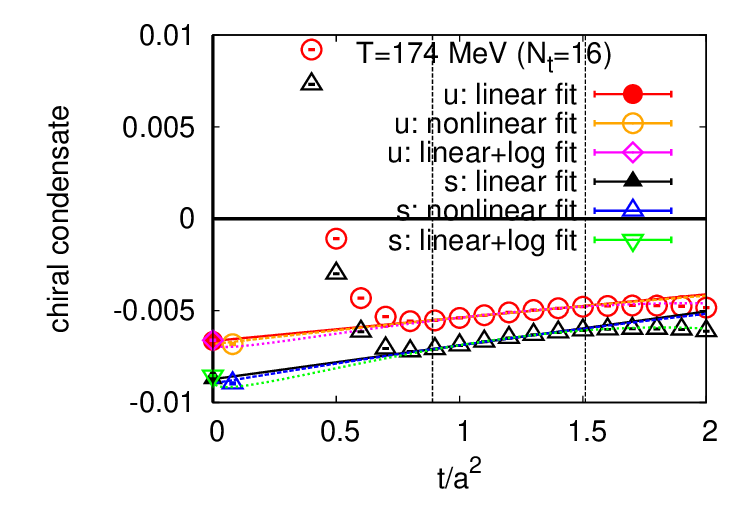}
\includegraphics[width=6.5cm]{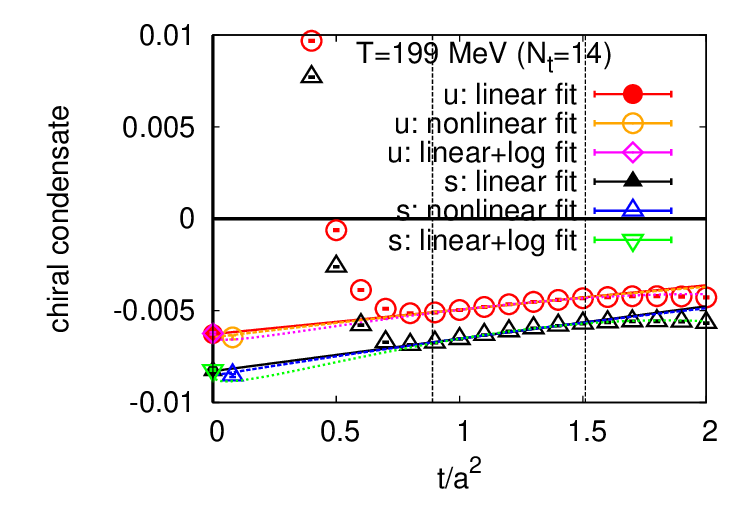}
\includegraphics[width=6.5cm]{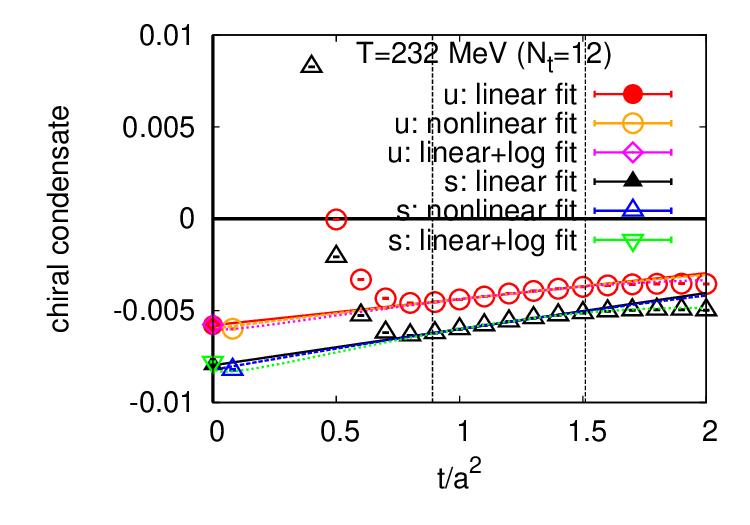}
\includegraphics[width=6.5cm]{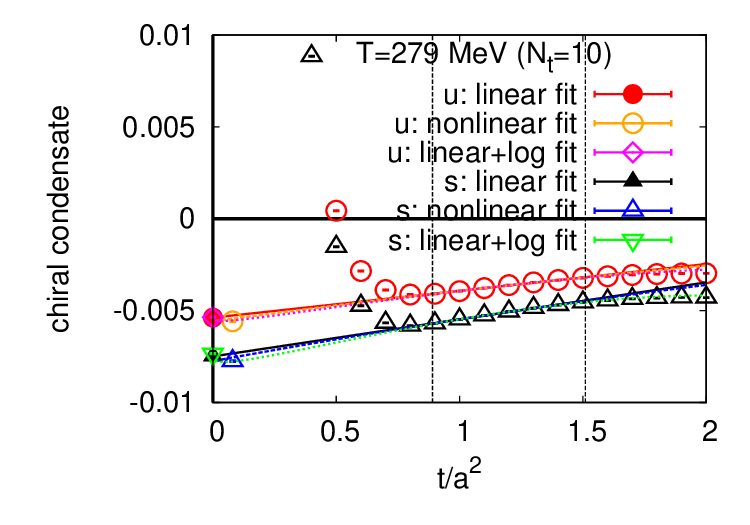}
\includegraphics[width=6.5cm]{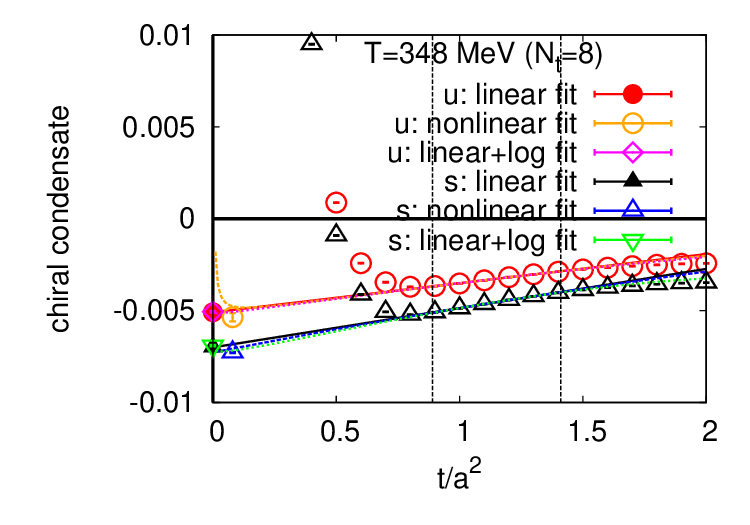}
\includegraphics[width=6.5cm]{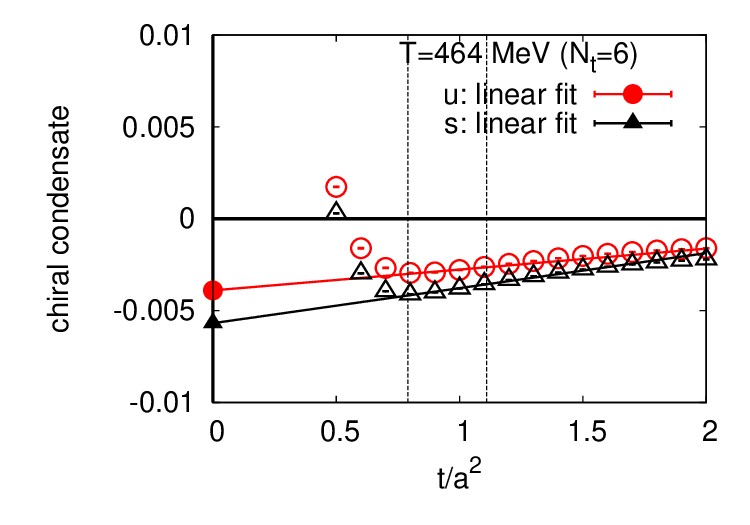}
\includegraphics[width=6.5cm]{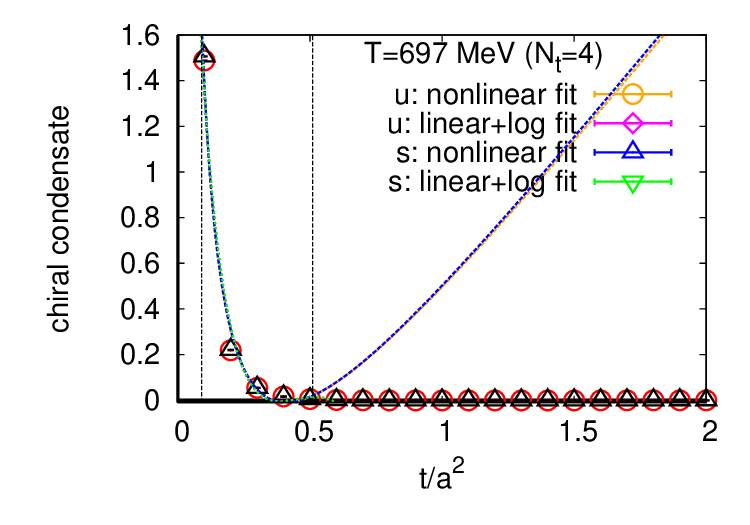}
\vspace{1cm}
\caption{
 The same as Fig.~\ref{fig:bpsipsi_subvev} but for the unsubtracted chiral
 condensate~$\left\langle\{\Bar{\psi}_f\psi_f\}^{(0)}\right\rangle$.
 From the top left to the bottom: $T\simeq0$, 174, 199, 232, 279, 348, 464, and 697 MeV 
 ($N_t=54$, 16, 14, 12, 10, 8, 6, and 4, respectively).
}
\label{fig:bpsipsi}
\end{figure}

Finally, we examine the effect of the VEV subtraction in the chiral condensate.
In Fig.~\ref{fig:bpsipsi}, we show the unsubtracted  chiral condensate 
$
\left\langle \left\{\Bar{\psi}_f\psi_f\right\}^{(0)}\!(t,x,a)\right\rangle 
$
averaged over lattice points.
Red open circles and black open triangles are for $f=u$ (or $d$) and $s$, respectively. 
As discussed in~Secs.~\ref{sec:2.3} and \ref{sec:finitet}, this quantity will have both $m_f/t$ and $a^2/t$ singularities towards $t\to0$. 
We note that the singularity of the subtracted chiral condensate shown in Fig.~\ref{fig:bpsipsi_subvev} at small $t$ is much milder than that of the unsubtracted chiral condensate shown in Fig.~\ref{fig:bpsipsi},
suggesting that the VEV subtraction not only removes the $m_f/t$ singularity but also reduces the $a^2/t$ singularity, as expected.

Adopting the same strategy as those for the energy-momentum tensor and the subtracted chiral condensate, 
we perform linear and nonlinear fits to the unsubtracted chiral condensate, 
to extract the renormalized chiral condensate in the $t\to0$ limit.
The linear windows determined by a study of $\chi^2/N_\textrm{dof}$ of the linear fits are shown by the pair of vertical dashed lines in Fig.~\ref{fig:bpsipsi}. 
We note that the values of $\chi^2/N_\textrm{dof}$ are in general worse than those for the subtracted chiral condensate, presumably due to the stronger singularities. 
Results of the linear fits for the renormalized chiral condensate are shown by filled red circles ($u$ or $d$ quark) and black triangles ($s$ quark) at $t=0$ in Fig.~\ref{fig:bpsipsi}.\footnote{$\chi^2/N_\textrm{dof}$ of the linear fits are less than 5, 
except for those for the strange quark condensate at $T=0$ and condensates at
$T=464$ MeV, for which $\chi^2/N_\textrm{dof}$ exceed 10.}
Corresponding results of the nonlinear fits are shown by green and blue open symbols at $t\sim0$ for $T\simle348$ MeV.
At $T\simeq464$ MeV, the nonlinear fit is not applicable because we do not have enough number of data points in the window.
At $T\simeq697$ MeV ($N_t=4$), though we attempt a nonlinear fit using the data in $t=[0.1,0.5]$, 
because the lattice artifact term is dominating in the fit, the results are not reliable for physical discussions.

\begin{figure}[ht]
\centering
\includegraphics[width=10cm]{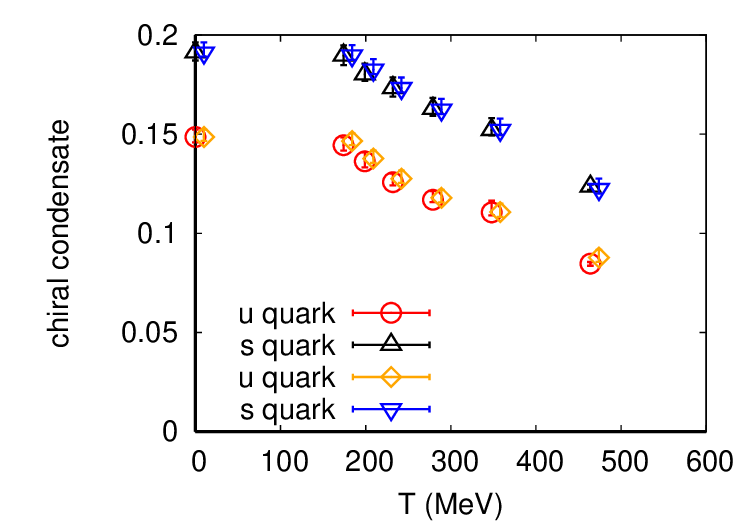}
\vspace{1cm}
\caption{Renormalized chiral condensate in $\overline{\mathrm{MS}}$ scheme,
$-\left\langle\{\Bar{\psi}_f\psi_f\}(x)\right\rangle^{(0)}_{\overline{\mathrm{MS}}}(\mu\!=\!2\,\mathrm{GeV})$,
as a function of temperature. 
Following a convention, the sign is flipped in the figure.
The vertical axis is in unit of ${\rm GeV}^3$.
Red circles and black upward triangles are $u$ (or $d$) and $s$ quark condensate extracted directly from  the unsubtracted chiral condensate shown in Fig.~\ref{fig:bpsipsi}.
Orange diamonds and blue downward triangles are $u$ and $s$ quark condensate obtained by adding the VEV of Eqs~(\ref{eq:vevud}) (\ref{eq:vevs}) to the subtracted chiral condensate shown in Fig.~\ref{fig:bpsipsi_subvev_T}.
Orange and blue symbols are slightly shifted in the horizontal direction for clarity.
Errors include the statistical error and the systematic error from the perturbative coefficients and fit Ansatz,
while, at $T\simeq464$MeV, the systematic error due to fit Ansatz was not estimated.
}
\label{fig:bpsipsiT}
\end{figure}

At $T\simle348$ MeV, the discrepancy between the two fit Ans\"atze turned out to be 2\%--5\%. 
This suggests that the singular terms in the unsubtracted chiral condensate 
are well controlled within the linear windows. 
As in the previous sections, we take the results of the linear fits as the central values and take the difference between the linear and nonlinear fits as an estimate of the systematic error due to the fit Ansatz.
The results for the renormalized chiral condensate without the VEV subtraction,
$\left\langle\{\Bar{\psi}\psi\}(x)\right\rangle^{(0)}_{\overline{\mathrm{MS}}}(\mu\!=\!2\,\mathrm{GeV})$,
are shown by red circles ($u$ or $d$ quark) and black upward triangles ($s$ quark)
in Fig.~\ref{fig:bpsipsiT}. 
At $T=0$, we obtain
\begin{eqnarray}
&&
\left. a^{3}\left\langle\{\Bar{\psi}_u\psi_u\}(x)\right\rangle^{(0)}_{\overline{\mathrm{MS}}} \right|_{T=0}
=-0.006841(33)(^{+82}_{-0})(^{+84}_{-170}),
\label{eq:vevud}
\\&&
\left. a^{3}\left\langle\{\Bar{\psi}_s\psi_s\}(x)\right\rangle^{(0)}_{\overline{\mathrm{MS}}}  \right|_{T=0}
=-0.008803(24)(^{+94}_{-0})(^{+159}_{-235}) .
\label{eq:vevs}
\end{eqnarray}
The first parenthesis is for the statistical error,  
the second is for the systematic error due to $a\Lambda_{\overline{\mathrm{MS}}}^{(3)}$ in the perturbative coefficients,
and the third is for the systematic error due to fit Ansatz 
estimated using nonlinear and linear+log fits.

To compare the results of the unsubtracted chiral condensate with those of the subtracted chiral condensate discussed in Sec.~\ref{sec:chiralcond-t}, we add back the VEV's given by Eqs.~(\ref{eq:vevud}) and (\ref{eq:vevs}) to the results of the subtracted chiral condensate shown in Fig.~\ref{fig:bpsipsi_subvev_T}.
Results are shown by orange diamonds ($u$ or $d$ quark) and blue downward triangles ($s$ quark) in Fig.~\ref{fig:bpsipsiT}.
We find that the results are completely consistent with those of direct fits to the unsubtracted chiral condensate.

\section{Conclusions and discussions}
\label{sec:conclusion}
In this paper we apply the gradient flow
method of Refs.~\cite{Suzuki:2013gza,Makino:2014taa} to calculate the 
energy-momentum tensor in $(2+1)$-flavor QCD. 
As the first test of energy-momentum tensor evaluation in
full QCD with the gradient flow method, we choose a simulation point
used in our previous study of the equation of state 
with degenerate heavy $u$ and $d$ quarks and almost physical $s$
quark: $m_\pi/m_\rho\simeq0.63$ and $m_{\eta_{ss}}/m_\phi\simeq0.74$ at a single but fine
lattice spacing $a\simeq0.07\,\mathrm{fm}$. 

The pressure, energy density, entropy density and trace anomaly are studied as a function of temperature. 
We found that the results of the gradient flow method are consistent
with those of the $T$-integration method at low
temperatures $T\simle280\,\mathrm{MeV}$ ($N_t \simge 10$). However, deviation is found at high
temperatures $T\simge350\,\mathrm{MeV}$ ($N_t \simle 8$). This may be due to a lattice
artifact of $O\left((aT)^2\right)=O\left(1/N_t^2\right)$ from the discretization of thermal modes, which becomes severe at high temperature
in the fixed-scale approach. 

Applying a similar idea using the gradient flow~\cite{Hieda:2016lly}, we also calculate the renormalized chiral
condensate in $\overline{\mathrm{MS}}$ scheme. 
Although the Wilson-type quarks violates the chiral symmetry explicitly, 
the gradient flow method enables us to directly evaluate 
the chiral condensate and its susceptibility on the lattice, 
without suffering from power divergences. 
We find that the chiral condensate 
starts to decrease just below~$T\simeq199$ MeV. This seems to be indicating the
nearby pseudocritical temperature corresponding to the chiral restoration crossover. 
Accordingly, we find that the disconnected chiral susceptibility 
shows a clear peak around~$T\simeq199\,\mathrm{MeV}$.
These results are consistent with a previous estimate
of $T_{\mathrm{pc}}\sim190\,\mathrm{MeV}$ for the chiral restoration crossover~\cite{Umeda:2012er}. 

Our study was made at a single lattice spacing and with heavy $u$ and $d$ quarks. 
A definite conclusion on physical observables can be made only after taking the continuum limit with physical quark masses. 
To carry out the continuum extrapolation, we are planning to repeat the study at different values of~$a$. 
Nevertheless, the good agreement of the equation of state with the conventional method at $N_t\simge10$ 
seems to be suggesting that, besides the $O(1/N_t^2)$ errors at $N_t\simle8$, our lattices are already close to the continuum limit for the quantities we studied.
We are thus planning to start another study just at the physical quark mass point.

\acknowledgments
This work is in part supported by JSPS KAKENHI Grants
No.\ 25800148, No.\ 26287040, No.\ 26400244, No.\ 26400251, No.\ 15K05041,
and No.\ 16H03982,
by the Large Scale Simulation Program of High Energy Accelerator
Research Organization (KEK) No.\ 14/15-23, 15/16-T06, 15/16-T-07, 15/16-25, 16/17-05 
and by Interdisciplinary Computational Science Program in CCS, University of
Tsukuba.
This work is in part based on Lattice QCD common code Bridge++ \cite{bridge}.



\clearpage

\clearpage

\clearpage

\begin{figure}[ht]
\centering
\includegraphics[width=6.5cm]{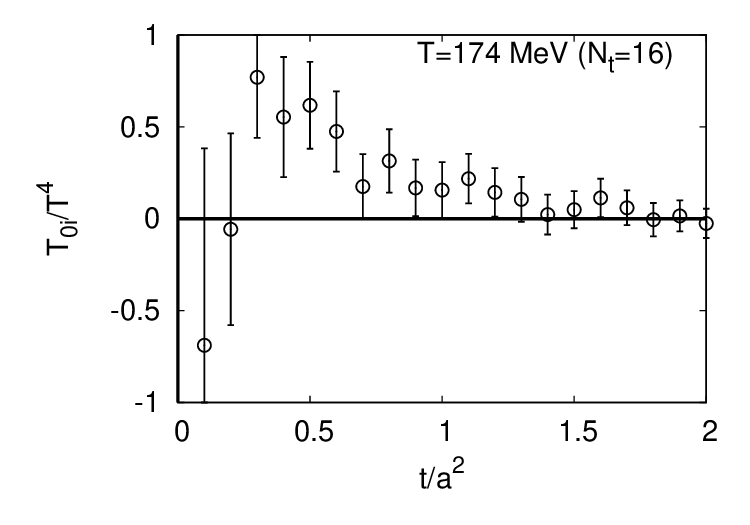}
\includegraphics[width=6.5cm]{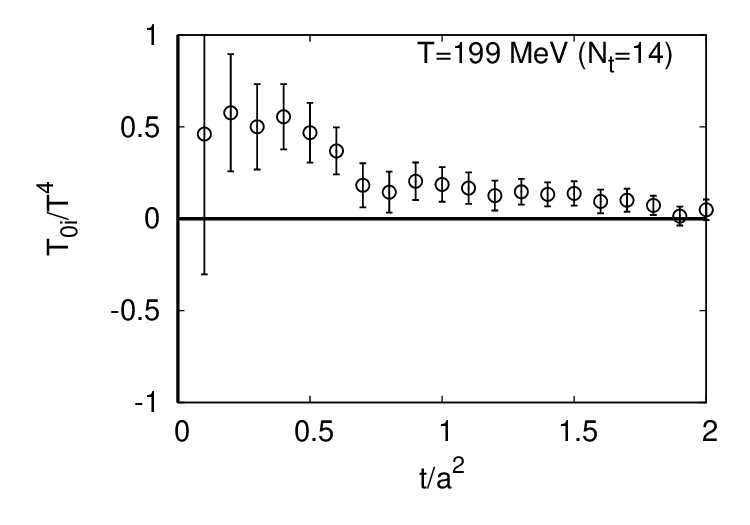}
\includegraphics[width=6.5cm]{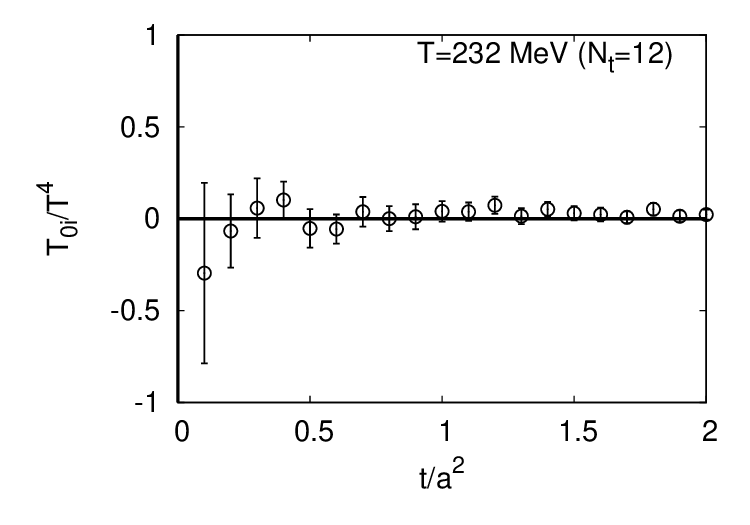}
\includegraphics[width=6.5cm]{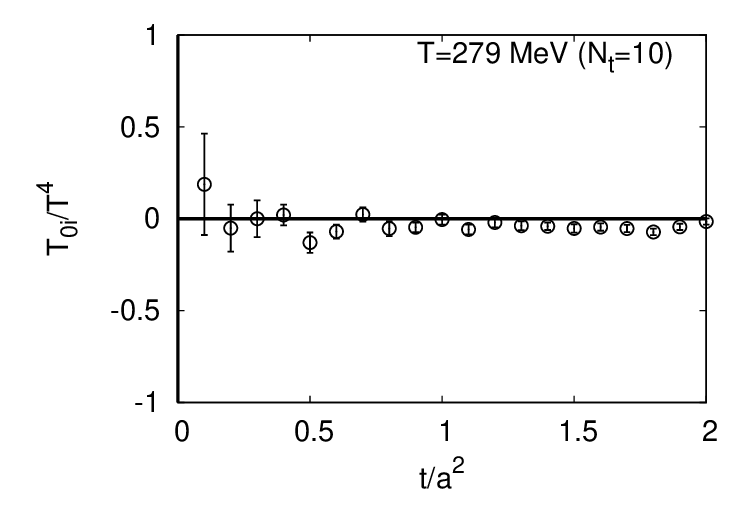}
\includegraphics[width=6.5cm]{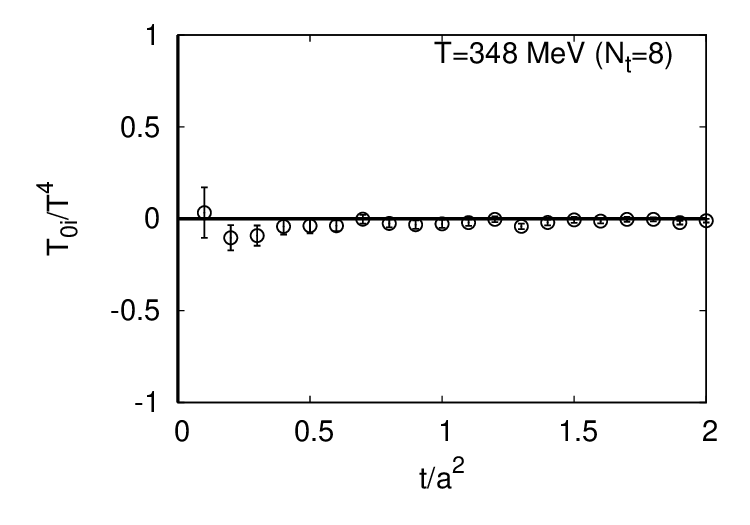}
\includegraphics[width=6.5cm]{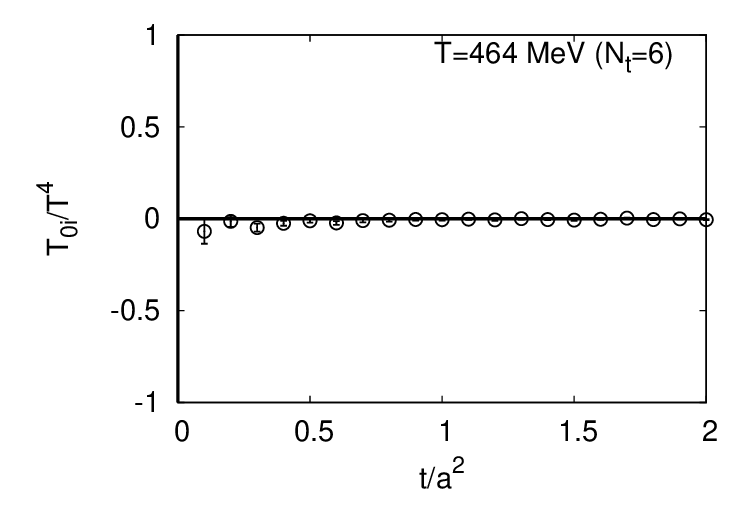}
\includegraphics[width=6.5cm]{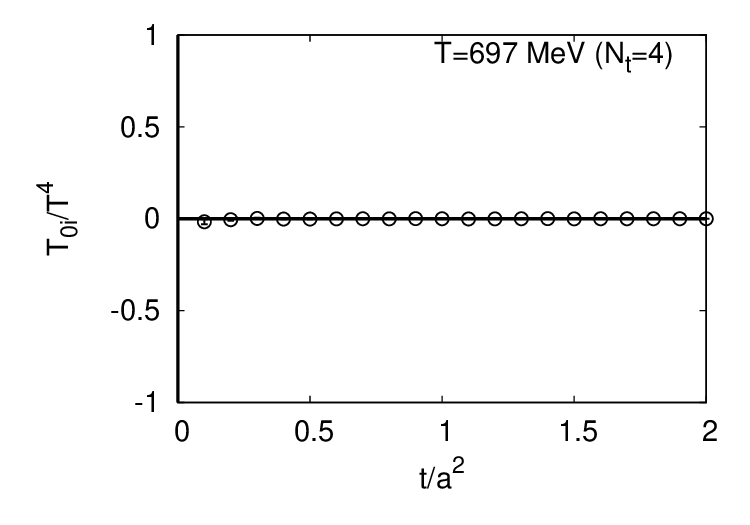}
\vspace{1cm}
\caption{Off diagonal component~$T_{i4}/T^4$, which corresponds to the momentum density, as a function of the flow
time~$t/a^2$. From the top left to the bottom: $T\simeq174$,
$199$, $232$, $279$, $348$, $464$, $697\,\mathrm{MeV}$.
Errors are statistical only.}
\label{fig9}
\end{figure}

\begin{figure}[ht]
\centering
\includegraphics[width=6.5cm]{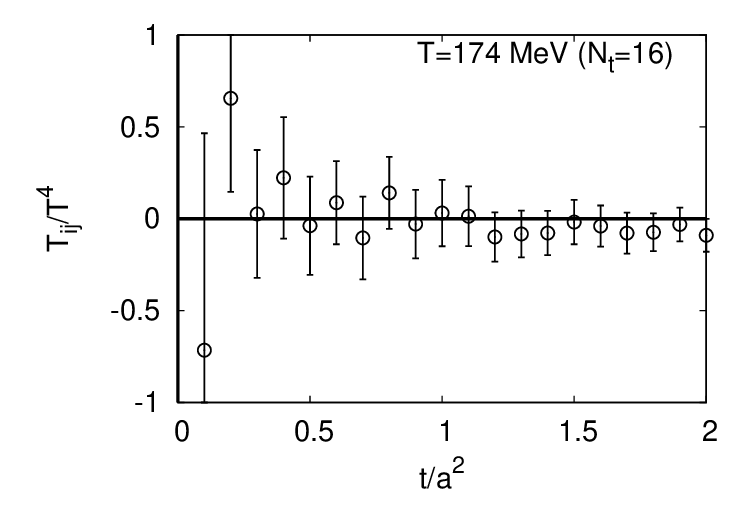}
\includegraphics[width=6.5cm]{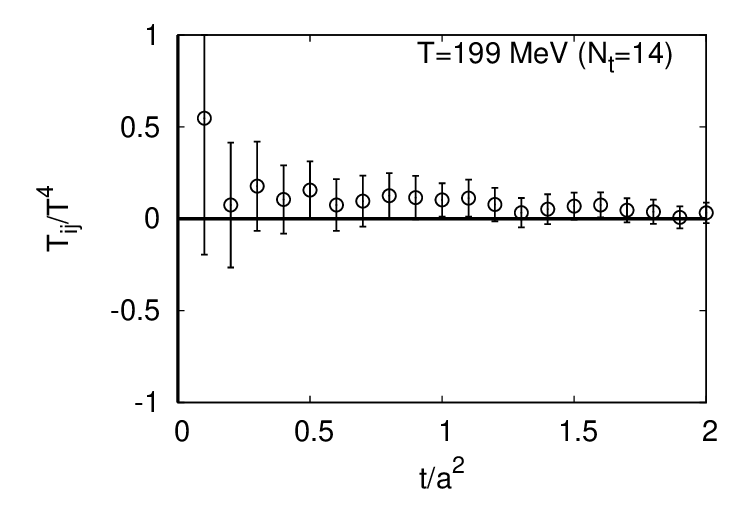}
\includegraphics[width=6.5cm]{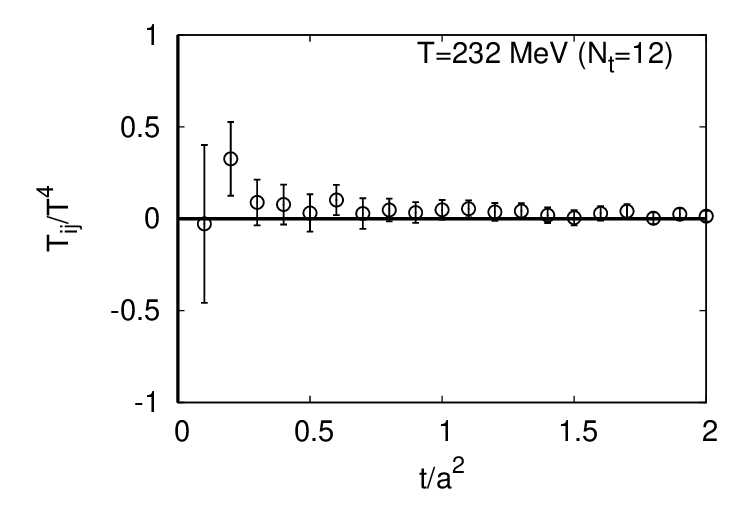}
\includegraphics[width=6.5cm]{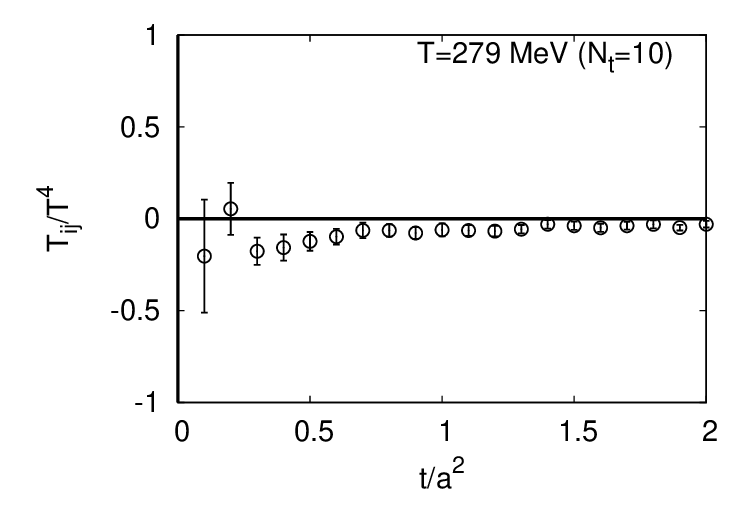}
\includegraphics[width=6.5cm]{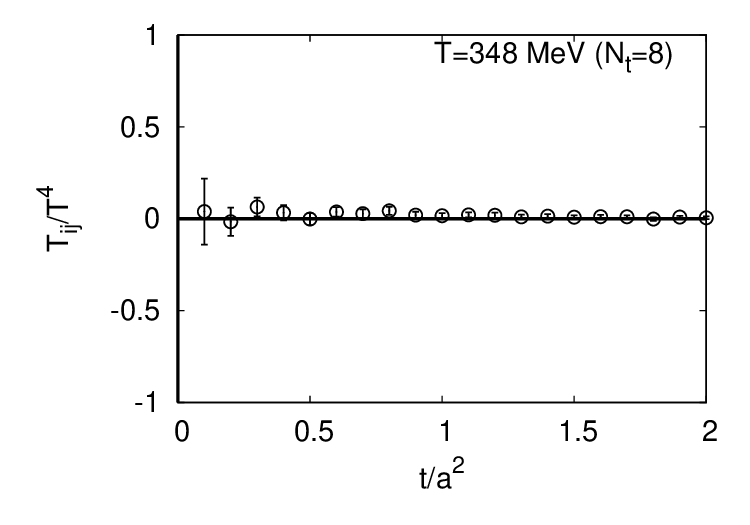}
\includegraphics[width=6.5cm]{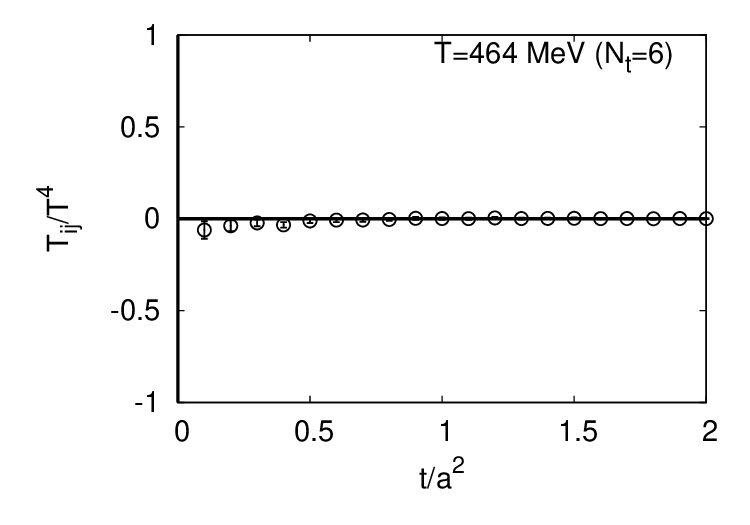}
\includegraphics[width=6.5cm]{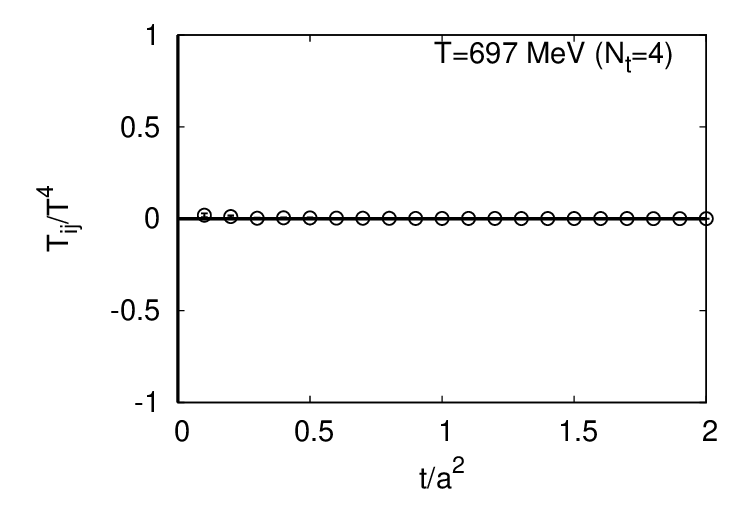}
\vspace{1cm}
\caption{The same as Fig.~\ref{fig9} but for the off diagonal component~$T_{i\neq j}/T^4$ corresponding to the stress density.}
\label{fig10}
\end{figure}

\clearpage

\begin{figure}[ht]
\centering
\includegraphics[width=6.5cm]{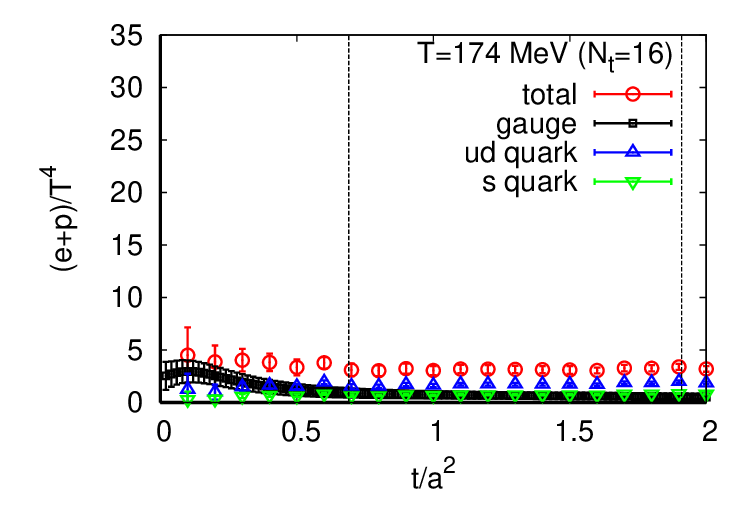}
\includegraphics[width=6.5cm]{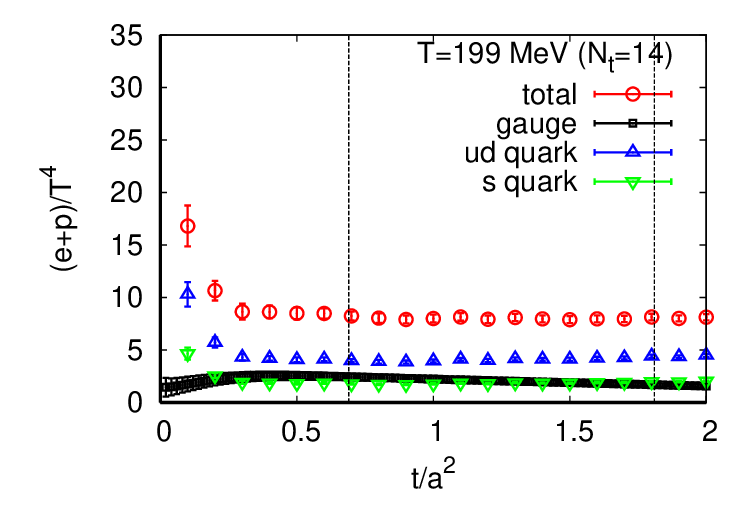}
\includegraphics[width=6.5cm]{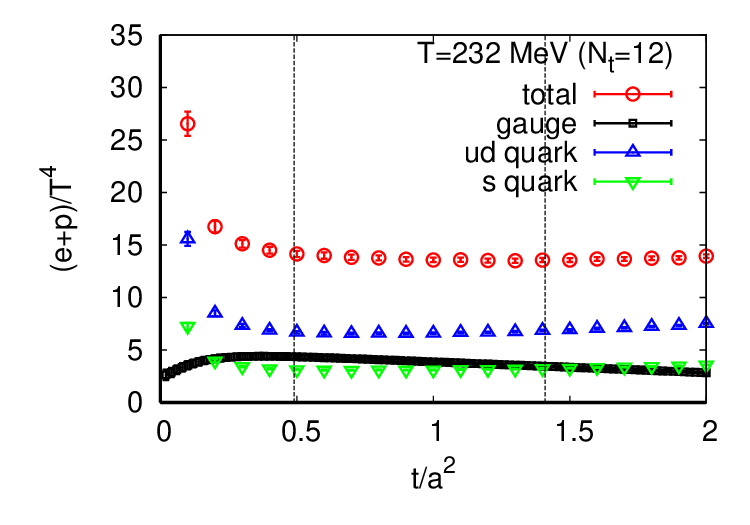}
\includegraphics[width=6.5cm]{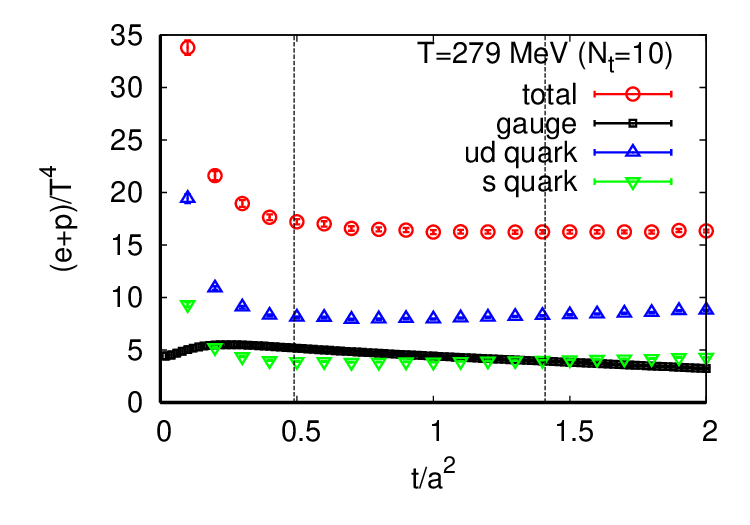}
\includegraphics[width=6.5cm]{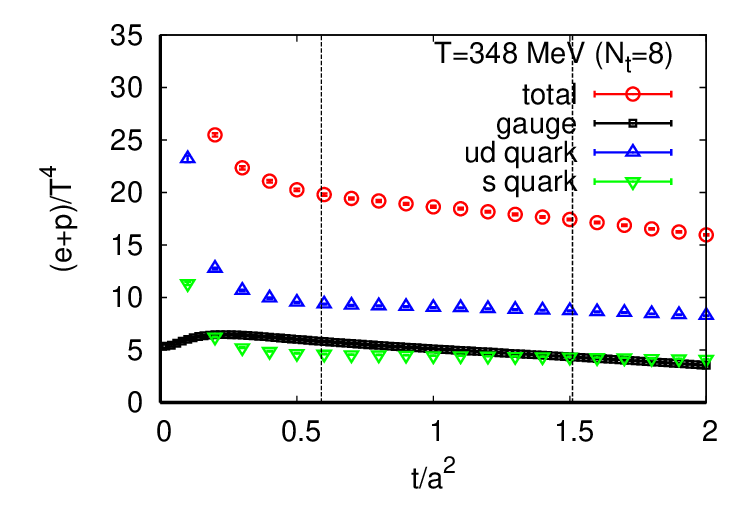}
\includegraphics[width=6.5cm]{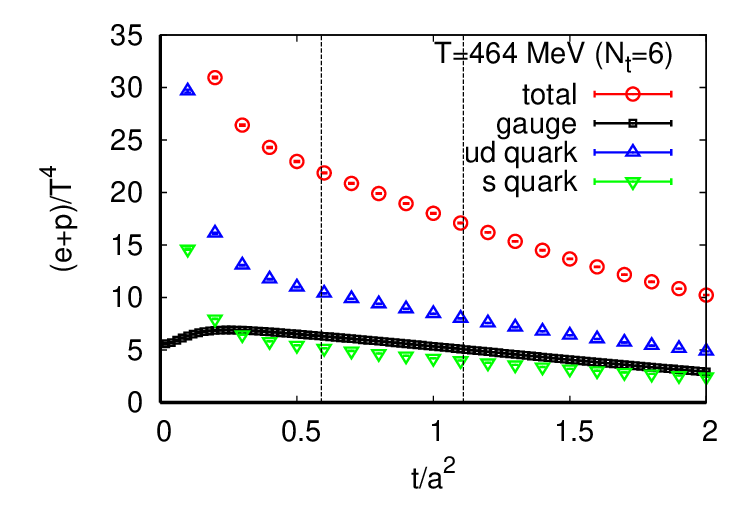}
\includegraphics[width=6.5cm]{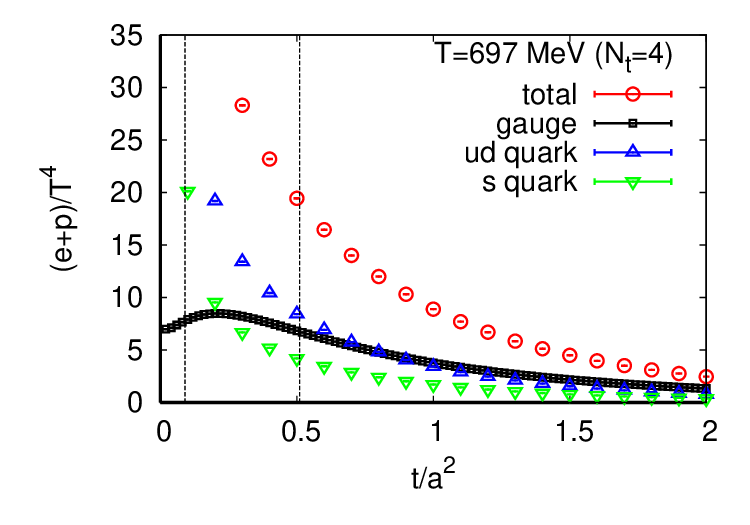}
\caption{Breakup of contributions from gauge and quark operators in the entropy density~$(\epsilon+p)/T^4$ 
as a function of the flow time~$t/a^2$. From the
top left: $T\simeq174$, $199$, $232$, $279$, $348$, $464$, $697\,\mathrm{MeV}$
($N_t=16$, 14, 12, 10, 8, 6, and 4, respectively).
Black squares are contribution from gauge operators (\ref{eq:(2.9)}) and (\ref{eq:(2.10)}).
Blue and green triangles are those from quark operators (\ref{eq:(2.11)}), (\ref{eq:(2.12)}) and (\ref{eq:(2.13)}) with $ud$ and $s$ quarks.
Red circles are the sum of all contributions.
Pair of dashed vertical lines indicates the window used for the fit.}
\label{fig:ep-each}
\end{figure}

\begin{figure}[ht]
\centering
\includegraphics[width=6.5cm]{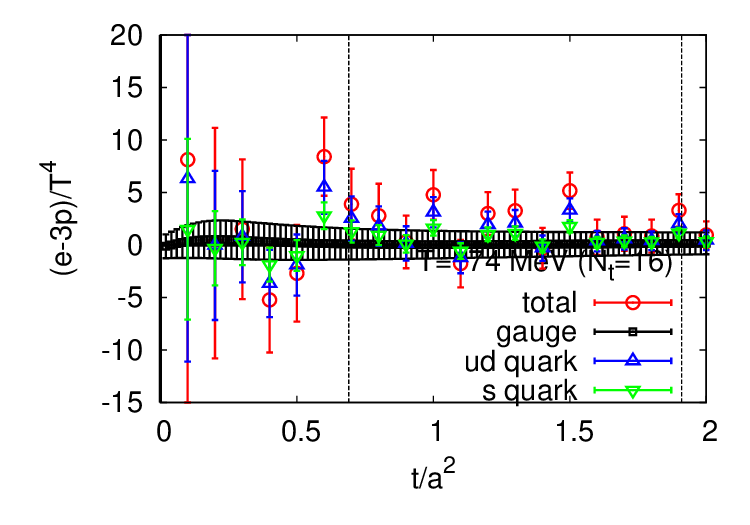}
\includegraphics[width=6.5cm]{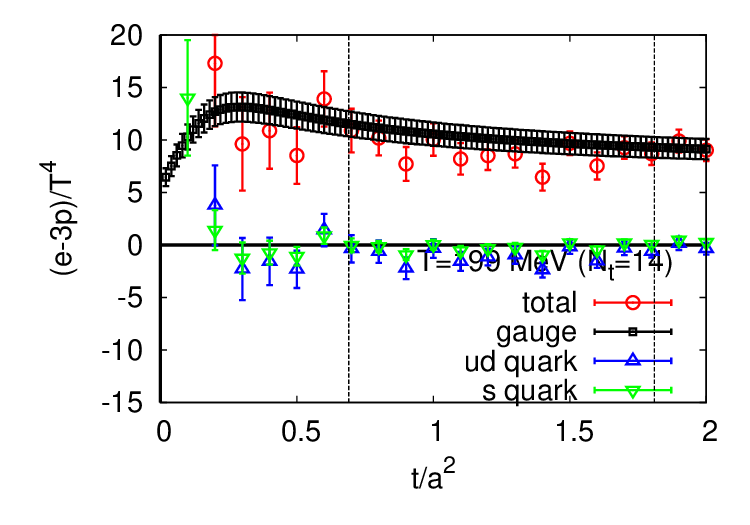}
\includegraphics[width=6.5cm]{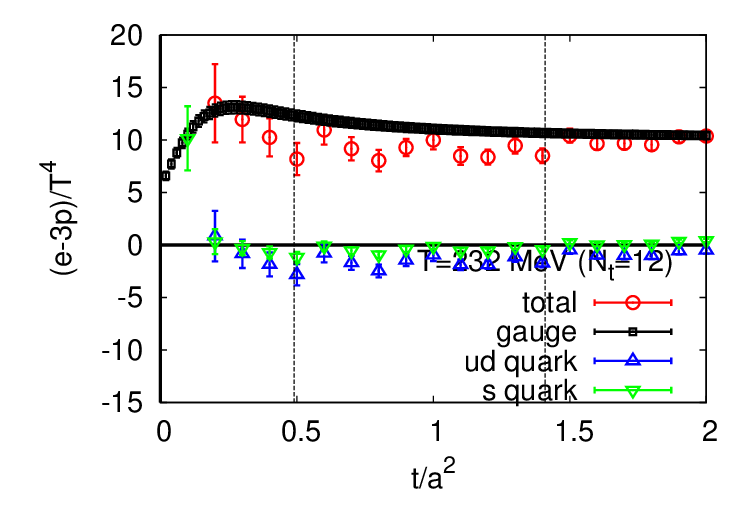}
\includegraphics[width=6.5cm]{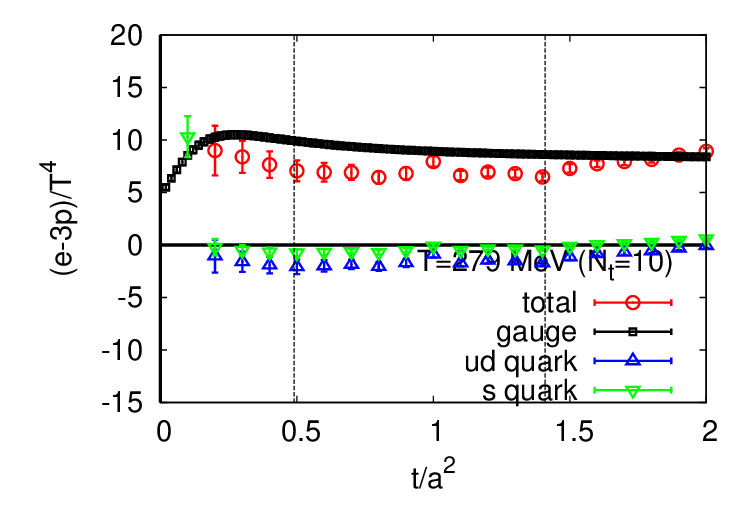}
\includegraphics[width=6.5cm]{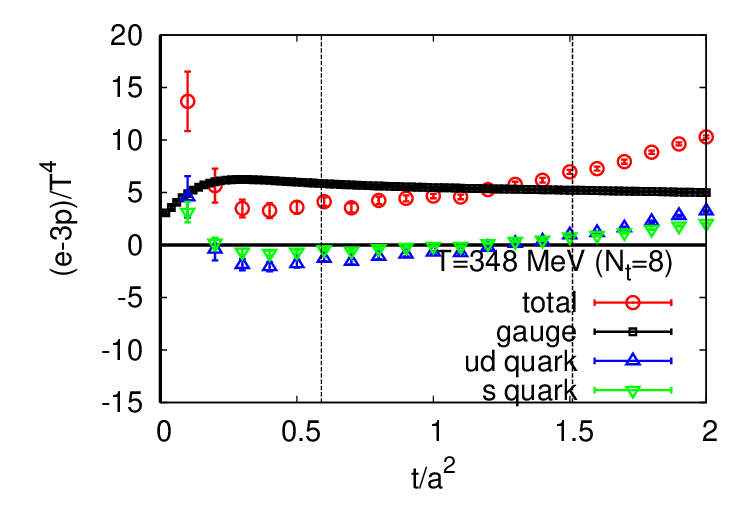}
\includegraphics[width=6.5cm]{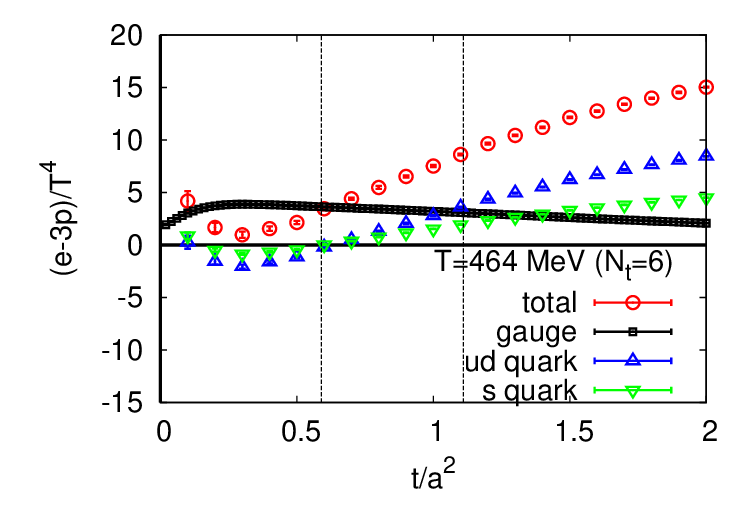}
\includegraphics[width=6.5cm]{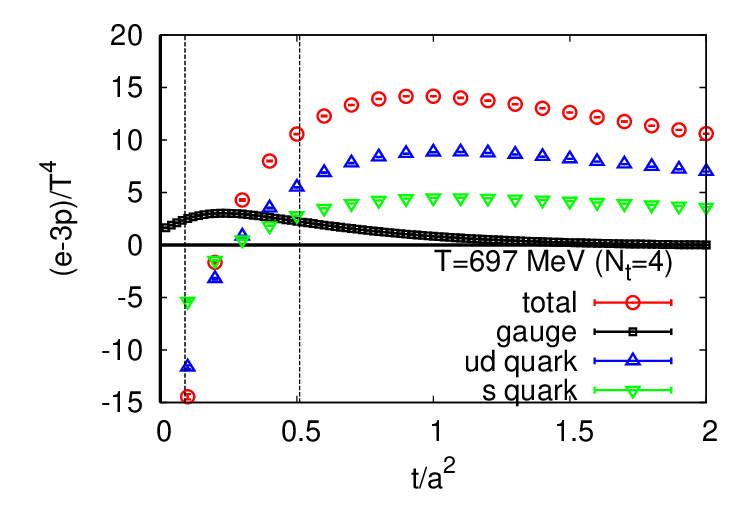}
\vspace{1cm}
\caption{The same as Fig.~\ref{fig:ep-each} but for the trace anomaly~$(\epsilon-3p)/T^4$.}
\label{fig:e3p-each}
\end{figure}

\begin{figure}[ht]
\centering
\includegraphics[width=10cm]{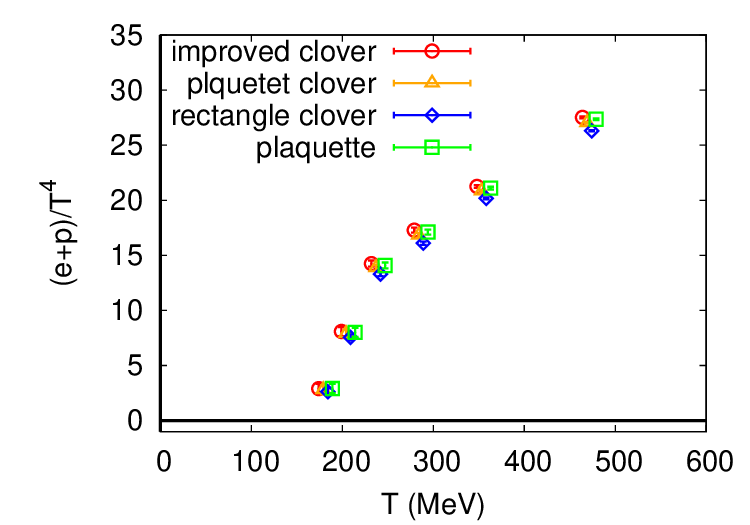}
\vspace{1cm}
\caption{Entropy density~$(\epsilon+p)/T^4$  as a function of temperature adopting four different operators 
for the field strength squared.
(i) Red circles are
results where the tree level improved combination of the clover terms is
used to define the field strength squared for the gauge contribution. 
Orange, blue, and green symbols are the results adopting (ii)~the clover term with four
plaquettes, (iii)~the clover term with eight $1\times2$ rectangle Wilson loops, and
(iv) the imaginary part of the plaquette to define the field strength. 
Symbols are slightly shifted in the horizontal direction for clarity of the figure. 
Errors are statistical only..}
\label{fig:eplusperra}
\end{figure}

\begin{figure}[ht]
\centering
\includegraphics[width=10cm]{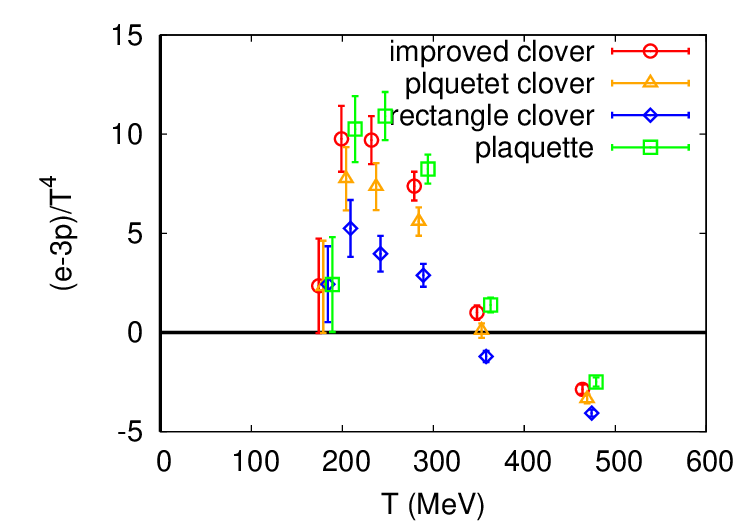}
\vspace{1cm}
\caption{The same as Fig.~\ref{fig:eplusperra} but for the trace anomaly~$(\epsilon-3p)/T^4$.}
\label{fig:eminus3perra}
\end{figure}

\appendix
\section{Numerical algorithm for flowed quark observables}
\label{sec:setup}

As Eqs.~\eqref{eq:(2.11)}-\eqref{eq:(2.13)} show, to compute the thermal
expectation value of the energy-momentum tensor, we need to compute
expectation values,
\begin{align}
   t_{\mu\nu}^f(t)
   &\equiv\frac{1}{N_\Gamma}\sum_x
   \left\langle\Bar{\chi}_f(t,x)\,\gamma_\mu
   \left(D_\nu-\overleftarrow{D}_\nu\right)\chi_f(t,x)\right\rangle,
\label{eq:(A.1)}
\\
   s^f(t)
   &\equiv\frac{1}{N_\Gamma}\sum_x
   \left\langle\Bar{\chi}_f(t,x)\,\chi_f(t,x)\right\rangle,
\label{eq:(A.2)}
\end{align}
where $N_\Gamma=\sum_x$ is the number of lattice points, both at finite and zero
temperatures. Note that the covariant derivatives in~Eq.~\eqref{eq:(A.1)}
refer to the flowed gauge field~$B_\mu(t,x)$ at the flow time~$t$.

The basic contraction of flowed quark fields is given by [see~Eq.~(6.8)
of~Ref.~\cite{Luscher:2013cpa}],
\begin{equation}
   \contraction{}{\chi}{_f(t,x)\,}{\Bar{\chi}}
   \chi_f(t,x)\,\Bar{\chi}_{f'}(s,y)
   =\delta_{ff'}\sum_{v,w}K(t,x;0,v)
   \left[S_f(v,w)-c_{\mathrm{fl}}\,\delta_{v,w}\right]K(s,y;0,w)^\dagger,
\label{eq:(A.3)}
\end{equation}
where $S_f(x,y)$ is the quark propagator with the bare mass~$m_{f0}$,
\begin{equation}
   \left(\Slash{D}+m_{f0}\right)S_f(x,y)=\delta_{x,y},
\label{eq:(A.4)}
\end{equation}
and $K(t,x;s,y)$ is the fundamental solution to the flow equation, defined by
\begin{equation}
   \left(\partial_t-\Delta\right)K(t,x;s,y)=0,\qquad
   K(t,x;t,y)=\delta_{x,y} ,
\label{eq:(A.5)}
\end{equation}
and $c_{\mathrm{fl}}$ is an improvement coefficient associated with the flowed
quark field~\cite{Luscher:2013cpa}. In Eq.~\eqref{eq:(A.3)} and in what
follows, the dagger ($\dagger$) implies the hermitian conjugation with respect
to the gauge and spinor indices only.

Carrying out the above contraction in~Eqs.~\eqref{eq:(A.1)}
and~\eqref{eq:(A.2)}, we have
\begin{align}
   t_{\mu\nu}^f(t)
   &=-\frac{1}{N_\Gamma}\sum_{x,y,v,w}\Biggl\{
   \left\langle
   \sum_{\alpha,i}\left[\gamma_\mu D_\nu^xK(t,x;0,v)
   S_f(v,w)K(t,y;0,w)^\dagger
   \right]_{\alpha i,\alpha i}\delta_{y,x}
   \right\rangle
\notag\\
   &\qquad\qquad\qquad\qquad{}
   -\left\langle\delta_{x,y}
   \sum_{\alpha,i}\left[K(t,y;0,v)
   S_f(v,w)
   K(t,x;0,w)^\dagger
   \overleftarrow{D}_\nu^x\gamma_\mu
   \right]_{\alpha i,\alpha i}\right\rangle
   \Biggr\},
\label{eq:(A.6)}
\\
   s^f(t)
   &=-\frac{1}{N_\Gamma}\sum_{x,y,v,w}
   \left\langle
   \sum_{\alpha,i}\left\{K(t,x;0,v)
   \left[S_f(v,w)-c_{\mathrm{fl}}\,\delta_{v,w}\right]
   K(t,y;0,w)^\dagger
   \right\}_{\alpha i,\alpha i}\delta_{y,x}
   \right\rangle,
\label{eq:(A.7)}
\end{align}
where $\alpha$ denotes the spinor index which runs over~$1$, $2$, $3$, and~$4$,
and $i$ denotes the color index running over~$1$, $2$, and~$3$. In writing down
Eq.~\eqref{eq:(A.6)}, we have used the fact that the term
in~Eq.~\eqref{eq:(A.3)} with the improvement coefficient~$c_{\mathrm{fl}}$ does
not contribute, because the trace of a single gamma matrix vanishes. Note that
$K$ and~$D_\nu$ have no spinor indices.

We evaluate the above trace over space-time points stochastically (i.e., by the
noise estimator). That is, we introduce a randomly generated complex scalar
field~$\eta(x)$ (noise field) which fulfills
\begin{equation}
   \left\langle\eta(x)\right\rangle_\eta=0,\qquad
   \left\langle\eta(x)\,\eta(y)^*\right\rangle_\eta=\delta_{x,y},
\label{eq:(A.8)}
\end{equation}
where expectation values refer to the average over~$\eta(x)$. Then the above
traces can be expressed as
\begin{align}
   t_{\mu\nu}^f(t)
   &=-\frac{1}{N_\Gamma}\Biggl\{
   \left\langle\left\langle
   \sum_{\alpha,i}\left[
   \sum_{v,w}\xi(t;0,v)^\dagger
   S_f(v,w)\,
   \psi_{\mu\nu}(t;0,w)\right]_{\alpha i,\alpha i}
   \right\rangle_{\!\!\eta}\,\right\rangle
\notag
\\
   &\qquad\qquad\qquad{}
   -\left\langle\left\langle
   \sum_{\alpha,i}\left[
   \sum_{v,w}\psi_{\mu\nu}(t;0,v)^\dagger
   S_f(v,w)\,
   \xi(t;0,w)\right]_{\alpha i,\alpha i}
   \right\rangle_{\!\!\eta}\,\right\rangle
   \Biggr\},
\label{eq:(A.9)}
\\
   s^f(t)
   &=-\frac{1}{N_\Gamma}
   \left\langle\left\langle
   \sum_{\alpha,i}\left\{
   \sum_{v,w}\xi(t;0,v)^\dagger
   \left[S_f(v,w)-c_{\mathrm{fl}}\,\delta_{v,w}\right]
   \xi(t;0,w)\right\}_{\alpha i,\alpha i}
   \right\rangle_{\!\!\eta}\,\right\rangle.
\label{eq:(A.10)}
\end{align}
Here, we have defined the combinations,
\begin{align}
   \xi(t;s,w)&\equiv\sum_xK(t,x;s,w)^\dagger\eta(x),
\label{eq:(A.11)}
\\
   \psi_{\mu\nu}(t;s,w)&\equiv
   \gamma_\mu\sum_xK(t,x;s,w)^\dagger D_\nu\eta(x).
\label{eq:(A.12)}
\end{align}
Finally, by noting
\begin{equation}
   S_f(v,w)=\gamma_5\,S_f(w,v)^\dagger\gamma_5,
\label{eq:(A.13)}
\end{equation}
we have
\begin{align}
   t_{\mu\nu}^f(t)
   &=\frac{2}{N_\Gamma}\re
   \left\langle\left\langle
   \sum_{\alpha,i}\left[\sum_v\psi_{\mu\nu}(t;0,v)^\dagger
   \sum_wS_f(v,w)\,
   \xi(t;0,w)\right]_{\alpha i,\alpha i}
   \right\rangle_{\!\!\eta}\,\right\rangle,
\label{eq:(A.14)}
\\
   s^f(t)
   &=-\frac{1}{N_\Gamma}
   \left\langle\left\langle
   \sum_{\alpha,i}\left[\sum_v\xi(t;0,v)^\dagger
   \sum_wS_f(v,w)\,
   \xi(t;0,w)\right]_{\alpha i,\alpha i}
   \right\rangle_{\!\!\eta}\,\right\rangle
\notag\\
   &\qquad\qquad\qquad{}
   +c_{\mathrm{fl}}\frac{1}{N_\Gamma}
   \left\langle\left\langle
   \sum_{\alpha,i}\left[\sum_v\xi(t;0,v)^\dagger
   \xi(t;0,v)\right]_{\alpha i,\alpha i}
   \right\rangle_{\!\!\eta}\,\right\rangle.
\label{eq:(A.15)}
\end{align}

So the procedure to compute the expectation values~\eqref{eq:(A.1)}
and~\eqref{eq:(A.2)} consists of following steps:
\begin{enumerate}
\item Take a gauge configuration.
\item Generate a random single component complex field~$\eta(x)$ which
satisfies Eq.~\eqref{eq:(A.8)}.
\item Multiply $\eta(x)$ by a unit vector whose nonzero spinor-color
component is~$(\alpha,i)$.
\item Compute $\xi(t;0,w)$~\eqref{eq:(A.11)}
and~$\psi_{\mu\nu}(t;0,w)$~\eqref{eq:(A.12)}. For this, we need to solve the
following ``adjoint flow equations'':
\begin{align}
   \left(\partial_s+\Delta\right)\xi(t;s,w)&=0,&
   \xi(t;t,w)&=\eta(w),
\label{eq:(A.16)}
\\
   \left(\partial_s+\Delta\right)\psi_{\mu\nu}(t;s,w)&=0,&
   \psi_{\mu\nu}(t;t,w)&=\gamma_\mu D_\nu\eta(w),   
\label{eq:(A.17)}
\end{align}
from $s=t$ to~$s=0$ \emph{backward\/} in the flow time.\footnote{Since $\Delta$
is the unit matrix in spinor space, we can avoid the reputation of
this integration over spinor indices.} This is the hardest part of the
computation and how to carry out this integration is described in Appendix~\ref{sec:aRK}.
\item Using $\xi(t;0,w)$ obtained above as the initial vector, compute a new
vector,
\begin{equation}
   \sum_wS_f(v,w)\,\xi(t;0,w),
\label{eq:(A.18)}
\end{equation}
by one of the standard methods. In the propagator defined
by~Eq.~\eqref{eq:(A.4)}, the gauge field is the gauge field without any flow
(i.e., original link variables).
\item Compute the following inner products:
\begin{align}
   &\frac{2}{N_\Gamma}\re\sum_v\psi_{\mu\nu}(t;0,v)^\dagger
   \sum_wS_f(v,w)\,
   \xi(t;0,w),
\label{eq:(A.19)}
\\
   &-\frac{1}{N_\Gamma}\sum_v\xi(t;0,v)^\dagger
   \sum_wS_f(v,w)\,
   \xi(t;0,w),
\label{eq:(A.20)}
\\
   &\frac{1}{N_\Gamma}\sum_v\xi(t;0,v)^\dagger
   \xi(t;0,v).
\label{eq:(A.21)}
\end{align}
\item Change $(\alpha,i)$ and go back to the step~(3) and repeat the above
procedures for $4\times3$~times.
\item Go back to the step~(2) and repeat the above procedures for enough numbers of
random fields.
\item Take a different gauge configuration and repeat the above procedures for
obtaining the Monte Carlo average.
\end{enumerate}

\section{Numerical algorithm for gradient flow}
\label{sec:flowalgo}

\subsection{Runge-Kutta integration for the gauge fields}
The Wilson flow of the lattice gauge field $U(x,\mu)$ is defined by
\begin{equation}
  \left( \partial_t V(t,x,\mu) \right) V(t,x,\mu)^{-1}=-g_0^2\,\partial_{x,\mu}
   S_w(V),\qquad
   V(t=0,x,\mu)=U(x,\mu),
\label{eq:(A.22)}
\end{equation}
where $S_w$ is the Wilson plaquette action and
\begin{equation}
   \partial_{x,\mu}^af(U)=\left.\frac{\mathrm{d}}{\mathrm{d}s}
   f(\mathrm{e}^{sX}U)\right|_{s=0},\qquad
   X(y,\nu)=\begin{cases}
   T^a&\text{if $(y,\nu)=(x,\mu)$,}\\
   0&\text{otherwise},\\
   \end{cases}
\label{eq:(A.23)}
\end{equation}
and
\begin{equation}
   \partial_{x,\mu}f(U)=T^a\qquad\partial_{x,\mu}^af(U).
\label{eq:(A.24)}
\end{equation}

It is convenient to write the flow equation~\eqref{eq:(A.22)} in the following
abstract form:
\begin{equation}
   \partial_t V_t=Z(V_t)\,V_t.
\label{eq:(A.25)}
\end{equation}
Then the third order Runge-Kutta integration which constructs $V_{t+\epsilon}$
from~$V_t$ proceeds as follows \cite{Luscher:2010iy}:
\begin{align}
   W_0&=V_t,
\notag\\
   W_1&=\exp\left(\frac{1}{4}Z_0\right)W_0,
\notag\\
   W_2&=\exp\left(\frac{8}{9}Z_1-\frac{17}{36}Z_0\right)W_1,
\notag\\
   W_3&=\exp\left(\frac{3}{4}Z_2-\frac{8}{9}Z_1+\frac{17}{36}Z_0\right)W_2,
\label{eq:(A.26)}
\end{align}
where $Z_i$ are given from the combination defined in~Eq~\eqref{eq:(A.25)} by
\begin{equation}
   Z_i=\epsilon\, Z(W_i),\qquad i=0,1,2,
\label{eq:(A.27)}
\end{equation}
and
\begin{equation}
   V_{t+\epsilon}=W_3.
\label{eq:(A.28)}
\end{equation}
With this integrator, the error in~$V_{t+\epsilon}$ turns out to
be~$O(\epsilon^4)$.

\subsection{Adjoint Runge-Kutta integration for the quark field}
\label{sec:aRK}

To compute the expectation value of composite operators containing flowed quark
fields, we need to solve the adjoint flow equations~\eqref{eq:(A.16)}
and~\eqref{eq:(A.17)}. Since the ``initial conditions'' in these equations are
given at the target flow time~$t$, we have to solve the flow equations
\emph{backward\/} in the flow time. The equation that we want to solve can be
written abstractly as
\begin{equation}
   \partial_s\xi_s=-\Delta(V_s)\,\xi_s.
\label{eq:(A.29)}
\end{equation}
Then the third order adjoint Runge-Kutta integrator that constructs $\xi_s$
from~$\xi_{s+\epsilon}$ is given by
\begin{align}
   \lambda_3&=\xi_{s+\epsilon},
\notag\\
   \lambda_2&=\frac{3}{4}\Delta_2\lambda_3,
\notag\\
   \lambda_1&=\lambda_3+\frac{8}{9}\Delta_1\lambda_2,
\notag\\
   \lambda_0&=\lambda_1+\lambda_2
   +\frac{1}{4}\Delta_0\left(\lambda_1-\frac{8}{9}\lambda_2\right),
\label{eq:(A.30)}
\end{align}
where
\begin{equation}
   \Delta_i=\epsilon\,\Delta(W_i),\qquad i=0,1,2,
\label{eq:(A.31)}
\end{equation}
and
\begin{equation}
   \xi_s=\lambda_0.
\end{equation}
The error in~$\xi_s$ is again~$O(\epsilon^4)$. For the derivation of
this procedure, see Appendix~E.1 of~Ref.~\cite{Luscher:2013cpa}.

In~Ref.~\cite{Luscher:2013cpa}, the author noted that the time direction to
which the Runge-Kutta integrator proceeds is quite important: One should use
the Runge-Kutta steps as indicated as above but not the reversed direction,
because the reversed direction is exponentially unstable. Thus, to carry out
the adjoint Runge-Kutta steps~\eqref{eq:(A.30)} from~$t$ to~$t-\epsilon$, we
have to compute Runge-Kutta steps for the gauge field from~$0$ to~$t$. Then,
for the next adjoint Runge-Kutta step from~$t-\epsilon$ to~$t-2\epsilon$, if
we do not keep \emph{any\/} intermediate flowed gauge-field configuration, we
have to evolve the gauge field anew from~$0$ to~$t-\epsilon$. In this way, to
integrate Eqs.~\eqref{eq:(A.16)} and~\eqref{eq:(A.17)} backward in time
from~$s=t$ to~$s=0$, we have to compute the flowed gauge field from the zero
flow time to intermediate flow times repeatedly. This large computational
burden for the adjoint Runge-Kutta calculations can be reduced by storing
intermediate flowed configurations, at the cost of the memory space.

\section{Running coupling and running masses}
\label{sec:running}
To use the coefficients~\eqref{eq:(2.17)}--\eqref{eq:(2.21)}, we need to have
the running coupling~$\Bar{g}\!\left(1/\sqrt{8t}\right)$ and the running
masses~$\Bar{m}_f\!\left(1/\sqrt{8t}\right)$.

The renormalization group invariant scale (the Lambda parameter) is defined by
\begin{equation}
   \frac{\Lambda}{\mu}=\left[b_0\bar{g}(\mu)^2\right]^{-b_1/(2b_0^2)}
   \exp\left[-\frac{1}{2b_0\Bar{g}(\mu)^2}\right]
   \exp\left\{
   -\int_0^{\Bar{g}(\mu)}dg\,\left[
   \frac{1}{\beta(g)}+\frac{1}{b_0 g^3}-\frac{b_1}{b_0^2g}
   \right]
   \right\},
\label{eq:(B.1)}
\end{equation}
where $\mu$ is the renormalization scale, while the running mass and the
renormalization group invariant mass~$M$ are related by
\begin{equation}
   \Bar{m}(\mu)
   =M\left[2b_0\bar{g}(\mu)^2\right]^{d_0/(2b_0)}
   \exp\left\{
   \int_0^{\Bar{g}(\mu)}dg\,\left[
   \frac{\tau(g)}{\beta(g)}-\frac{d_0}{b_0g}
   \right]
   \right\}.
\label{eq:(B.2)}
\end{equation}

The renormalization group functions, $\beta(g)$ and~$\tau(g)$, are known to the
four-loop order in the MS or $\overline{\mathrm{MS}}$
scheme~\cite{Czakon:2004bu}. For the $SU(N)$ gauge theory with $N_f$
fundamental fermions, setting
\begin{equation}
   \beta(g)=-g^3\sum_{k=0}^\infty b_kg^{2k},\qquad
   \tau(g)=-g^2\sum_{k=0}^\infty d_kg^{2k},
\end{equation}
the first two coefficients are given by~\cite{Caswell:1974gg,Jones:1974mm}
\begin{align}
   b_0&=(4\pi)^{-2}
   \left(\frac{11}{3}N-\frac{2}{3}N_f\right),
\\
   b_1&=(4\pi)^{-4}
   \left[\frac{34}{3}N^2-\left(\frac{13}{3}N-N^{-1}\right)N_f\right],
\end{align}
and~\cite{Tarrach:1980up,Nachtmann:1981zg}
\begin{align}
   d_0&=(4\pi)^{-2}
   \left(N-N^{-1}\right)3,
\\
   d_1&=(4\pi)^{-4}\left(N-N^{-1}\right)
   \left(\frac{203}{12}N-\frac{3}{4}N^{-1}-\frac{5}{3}N_f\right).
\end{align}
For higher orders ($k\geq2$), setting
\begin{equation}
   b_k=(4\pi)^{-2k-2}\sum_{l=0}^kb_{k,l}N_f^l,\qquad
   d_k=(4\pi)^{-2k-2}\left(N-N^{-1}\right)\sum_{l=0}^kd_{k,l}N_f^l,
\end{equation}
we have~\cite{Tarasov:1980au,Larin:1993tp,vanRitbergen:1997va}
\begin{align}
   b_{2,0}&=\frac{2857}{54}N^3,
\\
   b_{2,1}&=-\frac{1709}{54}N^2+\frac{187}{36}+\frac{1}{4}N^{-2},
\\
   b_{2,2}&=\frac{56}{27}N-\frac{11}{18}N^{-1},
\\
   b_{3,0}&=\frac{150473}{486}N^4-\frac{40}{3}N^2+\frac{44}{9}\zeta(3)N^4
   +352\zeta(3)N^2,
\\
   b_{3,1}&=-\frac{485513}{1944}N^3+\frac{58583}{1944}N-\frac{2341}{216}N^{-1}
   -\frac{23}{8}N^{-3}
\notag\\
   &\qquad{}
   -\frac{20}{9}\zeta(3)N^3-\frac{548}{9}\zeta(3)N+\frac{44}{9}\zeta(3)N^{-1},
\\
   b_{3,2}&=\frac{8654}{243}N^2-\frac{2477}{243}-\frac{623}{54}N^{-2}
   +\frac{28}{3}\zeta(3)N^2-\frac{64}{9}\zeta(3)+\frac{244}{9}\zeta(3)N^{-2},
\\
   b_{3,3}&=\frac{130}{243}N-\frac{77}{243}N^{-1},
\end{align}
and~\cite{Tarasov:1982gk,Larin:1993tq,Chetyrkin:1997dh,Vermaseren:1997fq}
\begin{eqnarray}
   d_{2,0}&=&\frac{11413}{108}N^2-\frac{129}{8}+\frac{129}{8}N^{-2},
\\
   d_{2,1}&=&-\frac{1177}{54}N+\frac{23}{2}N^{-1}-12\zeta(3)N-12\zeta(3)N^{-1},
\\
   d_{2,2}&=&-\frac{35}{27},
\\
   d_{3,0}&=&\frac{460151}{576}N^3-\frac{66577}{576}N+\frac{50047}{192}N^{-1}
   +\frac{1261}{64}N^{-3}
\notag\\
   &&\qquad{}
   +\frac{1157}{9}\zeta(3)N^3+104\zeta(3)N
   -47\zeta(3)N^{-1}+42\zeta(3)N^{-3}
\notag\\
   &&\qquad\qquad{}
   -220\zeta(5)N^3-220\zeta(5)N,
\\
   d_{3,1}&=&-\frac{23816}{81}N^2+\frac{10475}{108}+\frac{37}{3}N^{-2}
\notag\\
   &&\qquad{}
   -\frac{889}{3}\zeta(3)N^2-170\zeta(3)-111\zeta(3)N^{-2}+66\zeta(4)N^2
   +66\zeta(4)
\notag\\
   &&\qquad\qquad{}
   +160\zeta(5)N^2+100\zeta(5)-60\zeta(5)N^{-2},
\\
   d_{3,2}&=&\frac{899}{162}N-\frac{38}{27}N^{-1}+20\zeta(3)N+20\zeta(3)N^{-1}
   -12\zeta(4)N-12\zeta(4)N^{-1},
\\
   d_{3,3}&=&-\frac{83}{81}+\frac{16}{9}\zeta(3).
\end{eqnarray}

For $N=3$ and~$N_f=3$, we have
\begin{align}
   b_0&=(4\pi)^{-2}9,
\\
   b_1&=(4\pi)^{-4}64,
\\
   b_2&=(4\pi)^{-6}\frac{3863}{6},
\\
   b_3&=(4\pi)^{-8}\left[3560\zeta(3)+\frac{140599}{18}\right],
\end{align}
and
\begin{align}
   d_0&=(4\pi)^{-2}8,
\\
   d_1&=(4\pi)^{-4}\frac{364}{3},
\\
   d_2&=(4\pi)^{-6}\left[\frac{17770}{9}-320\zeta(3)\right],
\\
   d_3&=(4\pi)^{-8}\left[
   -\frac{297440}{27}\zeta(3)-\frac{16000}{3}\zeta(5)+48\pi^4
   +\frac{2977517}{81}\right].
\end{align}

Now, for our application, we adopt the $\overline{\mathrm{MS}}$ scheme and
set $\mu=1/\sqrt{8t}$. Then, the left-hand side of~Eq.~\eqref{eq:(B.1)} reads
\begin{equation}
   a\Lambda_{\overline{\mathrm{MS}}}\sqrt{8t/a^2}.
\end{equation}
Then by solving Eq.~\eqref{eq:(B.1)} with respect to~$\Bar{g}(\mu)$
numerically, we have the running coupling~$\Bar{g}\!\left(1/\sqrt{8t}\right)$ in the
$\overline{\mathrm{MS}}$ scheme. Another option (although we do not use it
in the present paper) is an approximate formula quoted in the Particle Data
Group~\cite{Agashe:2014kda},
\begin{align}
   \Bar{g}(\mu)^2
   &=\frac{1}{b_0 t}
   \Biggl[
   1-\frac{b_1}{b_0^2}\frac{\ln t}{t}
   +\frac{b_1^2(\ln^2t-\ln t-1)+b_0b_2}{b_0^4t^2}
\notag\\
   &\qquad\qquad{}
   -\frac{b_1^3\left(\ln^3t-\frac{5}{2}\ln^2t-2\ln t+\frac{1}{2}\right)
   +3b_0b_1b_2\ln t-\frac{1}{2}b_0^2b_3}{b_0^6 t^3}
   \Biggr],\qquad
   t\equiv\ln\left(\frac{\mu^2}{\Lambda^2}\right).
\end{align}

For $\Lambda_{\overline{\mathrm{MS}}}$, we use the value~\cite{Agashe:2014kda},
\begin{equation}
   \Lambda_{\overline{\mathrm{MS}}}^{(3)}=332(19)\,\mathrm{MeV}.
\end{equation}
Using~\cite{Ishikawa:2007nn}
\begin{equation}
   a(\beta=2.05)=0.0701(29)\,\mathrm{fm},
\end{equation}
we have
\begin{equation}
   a(\beta=2.05)\,\Lambda_{\overline{\mathrm{MS}}}
   =0.0701(29)\times332(19)/197.3269718.
\end{equation}

For the renormalization group invariant mass, we adopt~\cite{Ishikawa:2007nn,Aoki:2010wm},
\begin{align}
   a(\beta=2.05)\,M&=Z_m(\beta=2.05)\,a(\beta=2.05)\,m_{u,d}
\notag\\
   &=1.862(41)\times(
   0.02105\pm
   0.00017),
\end{align}
for $u$ and $d$ quarks, and
\begin{align}
   a(\beta=2.05)M&=Z_m(\beta=2.05)\,a(\beta=2.05)\,m_s
\notag\\
   &=1.862(41)\times(
   0.03524\pm
   0.00026),
\end{align}
for $s$ quark. Then the running masses
$a\Bar{m}_{ud}\!\left(1/\sqrt{8t}\right)$ and~$a\Bar{m}_s\!\left(1/\sqrt{8t}\right)$
are given by Eq.~\eqref{eq:(B.2)}.

\section{Additional tests on the energy-momentum tensor}
\label{sec:tests}

In this Appendix, we summarize our additional tests on our results of the energy-momentum tensor 
discussed in Sec.~\ref{sec:EMTresults}.

\subsection{Off diagonal components}
\label{sec:tests:od}

In order to check validity of the formulation, we calculate off diagonal
components of the energy-momentum tensor. In Fig.~\ref{fig9} and~\ref{fig10},
we plot the off diagonal components $T_{i4}/T^4$ and $T_{i\neq j}/T^4$,  
which correspond to the momentum and stress density, respectively, as
functions of $t/a^2$. 

We first confirm that the data are consistent with zero within $2\sigma$ in
the fit windows adopted in Sec.~\ref{sec:t0extrapolation}.
By identifying windows for the linear fit for each data, 
we find that the $t\to0$ extrapolation sometimes leads to a value which is slightly off the 0.
However, because the tendency as a function of $T$ is not uniform, we consider that 
this is caused by an accidental fluctuation due to insufficient statistics or an optimistic error estimation disregarding the correlation in $t/a^2$.

\subsection{Gauge and quark contributions}
\label{sec:tests:gq}

One may interested in how the gauge and quark operators contribute to the energy-momentum tensor quantitatively.
In Figs.~\ref{fig:ep-each} and \ref{fig:e3p-each}, the entropy density and trace anomaly are plotted as a function of the flow time $t/a^2$, respectively.
In these figures, we break up contributions from gauge operators (\ref{eq:(2.9)}) and (\ref{eq:(2.10)}) and
those from quark operators (\ref{eq:(2.11)}), (\ref{eq:(2.12)}) and (\ref{eq:(2.13)}) from $ud$ and $s$ quarks, 
where the $ud$ quark contribution is a mixture of those from $u$ and $d$ quark.

In general, the magnitude of contributions from the gauge and each quarks are almost the same for the energy and entropy density.
The trace anomaly is dominated by the gauge contribution at $T\simeq200$-$350\,\mathrm{MeV}$.
On the other hand, the quarks dominate at higher temperatures, which however is suspected to be contaminated by a lattice artifact.
For the pressure, gauge and quark contributions are similar in magnitude but opposite in sign at low temperatures.
At high temperatures, the quark contributions dominates but is also suspected to be contaminated by a lattice artifact.

We note that the singular behavior $a^{2}/t$ in the equation of state close to the origin comes dominantly from the quark operators.

\subsection{Lattice operators for the field strength}
\label{sec:tests:op}

For the quadratic terms of the field strength tensor $G_{\mu\nu}(x)$ in Eqs.~\eqref{eq:(2.9)} and \eqref{eq:(2.10)}, 
there are several alternative choices of lattice operators. 
In this study, we construct clover operators with four plaquette Wilson loops 
and with eight $1\times2$ rectangle Wilson loops. 
Combining these two clover operators, we get
the tree-level improved field strength squared~\cite{AliKhan:2001ym}. 

We also test a definition using the imaginary part of a plaquette Wilson loop. 
In summary, we study the following four
alternatives~\cite{Alexandrou:2015yba}:
\begin{itemize}
\item[(i)] the tree-level improved operator given by combining two clover term
contributions with four plaquette and eight $1\times2$ rectangle Wilson loops,
\item[(ii)] the clover term with four plaquette Wilson loops,
\item[(iii)] the clover term with eight $1\times2$ rectangle Wilson loops,
\item[(iv)] the imaginary part of the plaquette Wilson loop.
\end{itemize}
We adopt the first combination for the central value of our estimations
and estimate a part of the $O(a^2)$ lattice artifacts in the gauge operator 
by comparing the results of four alternatives.

In Figs.~\ref{fig:eplusperra}
and~\ref{fig:eminus3perra}, we plot the entropy
density and trace anomaly as functions of temperature by changing the
operator for the field strength squared. 
We confirm that the results are consistent
with each other, while that with the clover term with eight $1\times2$
rectangles shows slight deviation. 
This may be because the $O(a^2)$ lattice artifact is severer for that definition. 
Disregarding the data at $T\simeq697\,\mathrm{MeV}$ ($N_t=4$), we conclude that the systematic error from the choice of the operators for the gauge contribution is small.


\end{document}